\documentclass{article}
\usepackage{graphicx}
\def\ligne#1{\hbox to\hsize{#1}}
\def\leurre{\noindent\leftskip0pt\small\baselineskip 10pt}
\newtheorem{thm}{\textbf{Theorem}}
\newtheorem{fig}{\textbf{Figure}}
\newtheorem{tab}{\textbf{Table}}
\def\boxempty{\hbox{\vbox{\hsize=7pt\offinterlineskip
\ligne{
\vrule height 7pt depth 0pt width 0.6pt
\vbox to 7pt{\hsize=5.8pt
\hrule height 0pt depth 0.6pt width 5.8pt
\vfill
\hrule height 0.6pt depth 0pt width 5.8pt
}\hskip-0.5pt
\vrule height 7pt depth 0pt width 0.6pt
}}
}}

\author{Maurice {\sc Margenstern}}
\title{A weakly universal cellular automaton on the pentagrid with three states}
\begin{document}
\maketitle

\begin{abstract}
In this paper, we prove that there is a weakly universal cellular automaton on the pentagrid 
with three states which is rotation invariant and which uses \textit{\`a la Moore} neighbourhood.
Moreover, at each step of the computation, the set of non quiescent states has always
infinitely many cycles. 
\end{abstract}

\section{Introduction}

   The first paper about a universal cellular automaton in the pentagrid, the tessellation
$\{5,4\}$ of the hyperbolic plane, was \cite{fhmmTCS}. This cellular automaton was also rotation
invariant, at each step of the computation, the set of non quiescent states had infinitely many
cycles: we shall say that it is a truly planar cellular automaton. But, the neighbourhood
was \textit{\`a la von Neumann} and it had 22~states. 
This result was improved by a cellular automaton with 9~states in~\cite{mmysPPL}.
Recently, I improved this result with 5~states, see~\cite{mmpenta5st}. A bit later,
I proved that in the heptagrid, the tessellation $\{7,3\}$ of the hyperbolic plane,
there is a weakly universal cellular automaton with three states which is rotation invariant
and which is truly planar, \cite{mmhepta3st}. In a paper under printing, see~\cite{mm11.3.2st}, 
I proved that there is such a cellular automaton in the tessellation $\{11,3\}$ of the
hyperbolic plane.

   In the heptagrid and in the tessellation $\{11,3\}$, there is no difference between
a von Neumann neighbourhood and a Moore neighbourhood. In both case, a cell has 7~immediate
neighbours in the heptagrid and 11~of them in the tessellation $\{11,3\}$: a neighbour
which shares a vertex also shares an edge. The reason lies in the interior angle 
of the basic polygons: it is $\displaystyle{{2\pi}\over3}$ in both cases. In the pentagrid,
the situation is different because the interior angle is a right angle. In the pentagrid,
a cell has five neighbours which share an edge with it, and it also have five new neighbours
which only share a vertex with it. In the Euclidean plane, there is a difference between
cellular automaton with von Neumann or Moore neighbourhoods. Remember that the game of life
cellular automaton requires Moore neighbourhoods.

    And so the question was, to which point is it possible to reduce the number of states
for a weakly universal, rotation invariant and truly planar cellular automaton in the pentagrid
if we allow Moore neighbourhoods? This time we have ten neighbours, more than the seven neighbours
in the heptagrid, but less than the eleven ones in the tessellation $\{11,3\}$. But we have
the inconvenience of the right angle and we also have a slight advantage: we have
ten neighbours but only five admissible rotations. Indeed, rotating the sides induces a similar
rotation on the vertices and conversely. 

    In fact, I could prove the following result, which is stated in the abstract:

\begin{thm}\label{letheo}
There is a weakly universal cellular automaton on the pentagrid with Moore neighbourhoods
which is rotation invariant, truly planar and which has three states.
\end{thm}

   The paper is devoted to the proof of the theorem. The minimal introduction to hyperbolic 
geometry needed to understand the paper can be found in~\cite{mmbook3}. In order to
understand the constructions of the paper, it is important to remember that
in the pentagrid, there are lines of the hyperbolic plane that we call \textbf{lines of the
pentagrid} because they support the sides of infinitely many pentagons which pairwise share
a side which is orthogonal to that line. Such a situation is trivial in the Euclidean plane
with a grid made of squares. The difference, here is that two lines of the pentagrid
may be both perpendicular to the same side of a pentagon, but they are not parallel at all: they
go in different directions, on both sides of the considered edge. We shall use the fact
that the tiling is generated by a tree, see~\cite{mmbook1,mmbook3}, without further explanation.

In Section~\ref{hyprailway}, we sketchily remind the reader the model which is simulated by our
cellular automaton. In Section~\ref{scenar}, we describe how we implement the model on order
to observe the constraints of the three states. In Section~\ref{rules}, we give the
rules of the automaton, together with the figures illustrating the motion of the
device performing the computation of the automaton. That section will complete the proof
of Theorem~\ref{letheo}.

\section{Hyperbolic railway traffic}
\label{hyprailway}

     As the weakly universal cellular automata in the pentagrid described in the papers~\cite{fhmmTCS,mmysPPL,mmpenta5st}, the automaton which proves Theorem~\ref{letheo} simulates the same model of computation. We describe this model
in Sub-section~\ref{railway} and we consider its implementation in the pentagrid 
in~\ref{implement}.

\subsection{The railway model}
\label{railway}

   The railway model was introduced by Ian Stewart in~\cite{stewart}. It was initially devised
to simulate a Turing machine, but I used it to simulate a register machine. For technical
reasons, that latter model is easier to simulate as it is known that two registers are enough
to simulate any Turing machine on $\{0,1\}$, see~\cite{minsky}.

    The model consists of a railway circuit on which a single locomotive is running. The circuit
contains crosses and switches and the state of all switches of the circuit considered at
a given top of the clock constitutes the configuration of the circuit at that time. The circuit 
consists of tracks which are represented by assembling segments of straight lines, either 
horizontal or vertical ones and quarters of a circle. It contains crosses which allows 
tracks to intersect. It also contain switches in order to make suitable figures, in particular
cycles. There are three kinds of switches illustrated by Figure~\ref{switches}. From left to right
in the figure, we can see the \textbf{fixed switch}, the \textbf{flip-flop} and the 
\textbf{memory switch}. We can see that
a switch realizes the intersection of three tracks, we shall denote them by~$a$, $b$ and~$c$.
Say that~$a$ is the single track which arrives at the switch and that $b$ and~$c$ are
those from which the locomotive may leave the switch, either through~$b$ or through~$c$.
The track taken by the locomotive to leave the switch is called the \textbf{selected track}.
When the locomotive arrives to the track through~$a$ and leaves it through the selected track,
we say that it is an \textbf{active} crossing. If the locomotive arrives through~$b$ or through~$c$,
leaving the switch through~$a$, we say that it is a \textbf{passive} crossing. In the figure,
each kind of switch is represented in two forms. Both of them will be used in our further
illustrations.
\vskip 10pt
\vtop{
\ligne{\hfill
\includegraphics[scale=0.8]{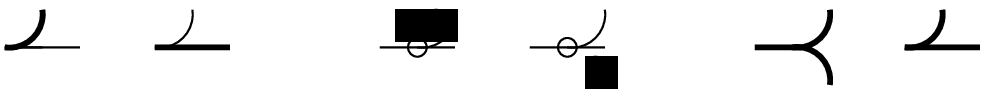}
\hfill}
%\vspace{-20pt}
\begin{fig}\label{switches}
\leurre
The switches used in the railway circuit of the model.
\end{fig}
}

   In the fixed switch, the selected track is determined once and for all. The switch can be
crossed both actively or passively. The flip-flop may be crossed actively only. When
it leaves the switch, the selected track changes: the previous non-selected track becomes
selected and the previously selected becomes non-selected. The memory switch can also be crossed
either actively or passively. The selected track is determined by the last passive crossing.
These switches are exactly those defined in~\cite{stewart}. We also take from 
paper~\cite{stewart} the configuration illustrated by Figure~\ref{basicelem} which contains 
one bit of information.

\vtop{
\ligne{\hfill
\includegraphics[scale=0.6]{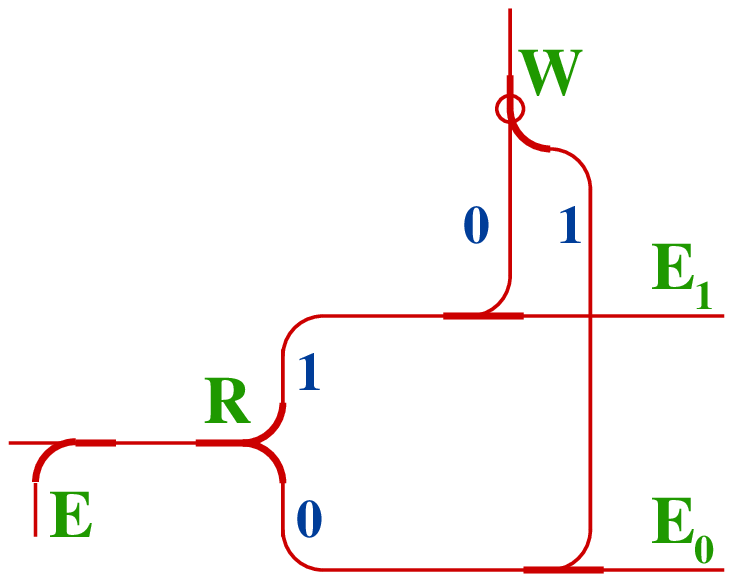}
\hfill}
%\vspace{-20pt}
\begin{fig}\label{basicelem}
\leurre
The basic element containing one bit of information.
\end{fig}
}

This configuration, called the \textbf{basic element}, is used as follows. If the locomotive 
arrives through~$R$, it goes either through track~0 or through track~1. If the locomotive goes 
through the track~$\alpha$ with 
$\alpha\in\{0,1\}$, it leaves the configuration through~$E_\alpha$. We say that the locomotive
\textbf{has read~$\alpha$}. If the locomotive arrives through~$W$, it goes through track~0 or~1.
Now, if it arrives through the track~$\alpha$, it passively crosses~$R$ through the track~$\beta$,
with $\beta\in\{0,1\}$ and \hbox{$\alpha+\beta=1$}. Eventually, the locomotive leaves the element 
through~$E$. Now, as the locomotive crossed~$W$,
the selected track at~$W$ is now~$\beta$. Consequently, we can say that the selected tracks 
at~$R$ and~$W$ are always the same when the locomotive is arriving at the configuration and
also when it leaves it. Accordingly, the locomotive may read~0 or~1 and also it may rewrite~$\alpha$
into~$\beta$.

   Figure~\ref{unitreg} illustrates how basic elements can be assembled in order to constitute
a unit for registers, say a \textbf{register unit}. Each unit consists of two basic elements 
which contain the same value, see Figure~\ref{unitreg}. The leftmost element 
is read by an instruction which perform the incrementation of the register, the other is read by 
an instruction performing the decrementing on it. We leave the
reader to check on Figure~\ref{unitreg} that a locomotive arriving from the left through the 
track~$i$ reads the left-hand side basic element: if it reads~1, it goes back to~$i$ a bit 
further in order to read the next register unit. When it reads~0, it is sent to the $W$-entry 
of the basic element in order to rewrite~0 to~1, it is sent to the second element in order to 
set it to~1 too. The number of units set to~1 has been increased by~1. From there it is sent
to the track~$r$ which leads the locomotive out of the register.

\vtop{
\vspace{-15pt}
\ligne{\hfill
\includegraphics[scale=0.6]{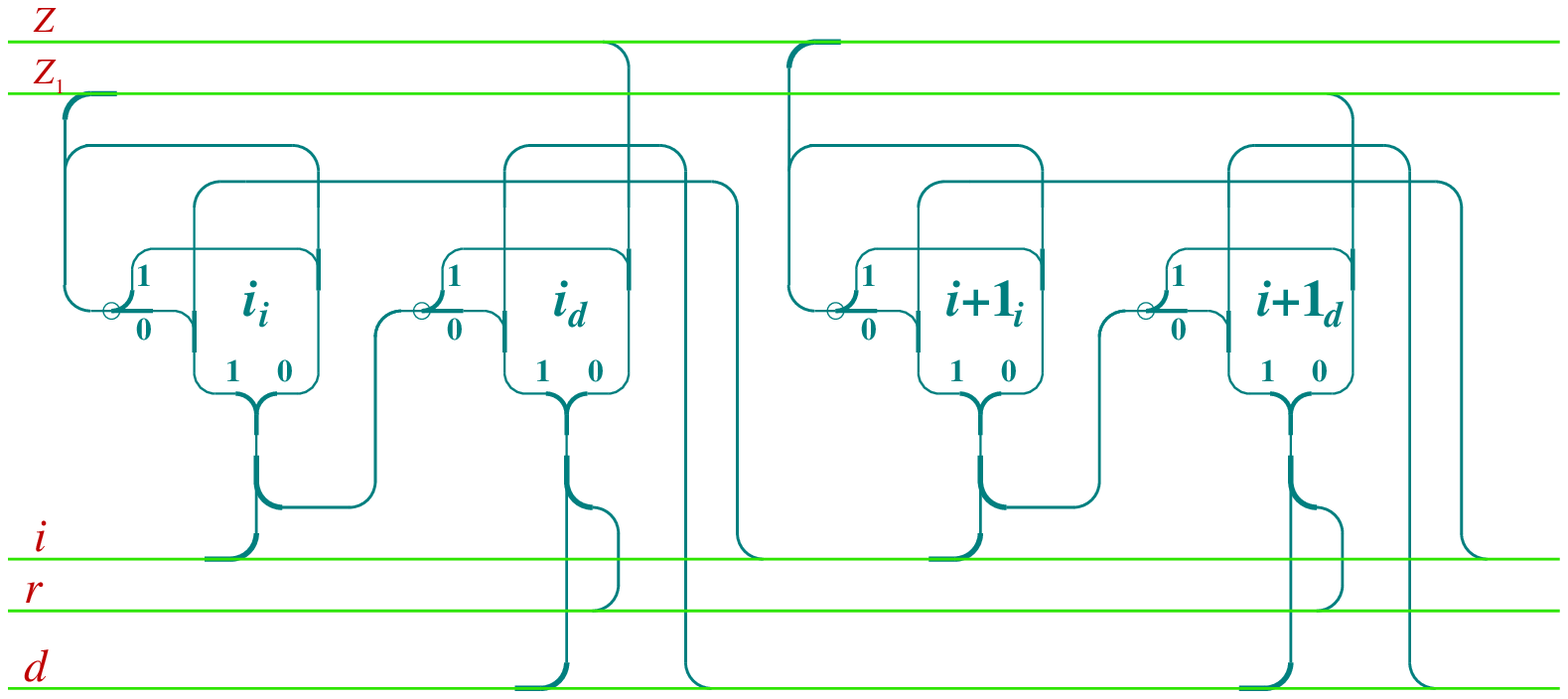}
\hfill}
\vspace{5pt}
\begin{fig}\label{unitreg}
\leurre
Two consecutive units of a register.
\end{fig}
}

     The reader can also check that if the locomotive arrives through the track~$d$, it
reads the second basic element of the register unit. If it reads~1, as previously, it goes further,
again through~$r$. If it reads~0, it goes back to the previous register unit through the track~$z$
or through the track~$z_1$. If this happens with the first unit of the register, this
means that the register contains~0 and the locomotive goes back from the register through~$z$.
If this happens on a further register unit, the locomotive goes back to the previous register
unit, through~$z$ or through~$z_1$~: it is sent to the first basic element, so that there it 
rewrites~1 to~0 and it performs the same for the second basic element of the unit. The number
of units set to~1 has been lowered by~1. The locomotive is send outs side the register
through the track~$r$.

    When the locomotive comes back from the register where it performed the required
operation, an appropriate structure allows it go back to the instruction which follows 
that which had sent it to the register, or to jump to the next one in the program
if the decrementing could not be performed. The details of the structures allowing the
locomotive to perform these tasks can be found in~\cite{mmbook3}.

As explained in~\cite{mmbook3}, using the basic unit illustrated by Figure~\ref{basicelem} 
and crossing with switches as in Figure~\ref{switches}, it is possible to build a circuit which
simulates the working of a register machine. Figure~\ref{toyexample} illustrates how such
a construction can be performed by implementing in this setting a small program of a register
machine involving three registers. The program performs the action denoted by the instruction
\hbox{$X := Y;$} of many programming languages. At any time, a configuration of the computation
consists in the position of all the switches of the circuit. However, note that at each time,
only finitely many of them are changed so that at each time, the computation can be described
with a finite word, despite the fact that the initial configuration is infinite. 

\vtop{
\ligne{\hfill
\includegraphics[scale=0.45]{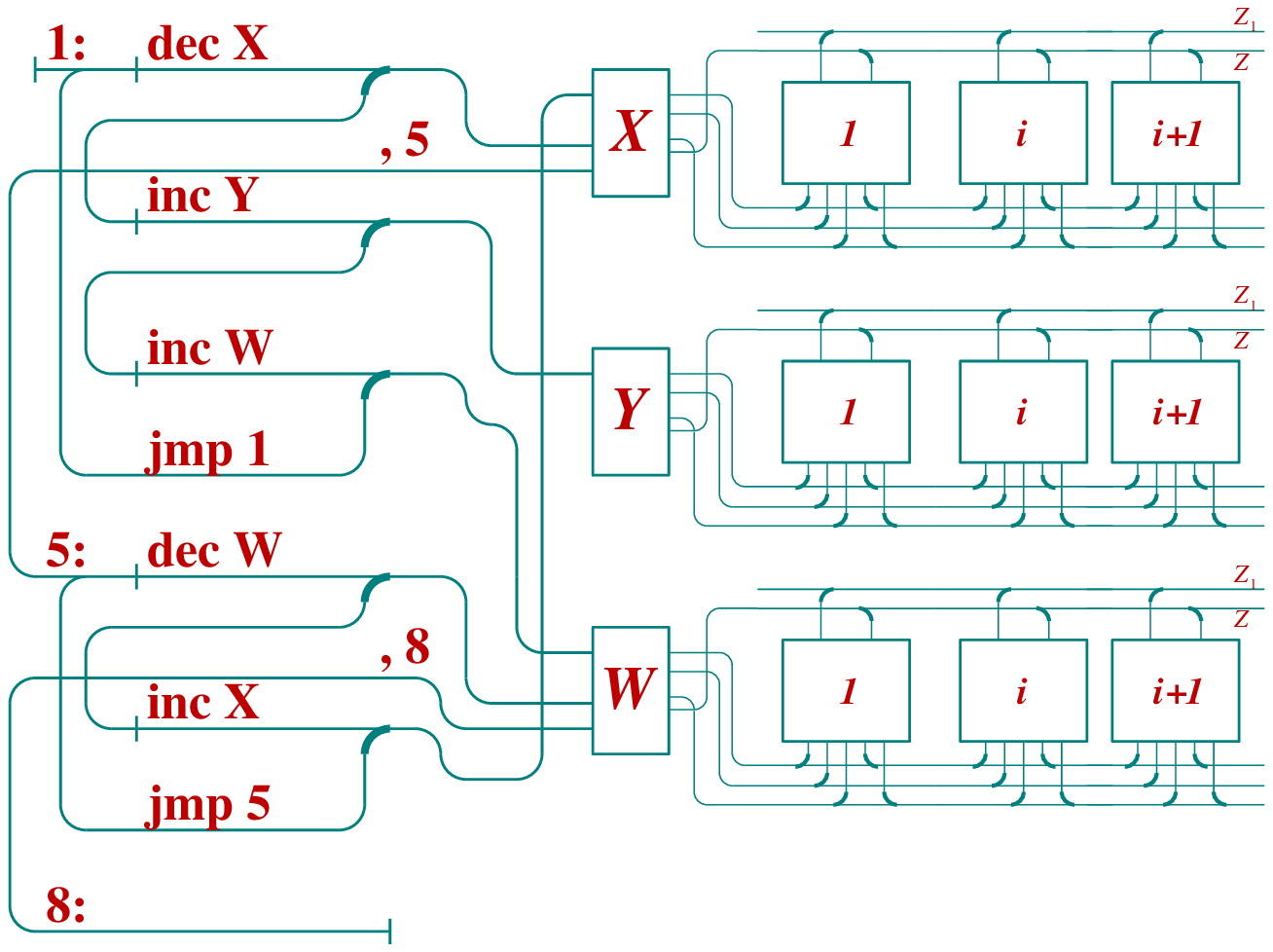}
\hfill}
%\vspace{-20pt}
\begin{fig}\label{toyexample}
\leurre
Implementation of a toy-register machine associated to a toy-program.
\end{fig}
}

\subsection{Hyperbolic implementation of the model}
\label{implement}

    The model of computation described in Sub-section~\ref{railway} is described in~\cite{stewart} as a circuit lying in the Euclidean plane. It cannot been transported into
the hyperbolic plane without detailed explanations. There are two steps in this translation. First, we need to have a global view of how to implement the circuitry which will simulate
the computation. This is what is discussed in Sub-section~\ref{global}. Once this question
is solved, we can look at the details which is the purpose of Sub-section~\ref{basics}.

\subsection{Global view}
\label{global}

    In order to fix things, we consider the Poincar\'e's disc model as the frame of our
investigation. We take this model for two reasons. The first one is that it preserves angles, 
which allow us to see something. The second reason is that the model gives us a global view on
the hyperbolic plane. However, we must be aware of two big restrictions. The first restriction
is that the model imposes a distortion on distances: when we are close to the centre of the 
disc, the distances are small and the closer we go to the border, the bigger are the distances.
Also, second restriction, there is no similarity in the hyperbolic geometry. There, distances
are absolute. This means that if we change our unit to measure distances by another one,
the new distances are not simply multiplied by a simple factor. 

    A good image is the following one. We have to see Poincar\'e's disc as a window over the
hyperbolic plane, as if we were flying over that plane in some spacecraft. Accordingly, the 
centre of Poincar\'e's disc is not the centre of the hyperbolic plane: that plane has no 
centre. The centre of the disc is the centre of our window in the spacecraft: it is the 
point of the plane on which our attention is focusing. %at which we are looking. 
But what is close to the border is not affected by changing the point at which we are looking. If
we change of unit, this amounts to change our altitude over the hyperbolic plane in our image.
We shall have a larger view close to the centre but what is close to the border still remains
out of our view. 

   Figure~\ref{uhcaglobal} shows how to implement the toy-example given in 
Figure~\ref{toyexample}.  
    
\vtop{
\ligne{\hfill
\includegraphics[scale=0.2]{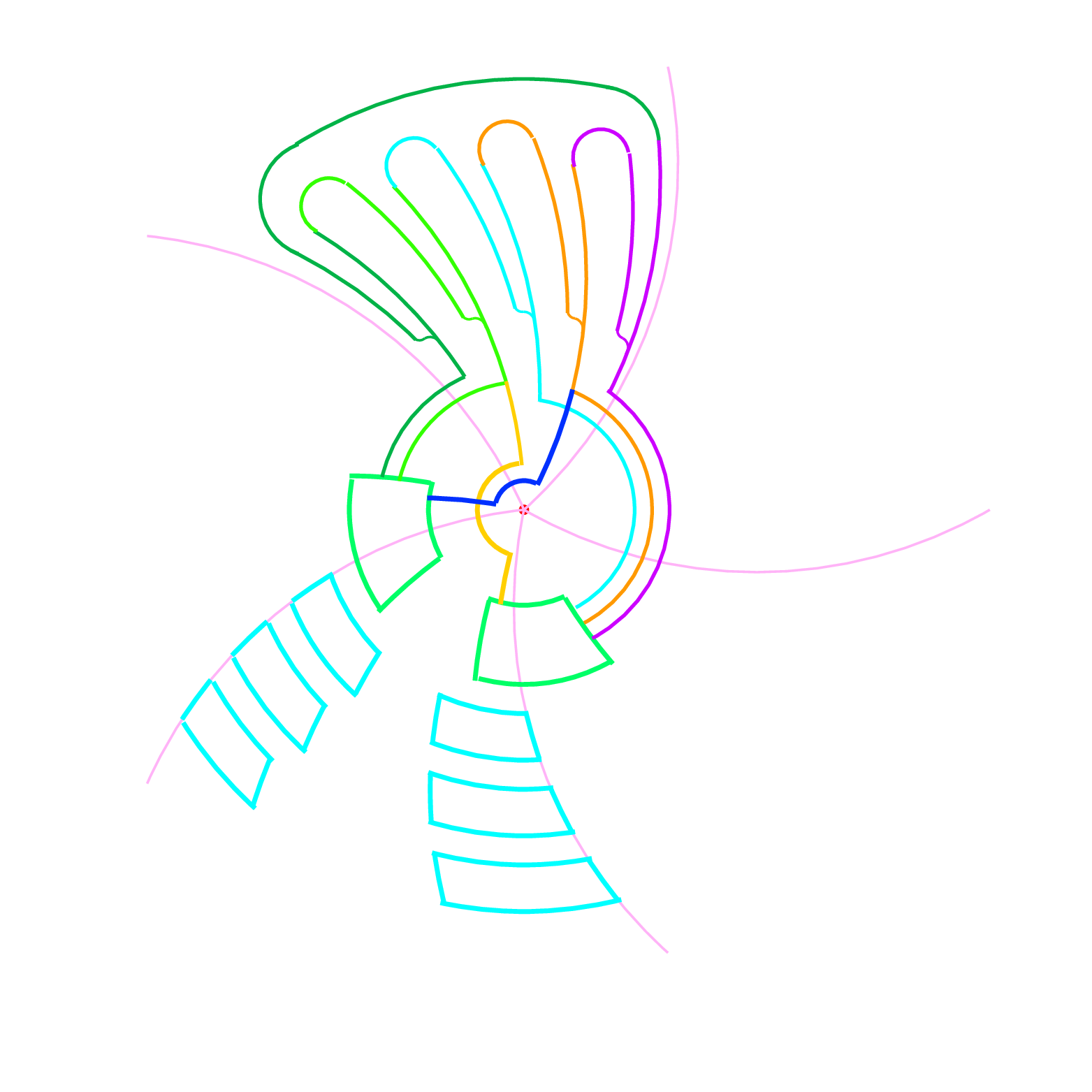}
\hfill}
\vspace{-20pt}
\begin{fig}\label{uhcaglobal}
\leurre
Implementation of the toy-example from Figure~{\rm\ref{toyexample}} in the pentagrid.
\end{fig}
}

The vertical lines of Figure~\ref{toyexample} are replaced by rays issued from the centre of 
the disc and horizontal lines of that figure are replaced by arcs of circles centred at the 
centre of Poincar\'e's disc.

In a similar way, we can implement the basic element on which all the other structures are
built, see Figure~\ref{hca54elem}. 

   However, our implementation is not directly performed in the hyperbolic plane, but in
a tiling of this plane, more precisely the \textbf{pentagrid}, the tessellation $\{5,4\}$
of the hyperbolic plane which is illustrated by the left-hand side part
of Figure~\ref{penta}. 

\vtop{
\ligne{\hfill
\includegraphics[scale=0.3]{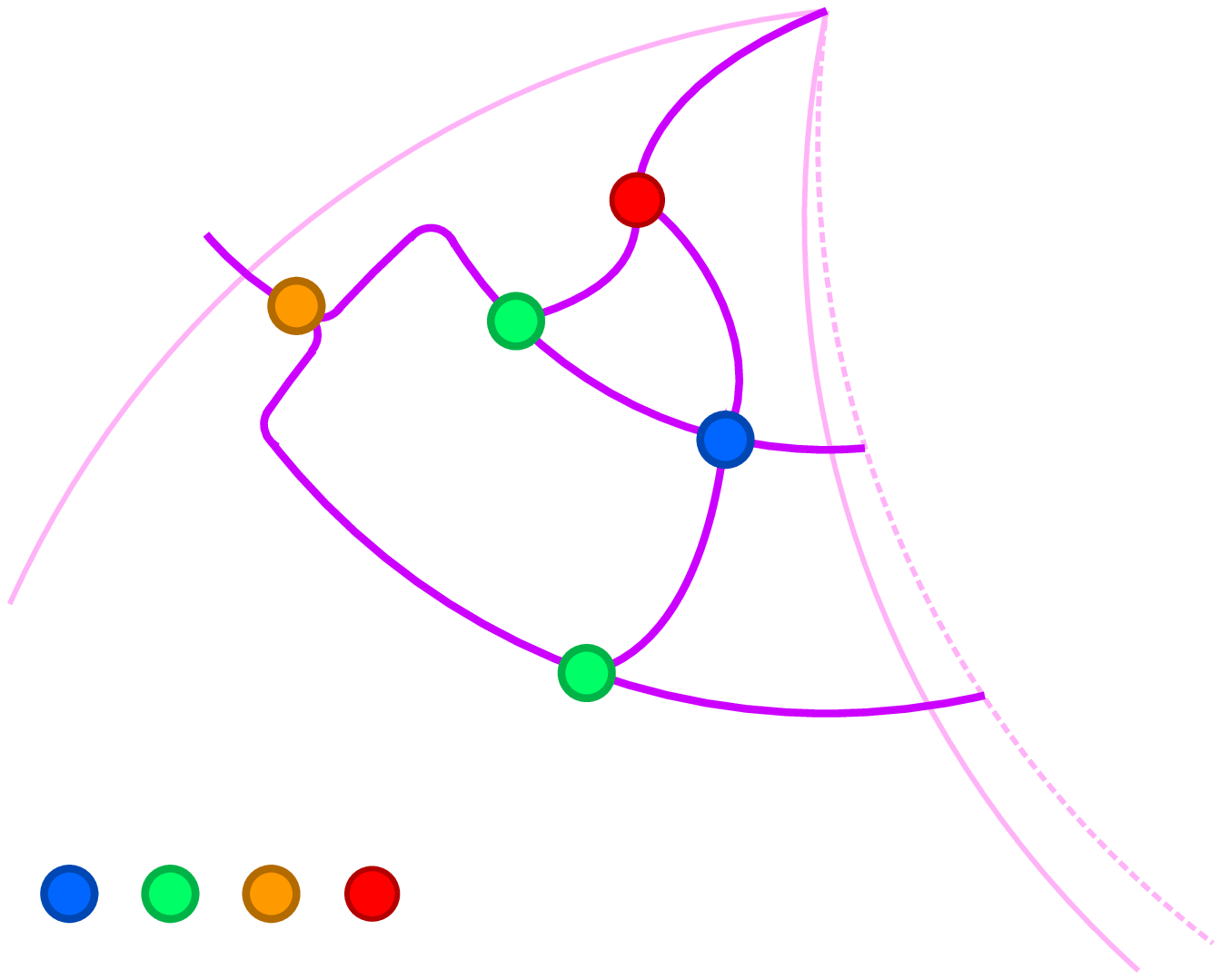}
\hfill}
\vspace{-5pt}
\begin{fig}\label{hca54elem}
\leurre
Implementation of the basic element of Figure~{\rm\ref{basicelem}} in the 
pentagrid.
\end{fig}
}

It is important to know that the tiling can be generated by a tree. From the tree, we can get 
a numbering of the tiles which is illustrated by Figure~\ref{penta}, on its right-hand side part.
On Figure~\ref{penta}, we can see the \textbf{lines of the pentagrid}: they are the lines
constituted by sides of tiles which lie on them, taking into account that, in Poincar\'e's
disc model, lines are represented by traces in the disc of diameters or circles which are
orthogonal to the border of the disc. A particular tree plays a key role. It generates the
tiles which are inscribed in a quarter of the hyperbolic plane. There are five such quarters
around the central tile of Figures~\ref{penta}. Such a quarter can be seen on the right-hand part
of the figure: it is defined by two lines of the pentagrid which meet at a vertex of tile~1
and which contain sides of tile~2 and of tile~4. We shall use the tiles which are along such lines
as, for instance tiles~1, 2, 5 and~13 or tiles~2, 7 and~20. Such sequences of tiles will be
our verticals for the implementation of the railway model we consider in the paper. The analog 
of circles lie on tiles which are at the same distance from the central tile. In 
Sub-section~\ref{trackrules} we have a detailed look at this problem.
\vskip 10pt
\vtop{
\ligne{\hfill
\includegraphics[scale=0.6]{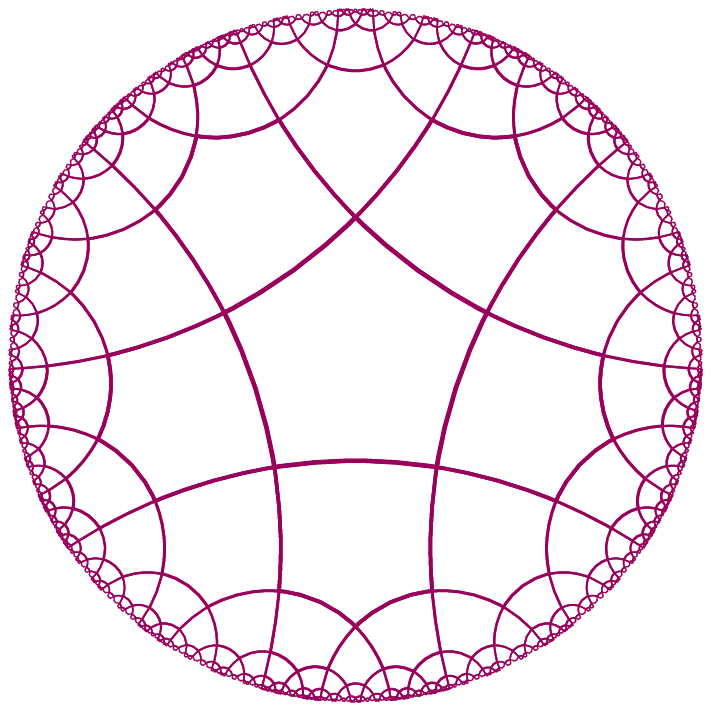}
\input
\includegraphics[scale=0.6]{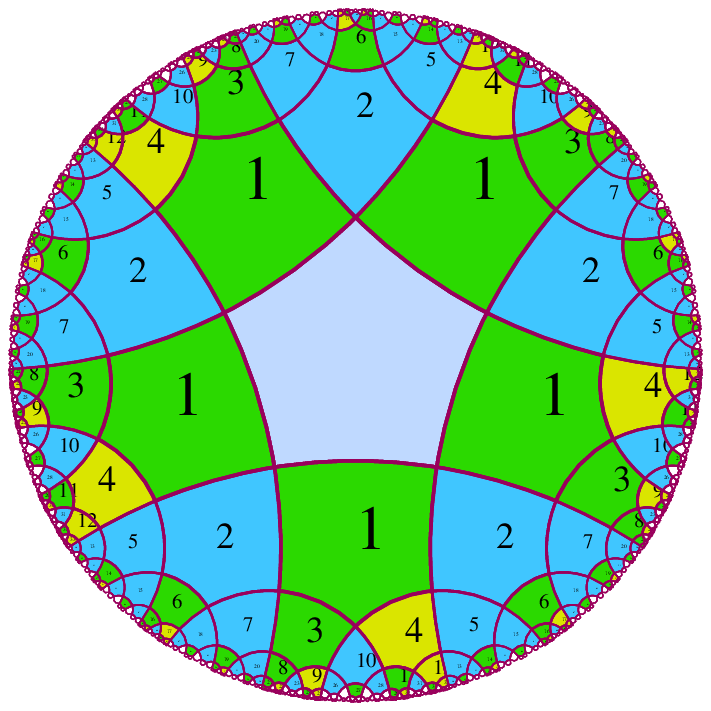}
\hfill}
\vspace{-5pt}
\begin{fig}\label{penta}
\leurre
To left: the pentagrid. To right: numbering the tiles.
\end{fig}
}

\subsection{Basic principles}
\label{basics}

    Among the details we have now to look at, the first task in the 
implementation is that of the tracks. The main reason is that without tracks, our locomotive can go
nowhere, so that there cannot be any simulation. The second reason is that it may happen that we 
can construct the switches of the circuit with say $k$~states but that we need a $k$+1$^{\rm th}$ 
one to build the tracks. Subsection~\ref{tracks} is devoted to that problem. Once this problem
is temporarily solved, we shall see the main features of the crossings and of each kind
of switches in Sub-sections~\ref{roundabout}, \ref{fix}, \ref{doublesel} and~\ref{forcontrol}.

\subsection{The tracks}
\label{tracks}

What we said about the tracks tells us that we should define them as simple as possible,
\textit{i.e.} with a single state outside the quiescent one which we called white, as already
mentioned.  As in~\cite{mmpenta5st}, we implement the tracks with \textbf{milestones} which
are cells in a non-quiescent state, the locomotive going through the white cells delimited
by the milestones. We shall call \textbf{blue} the state we use in order to mark the tracks.
Contrarily to what is done in~\cite{mmpenta5st}, here we can construct the tracks with
a single element which is illustrated by Figure~\ref{elemtrack}.
   
\vtop{
\ligne{\hfill
\includegraphics[scale=1.2]{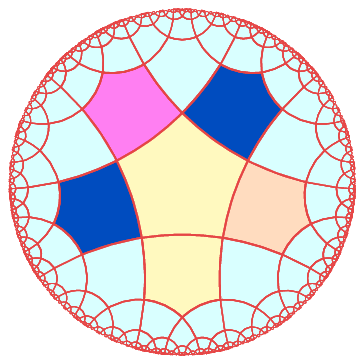}
\hfill}
\vspace{-20pt}
\begin{fig}\label{elemtrack}
\leurre
Element of the tracks.
\end{fig}
}

Assuming that the milestones are touching the cell of the track through its sides~2 and~5
in Figure~\ref{elemtrack}, the locomotive crosses the element by visiting the cells~3, 0, 1
or~4, 0, 1{} in this order exactly. The cells 0, 1, 3 and~4 are given light colours in 
Figure~\ref{elemtrack} in order to easily identify them. The mauve colour is that of the exit.
The side of the exit is the single side which belongs to the cell~0 and which touches both 
milestones.

   Later, in Sub-section~\ref{trackrules}, we shall assemble copies of the element illustrated
by Figure~\ref{elemtrack} in order to define the verticals and the horizontal required by our 
implementation of the model.

\section{The scenario of the simulation}
\label{scenar}

    With only three states at my disposal, I do not know how to directly 
implement switches where the needed sensors and controls are immediate neighbours of the 
cell where the tree tracks arriving at the switch meet. The simplest solution for that 
purpose is to devote a 
state for the tracks, different from the blank and then to give a series of states for 
each type of switch and for each step required for the crossing of the switch by the 
locomotive. This may require more states than the 22 states of the automaton 
in~\cite{fhmmTCS}.
 
   We go on the idea of the previous Sub-section where tracks are blank cells marked
by appropriate milestones. We even go further, following the new way opened  
in~\cite{mm13_3}. % {13,3}
The analysis of previous implementations showed me that the crossing is the most expensive
structure. This is why in~\cite{mm13_3} crossings are replaced by a new structure, the
round-about, see Sub-section~\ref{roundabout}. Later, in~\cite{mmhepta3st}, we decomposed
two switches in simpler structures, extending that new implementation to the passive memory
switch in~\cite{mmpenta5st}, see Sub-section~\ref{forcontrol}.  
 
\subsection{The round-about}
\label{roundabout}

    The round-about replaces the crossing, a railway structure, by a structure inspired by
road traffic. At a round-bout where two roads are crossing, if you want to keep the 
direction arriving at the round-about, you need to leave the round-about at the second 
road. Figure~\ref{roundaboutfig} illustrates this features.

\vtop{
\ligne{\hfill
\includegraphics[scale=0.6]{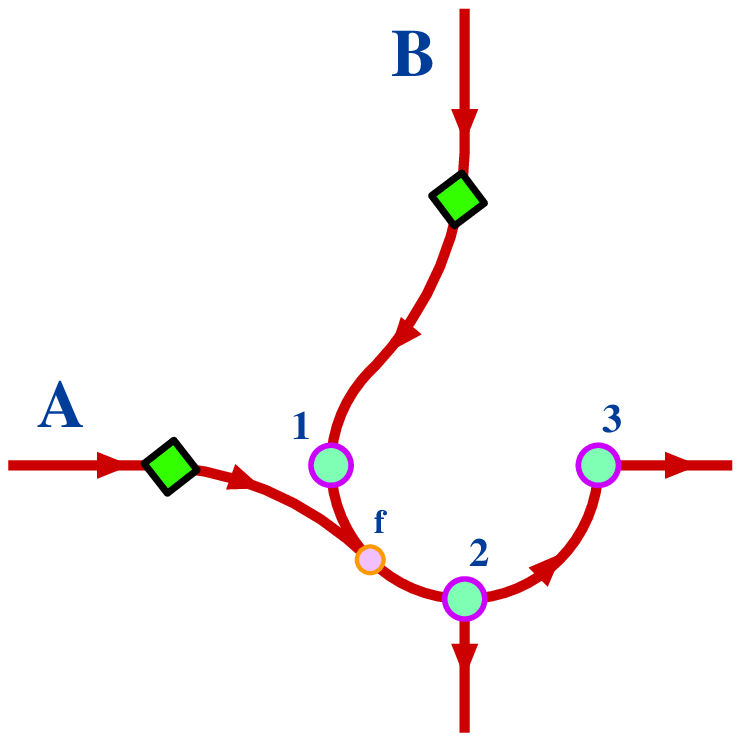}
\hfill}
\vspace{-5pt}
\begin{fig}\label{roundaboutfig}
\leurre
Implementation scheme for the round-about.
\end{fig}
}

Now, this strategy requires the cellular automaton knows how to count up to two. We already
assigned the blank for the space itself and the cells where the locomotive runs. We assigned
the blue to the milestones delimiting the tracks. It will also be used to delimit other
structure. I needed a third state, the red, to mark the locomotive itself which is reduced to
a single cell. This is the main reason why the tracks are one-way: by itself, the locomotive 
cannot distinguish between its front and its rear. As we have no additional state and as 
I could not handle a blue locomotive, counting to two requires that two contiguous 
locomotives travel together. In the sequel, we shall say \textbf{simple locomotive}
when the locomotive consists of a single red cell. We shall say \textbf{double locomotive}
when it consists of two contiguous red cells. 
To manage these features, we need the three auxiliary
structures which are represented in Figure~\ref{roundaboutfig} by a rhombus, a small circle
and a bit larger ones.

   These auxiliary structures are~: the fixed switch,
see Sub-subsection~\ref{fix}, the doubler and the selector, see Sub-subsection{doublesel}. 
When the locomotive arrives close to the round-about, it first meet the doubler: it 
transforms the simple locomotive a double one. Then, the locomotive may arrive at the 
fixed switch, depending on whether it arrived from~\textbf{A} or from\textbf{B}. Then,
the double locomotive arrives at the first selector: the structure recognizes a double 
locomotive. It kills one red cell and the surviving simple locomotive is sent further
on the round-about. When it meets the second selector, the structure recognizes a simple
locomotive. Accordingly, it send it on the track which leaves the round-about at that point.

\subsubsection{The fixed switch}
\label{fix}

    As the tracks are one-way and as an active fixed switch always sends the locomotive in
the same direction, there is no need of the other direction: there is no active fixed switch.
Now, passive fixed switch are still needed as just seen in the previous paragraph. 
Figure~\ref{stabfix} illustrates the passive fixed switch when there is no locomotive around. 
We can see that it consists of elements of the tracks which are simply assembled in the 
appropriate way in order to drive the locomotive to the bottom direction in the picture, 
whatever upper side the locomotive arrived at the switch. 
From our description of the working of the round-about, a passive fixed switch must be 
crossed by a double locomotive as well as a simple one. 

\vtop{
\ligne{\hfill
\includegraphics[scale=1]{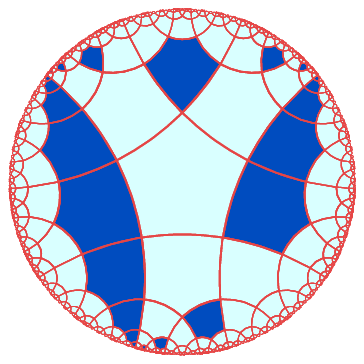}
\hfill}
\begin{fig}\label{stabfix}
The passive fixed switch in the pentagrid.
\end{fig}
}

   Later, in Section~\ref{fixedswitch}, we shall check that the structure illustrated by
Figure~\ref{stabfix} allows these crossings.

\subsubsection{The doubler and the selector}
\label{doublesel}

    The doubler  is illustrated by Figure~\ref{stab_doublsel}. Its structure again consists
of blue milestones only. However, the neighbourhood of the central cell~$C_0$ of the picture
is not that of an ordinary element of the tracks. In order to understand the difference,
we number the neighbours of a cell as follows: the sides are numbered from~1 to~5 while 
counter-clockwise turning around the cell and the vertices are also numbered while 
counter-clockwise turning around the cell, vertex~6 being shared by sides~1 and~5. 
The neighbours sharing a side are numbered after that side. The neighbours sharing a
vertex only are also numbered after that vertex. We also called these neighbours as
\textbf{vertex-neighbours}. In the 
figure, the neighbours~2, 4 and~5 of~$C_0$ are blue, which is enough to fix the numbering 
for~$C_0$. Now, $C_0$ has an additional blue milestone, its 10$^{\rm th}$ neighbour 
and its second neighbour has an additional neighbour which is not a milestone of the 
tracks. This allows the second neighbour to have a particular behaviour: that will 
produce the second cell which transforms the simple locomotive into a double one.
\vskip 10pt
\vtop{
\ligne{\hfill
\includegraphics[scale=1]{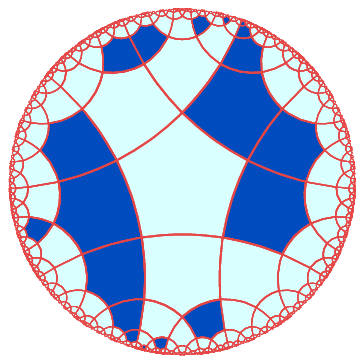}
\includegraphics[scale=1]{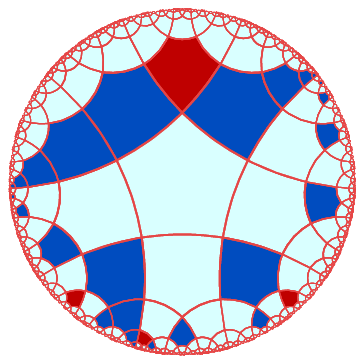}
\hfill}
\begin{fig}\label{stab_doublsel}
\leurre
To left: the doubler. To right: selector.
\end{fig}
}

   In the right-hand side of Figure~\ref{stab_doublsel}, we have an illustration of the
selector. The locomotive, a simple or double one, arrives through the neighbour~3 of~$C_0$,
again the central cell of that figure. It is plain that the neighbours~8 and~9 of~$C_0$
can see whether the locomotive is simple or double when a red cell arrives at~$C_0$ when it
is white. Note that these neighbours are not ordinary milestones of the tracks: they have 
an additional red neighbour as well as an additional blue one. When it is a simple 
locomotive, the neighbour~8 of~$C_0$ becomes red and the neighbour~9 remains blue. 
This prevents a locomotive from entering the neighbour~2 of~$C_0$ and the simple locomotive
enters the neighbour~4 of~$C_0$. When it is a double locomotive, the neighbour~8 of~$C_0$
remains blue and the neighbour~9 becomes red. Now, the red cell in~$C_0$ goes to its 
neighbour~2, going on the way which leads to the second selector, and the red neighbour~9  
prevents the red cell in~$C_0$ to enter its neighbour~4.

\subsection{The fork, the controller and the controller-sensor}
\label{forcontrol}

    In this Sub-section, we look at the decomposition of two active switches: the flip-flop 
and the active part of the memory switch. In both cases, we can split the working of the
switch by separating at two different stages, the bifurcation and the fact that the passage
in one direction is forbidden. It is not mandatory that both acts should happen 
simultaneously, as might be suggested by the definition of Sub-section~\ref{railway}. A fork
may dispatch two simple locomotives and, later, on one track, a controller let the locomotive
go on its way on the track and, on the other track, another controller stops the locomotive
and destroys it. The difference between the switches is the way in which the change of
selection is performed. Figure~\ref{activswitches} illustrates how to assemble forks
and controllers in order to obtain either a flip-flop, left-hand side of the figure, or
an active memory switch, right-hand side of the picture. The difference is important: the 
flip-flop immediately changes the selection once it was crossed by the locomotive. The active
memory switch changes the selection if and only if the passive switch ordered it to to do
so through a signal sent to the active switch. This is realized in 
Figure~\ref{activswitches}.

\vtop{
\ligne{\hfill
\includegraphics[scale=0.35]{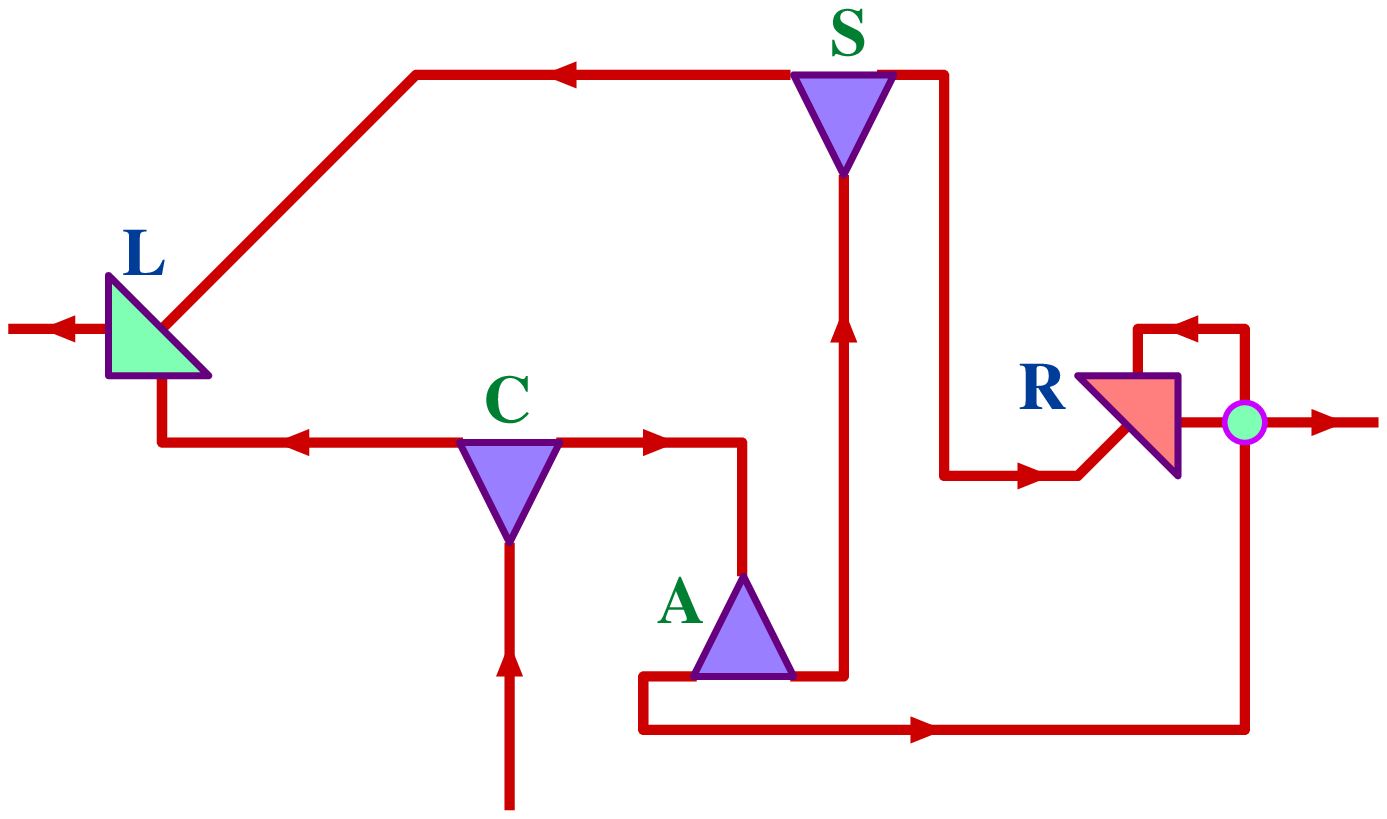}
\includegraphics[scale=0.35]{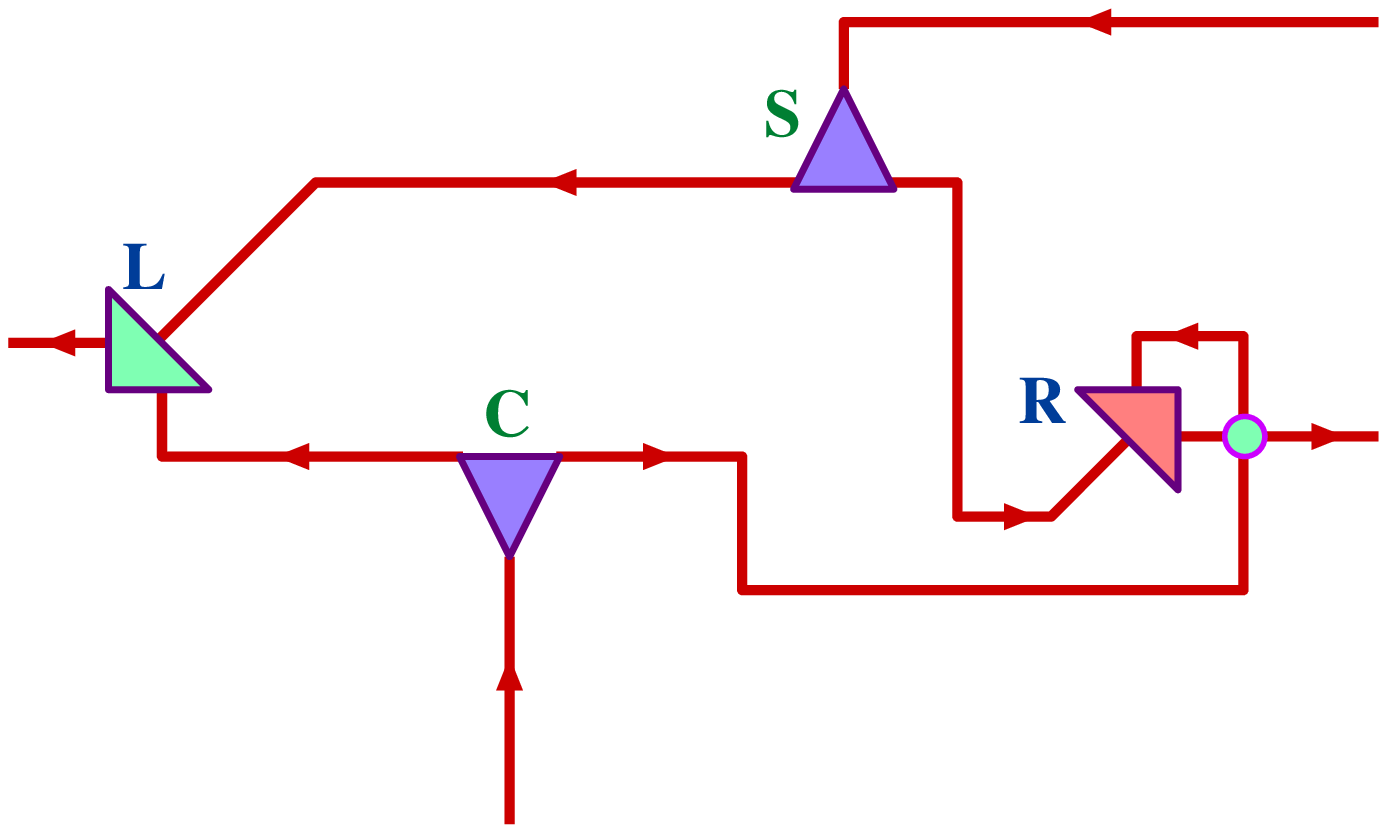}
\hfill}
\begin{fig}\label{activswitches}
\leurre
To left: the flip-flop. To right: the active memory switch.
\end{fig}
}

Indeed, in the active memory switch, when the simple locomotive arrives at the fork~$C$, 
it is duplicated into two simple locomotives, each one following its own path. One of these 
locomotives goes on its way to the controller, and the other is now a third locomotive which 
is sent to a second fork, labelled 
with~$A$ in the figure. When it is crossed by that locomotive, $A$~sends two locomotives~:
one to the other controller and one to a third fork~$S$. 
The controllers are blue and red. The blue one let the locomotive
go its way. The red controller stops the locomotive, which kills it. 
The locomotives sent by~$S$ go both to a controller, one to the blue controller, the other
to the blue one. Now, when a locomotive sent by~$S$ arrives at a controller, it changes the 
blue one into a red one and the red one into a blue one, so that what should be performed 
by a flip-flop is indeed performed. It is enough to manage things in such a way that the 
locomotives arriving to the controllers from~$S$ arrive later than those sent by~$C$ 
and by~$A$.

\vtop{
\ligne{\hfill
\includegraphics[scale=1]{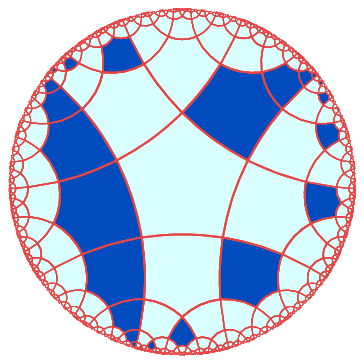}
\hskip 10pt
\includegraphics[scale=1]{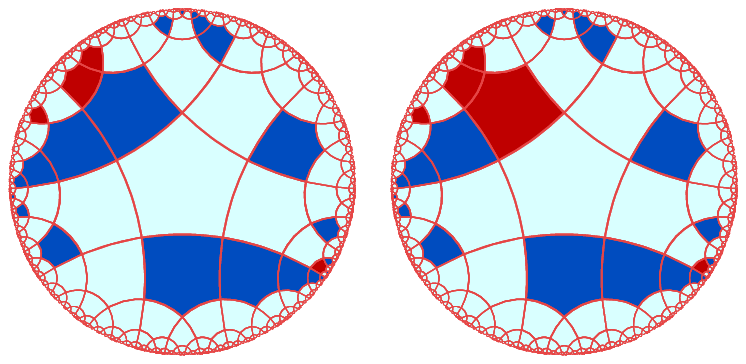}
\hfill}
\begin{fig}\label{stab_forkctrl}
\leurre
To left: the fork. To right, two configurations: the controller of the flip-flop and of the
active memory switch.
\end{fig}
}

In the case of an active memory switch, the situation is simpler, as shown by the 
right-hand side of Figure~\ref{activswitches}. From the for~$S$, the locomotives each
locomotive directly goes to the controller. But the change of signal in the controllers
is performed by a new locomotives sent by~$S$ which is triggered by a locomotive sent from
the passive memory switch. In this sense, the passive memory switch is active while the
active one is passive. 

Figure~\ref{stab_forkctrl} illustrates the configurations which implement the fork and the 
controller. The first picture represents the fork whose milestones are all blue. The second
picture is a blue controller while the third picture is a red one. The fork is implemented as
a mere appropriate assembling of elements of the tracks. Note the red neighbours of the
first neighbour of the central cell: they characterize the switch. The track goes 
'horizontally' from right to left on these pictures of the controller. The signal for changing
the colour of the control arrives from above. 

   Figure~\ref{memopass} illustrates the construction of the passive memory switch with
the help of forks and controllers. However, the controllers of Figure~\ref{memopass} are not
the same as those of Figure~\ref{activswitches}.  The difference lies in the different role
of the controllers in this case. As the switch is passive, even if the locomotive passes 
through the non-selected track, it should not be stopped by the controller. The controller
should let it go but, afterwards, it should change its colour from red to blue and it 
should send a message to the active memory switch in order that switch should also change
the selected track. The switch should also send a message to its other controller which
would change from~blue to~red.  This is performed by the scheme depicted by 
Figure~\ref{memopass}. When a simple locomotive arrives through the selected track, on the 
left-hand side in the figure, it crosses a blue controller and then passively crosses 
the fixed switch~$F_1$ and nothing happens. When a simple locomotive arrives through the 
non-selected track, it first crosses a red controller. The controller becomes blue and it 
sends a locomotive to the fork~$S_1$. The fork sends two simple locomotives, one to the 
fixed switch $F_2$ and then to the active memory switch and the second locomotive to the blue
controller which after this visit by a locomotive changes from blue to red. Note that here,
when the locomotive sent from~$S_1$ arrives to the blue controller, the locomotive which
crossed~$R$ arrived to~$F_1$. Of course, a similar sequence of actions is performed when
the non-selected track is on the left-hand side. 

\vtop{
\vspace{5pt}
\ligne{\hfill
\includegraphics[scale=0.45]{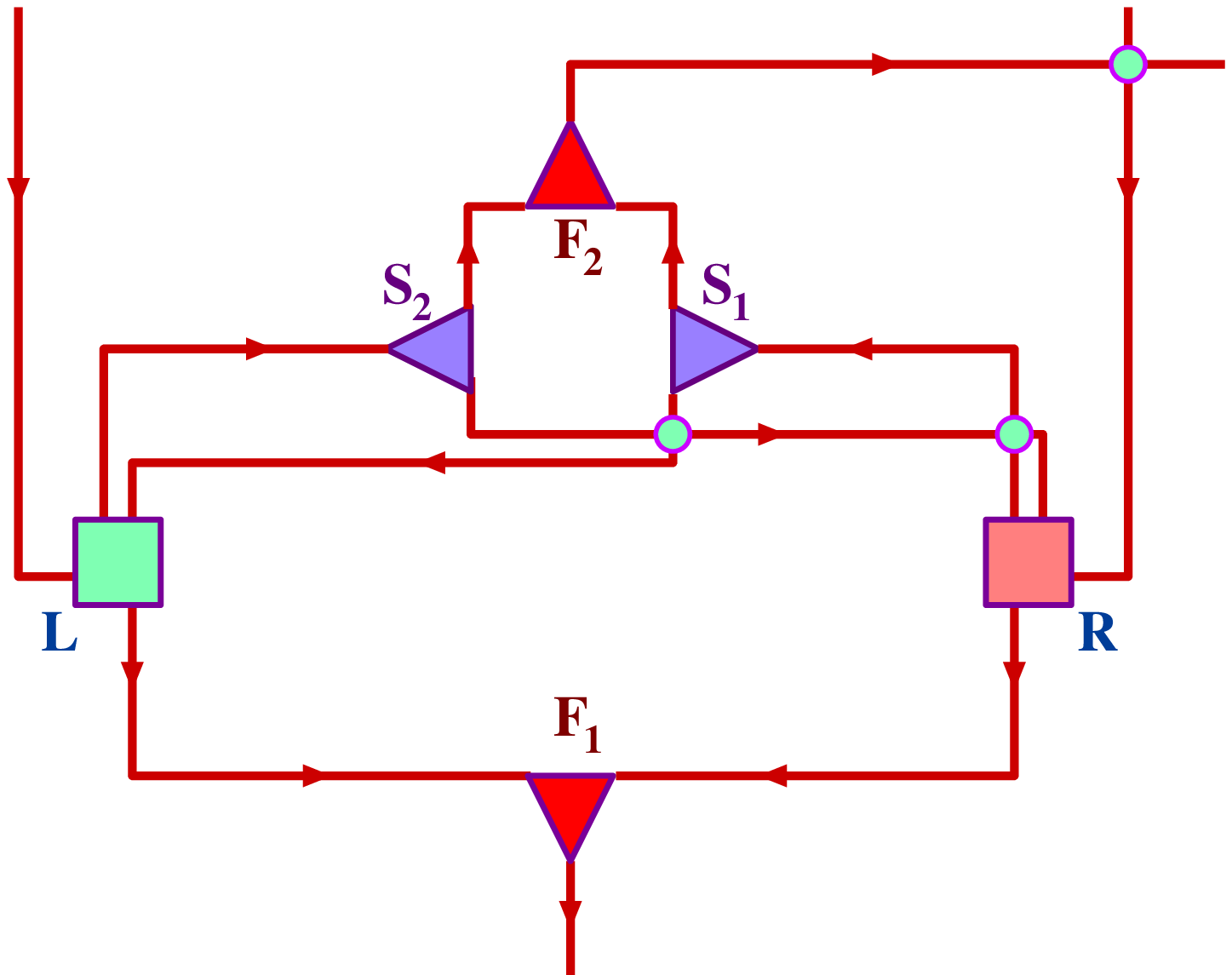}
\hfill}
\begin{fig}\label{memopass}
\leurre
Organization of the passive memory switch with forks and sensors. Note that the sensors
are not represented with the same symbol as the controllers in 
Figure~{\rm\ref{activswitches}}.
\end{fig}
}

   Figure~\ref{stab_ctrlsgn} illustrates the controller of the passive memory switch. It is
presented when no locomotive is around and it presents the structure in its two basic 
configurations: when it is blue and when it is red. The difference is shown by the colour of 
the second neighbour of the central cell of the figure. The track of the circuit is a 
'horizontal' line from right to left, in the upper part of the disc. The signal arrives 
close to the second neighbour of the central cell from the left-had side lower part of the
disc.

\vtop{
\ligne{\hfill
\includegraphics[scale=1]{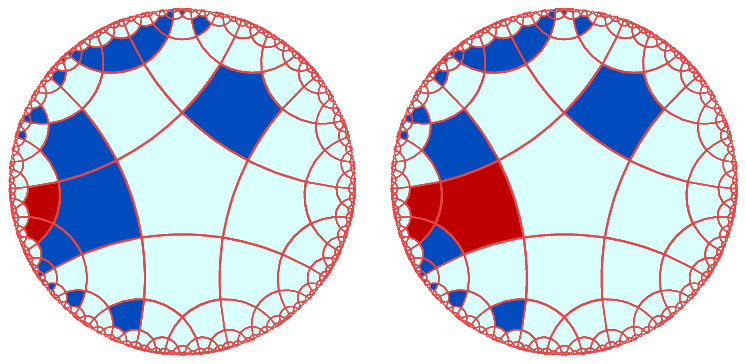}
\hfill}
\begin{fig}\label{stab_ctrlsgn}
\leurre
The controller-sensor of the passive memory switch.
\end{fig}
}

\section{Rules and figures}
\label{rules}

    The figures of Section~\ref{scenar} help us to establish the rules. The rules and the 
figures were established with the help of a computer program which checked the rotation 
invariance of the rules and which wrote the PostScript files of the pictures from the 
computation of the application of the rules to the configurations of the various type 
of parts of the circuit. The computer program also established the traces of execution which
allow the reader to check the application of the rules.

    We have to keep in mind that there are two types of rules. Those
which keep the structure invariant when the locomotive is far from them, we call this
type of rules \textbf{conservative}, and those which control the motion of the locomotive.
These latter rules, which we call \textbf{motion rules}, are the rules applied to the
cells of  the tracks as well as their milestones and, sometimes to the cells of the structures
which may be affected by the passage of the locomotive.

   Before establishing the rules, we indicate their format. A rule has the form
\hbox{\footnotesize\tt$\underline{\hbox{X}}$$_o$X$_1$..X$_{12}$$\underline{\hbox{X}}$$_n$},
where \hbox{\footnotesize\tt X$_o$} is the state of the cell~$c$,
\hbox{\footnotesize\tt X$_i$} is the state of the neighbour~$i$ of~$c$ 
and \hbox{\footnotesize\tt X$_n$} is the new state of~$c$ once the rule
has been applied. As the rules must be rotation invariant, the only constraint we set on the
numbering is to be increasing when counter-clockwise turning around the cell. However, in many 
cases we shall consider that neighbour~1 is that we indicated in Figure~\ref{stab_doublsel}
for the central cell and, for the others, neighbour~1 is the side turned to the father of 
the cell in the tree to which it belongs. In the sequel, we shall also consider the word
\hbox{\footnotesize\tt X$_1$..X$_{12}$} which we call the \textbf{neighbourhood} of the 
rule.

   In comments of the figures, the central tile is denoted by~0(0).  The cells which are
neighbours of~0(0) by sharing an edge with that cell are numbered 1($i$), with 
\hbox{$i\in\{1..5\}$}, increasingly while counter-clockwise turning around 0(0). The cell 1(1)
is upwards and to left. The vertex-neighbours of~0(0) are numbered 2($i$), again
with \hbox{$i\in\{1..5\}$} and also increasingly while counter-clockwise turning around 0(0).
The cell 2(1) is a neighbour of both 1(1) and 1(5), see Figure~\ref{penta}. Figure~\ref{penta}
clearly shows how to number the cells up to~12. For further numbers, the reader is invited to
look at~\cite{mmbook1}.

   Table~\ref{consrules} provides us with rules which concern the white cells which
are not on the track and who have a single or two non white neighbours. It also concerns
blue or red cells which have a single or two non white neighbours. A few other conservative
rules of the same kind also appear in further tables.

\def\aff #1 #2 #3 #4 {\ligne{\hfill\footnotesize\tt\hbox to 13pt{\hfill#1}
\hskip 5pt$\underline{\hbox{\tt#2}}$#3$\underline{\hbox{\tt#4}}$\hfill}\vskip-4pt
}
\vtop{
\begin{tab}\label{consrules}
\leurre
Conservative rules.
\end{tab}
\ligne{\hfill
\vtop{\leftskip 0pt\parindent 0pt\hsize=78pt
\aff {  1} {W} {WWWWWWWWWW} {W}
\aff {  2} {B} {WWWWWWWWWW} {B}
\aff {  3} {R} {WWWWWWWWWW} {R}
\aff {  4} {W} {BWWWWWWWWW} {W}
}
\hfill
\vtop{\leftskip 0pt\parindent 0pt\hsize=78pt
\aff {  5} {W} {RWWWWWWWWW} {W}
\aff {  6} {B} {BWWWWWWWWW} {B}
\aff {  7} {B} {RWWWWWWWWW} {B}
\aff {  8} {R} {BWWWWWWWWW} {R}
}
\hfill
\vtop{\leftskip 0pt\parindent 0pt\hsize=78pt
\aff {  9} {R} {RWWWWWWWWW} {R}
\aff { 10} {W} {WWWWWBWWWW} {W}
\aff { 11} {W} {WWWWWRWWWW} {W}
\aff { 12} {B} {WWWWWBWWWW} {B}
}
\hfill
\vtop{\leftskip 0pt\parindent 0pt\hsize=78pt
\aff { 13} {B} {WWWWWRWWWW} {B}
\aff { 14} {R} {WWWWWBWWWW} {R}
\aff { 15} {R} {WWWWWRWWWW} {R}
\aff { 16} {W} {WWWWBWBWWW} {W}
}
\hfill}
}

\subsection{Defining tracks and their rules}
\label{trackrules}

   As explained in Section~\ref{scenar}, the construction of the tracks is very important. 
As mentioned in Section~\ref{scenar}, the figures we shall consider give only a local view.
Nonetheless, they allow us to define the rules for the motion of the locomotive. Indeed,
three consecutive elements are enough for that purpose and the window on the hyperbolic
plane given by the Poincar\'e's disc offers this possibility.

   As indicated in Section~\ref{scenar}, we mainly use verticals and horizontals in order
to construct the tracks. We can choose the branches on which the vertical are built in such a
way that a big section of them goes along a line of the pentagrid. Figures~\ref{voievertd}
and~\ref{voievertm} illustrate the construction of the tracks which use a single element
as indicated in Section~\ref{scenar}.

\newdimen\largesept\largesept=21.5pt
\def\execneuf #1 #2 #3 #4 #5 #6 #7 #8 #9 {
\hbox{
\hbox to 25pt{\hfill#1\hskip 4pt}
\hbox to \largesept{\hfill#2\hskip 4pt}
\hbox to \largesept{\hfill#3\hskip 4pt}
\hbox to \largesept{\hfill#4\hskip 4pt}
\hbox to \largesept{\hfill#5\hskip 4pt}
\hbox to \largesept{\hfill#6\hskip 4pt}
\hbox to \largesept{\hfill#7\hskip 4pt}
\hbox to \largesept{\hfill#8\hskip 4pt}
\hbox to \largesept{\hfill#9\hskip 4pt}
}
}

\def\exechuit #1 #2 #3 #4 #5 #6 #7 #8 {
\hbox{
\hbox to 25pt{\hfill#1\hskip 4pt}
\hbox to \largesept{\hfill#2\hskip 4pt}
\hbox to \largesept{\hfill#3\hskip 4pt}
\hbox to \largesept{\hfill#4\hskip 4pt}
\hbox to \largesept{\hfill#5\hskip 4pt}
\hbox to \largesept{\hfill#6\hskip 4pt}
\hbox to \largesept{\hfill#7\hskip 4pt}
\hbox to \largesept{\hfill#8\hskip 4pt}
}
}

\def\execsept #1 #2 #3 #4 #5 #6 #7 {
\hbox{
\hbox to 25pt{\hfill#1\hskip 4pt}
\hbox to \largesept{\hfill#2\hskip 4pt}
\hbox to \largesept{\hfill#3\hskip 4pt}
\hbox to \largesept{\hfill#4\hskip 4pt}
\hbox to \largesept{\hfill#5\hskip 4pt}
\hbox to \largesept{\hfill#6\hskip 4pt}
\hbox to \largesept{\hfill#7\hskip 4pt}
}
}

\def\execsix #1 #2 #3 #4 #5 #6 {
\hbox{
\hbox to \largesept{\hfill#1\hskip 4pt}
\hbox to \largesept{\hfill#2\hskip 4pt}
\hbox to \largesept{\hfill#3\hskip 4pt}
\hbox to \largesept{\hfill#4\hskip 4pt}
\hbox to \largesept{\hfill#5\hskip 4pt}
\hbox to \largesept{\hfill#6\hskip 4pt}
}
}

\def\execcinq #1 #2 #3 #4 #5 {
\hbox{
\hbox to \largesept{\hfill#1\hskip 4pt}
\hbox to \largesept{\hfill#2\hskip 4pt}
\hbox to \largesept{\hfill#3\hskip 4pt}
\hbox to \largesept{\hfill#4\hskip 4pt}
\hbox to \largesept{\hfill#5\hskip 4pt}
}
}

\def\execquat #1 #2 #3 #4 {
\hbox{
\hbox to \largesept{\hfill#1\hskip 4pt}
\hbox to \largesept{\hfill#2\hskip 4pt}
\hbox to \largesept{\hfill#3\hskip 4pt}
\hbox to \largesept{\hfill#4\hskip 4pt}
}
}

\def\exectroi #1 #2 #3 {
\hbox{
\hbox to \largesept{\hfill#1\hskip 4pt}
\hbox to \largesept{\hfill#2\hskip 4pt}
\hbox to \largesept{\hfill#3\hskip 4pt}
}
}

\vtop{
%\ligne{\hrulefill}
%\vskip 3pt
\ligne{\hfill
%\phantom{
\includegraphics[scale=0.4]{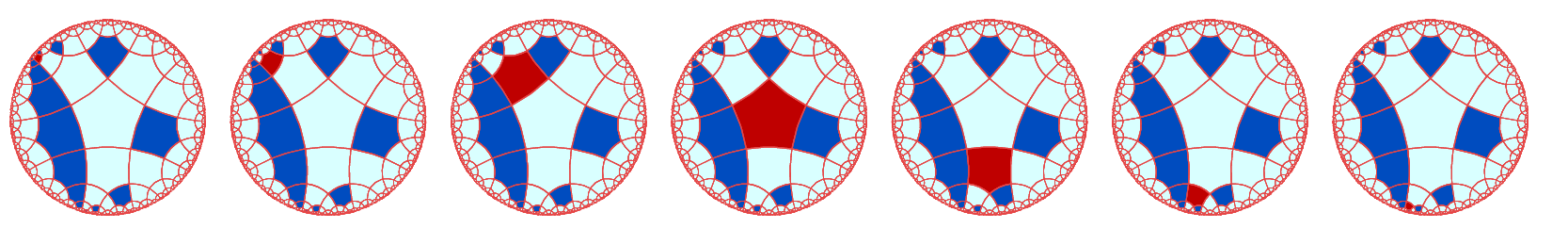} %}
\hfill}
\ligne{\hfill
%\phantom{
\includegraphics[scale=0.4]{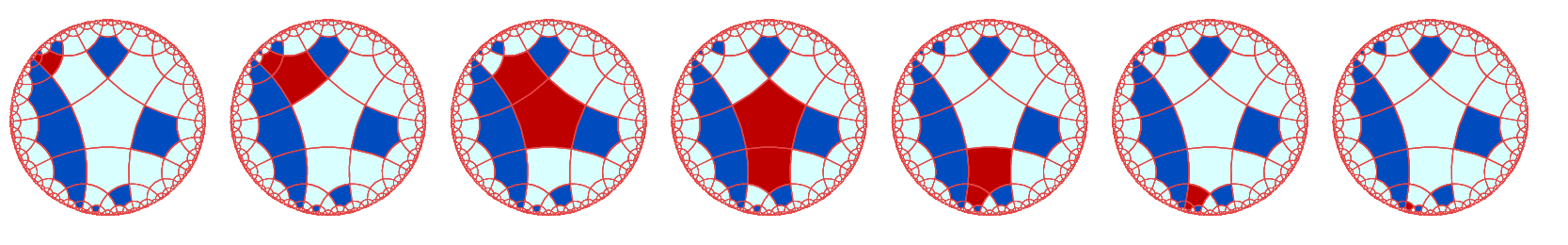} %}
\hfill}
\begin{fig}\label{voievertd}
\leurre
Above: a single locomotive going down a vertical track.
Below: a double locomotive going down a vertical track.
\end{fig}
}

\vtop{
%\ligne{\hrulefill}
%\vskip 3pt
\ligne{\hfill
%\phantom{
\includegraphics[scale=0.4]{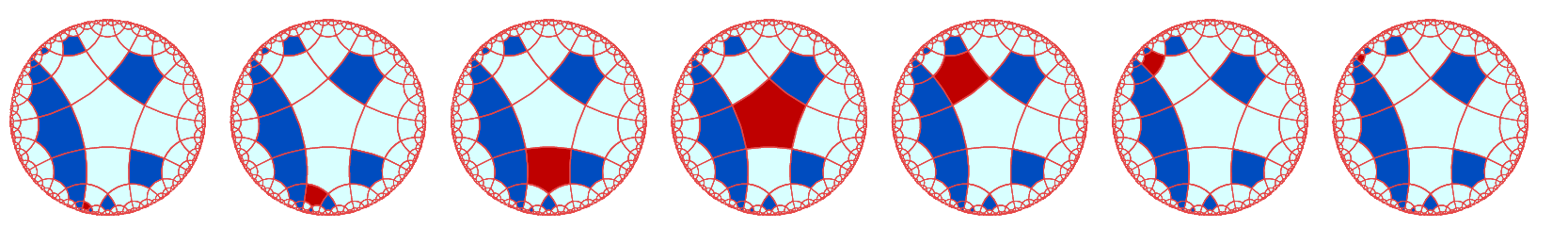} %}
\hfill}
\ligne{\hfill
%\phantom{
\includegraphics[scale=0.4]{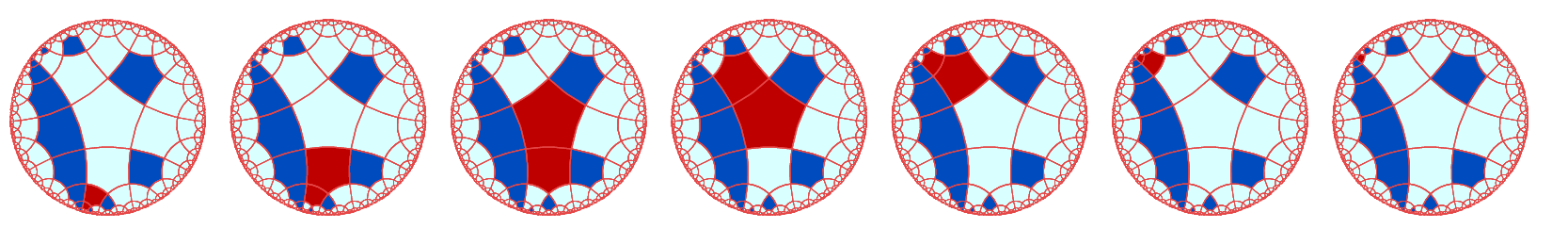} %}
\hfill}
\begin{fig}\label{voievertm}
\leurre
Above: a single locomotive going up a vertical track.
Below: a double locomotive going up a vertical track.
\end{fig}
}
\vskip 5pt
  Figure~\ref{voievertd} shows us a track which \textit{goes down} along a line, while
figure~\ref{voievertm} shows us a track which \textit{goes up } along the same line.
These figures consists in assembling basic elements only. The reader may notice that for 
each element, one milestone touches the same line of the pentagrid, say $\delta$, while the 
second milestone is on the same side of~$\delta$ for all elements of this part of the 
tracks. What happens if the second milestone is on the other side of~$\delta$ for all 
elements of the tracks? In fact, the motion in the opposite direction given in the figures 
answers the question. Indeed, denote by~$(0)0$ the central tile in Figures~\ref{voievertd}
and~\ref{voievertm}. Then, note that $0(0)$ has a side on~$\delta$, and that $0(0)$ 
belongs to the track along~$\delta$. Apply to~$\delta$ a reflection in the center 
of~$0(0)$ and let~$\eta$ be its image. Then, by a suitable rotation, $\eta$ can be mapped 
onto a line~$\rho$ of the pentagrid which passes along another side of~$0(0)$, so that 
$\rho$ and~$\delta$ have a common perpendicular which passes through another side~$\sigma$ 
of~$0(0)$. Then the shift along~$0(0)$ maps~$\rho$ onto~$\delta$ and, in this 
transformation, the track which is along~$\delta$ is mapped onto the track along $\delta$ 
which lies on the side which is opposite to that of~$0(0)$. And so we got a going down 
track which is supported by~$\delta$ but which is on the other side with respect to the one 
we can see on Figure~\ref{voievertd}. This is why the rules of 
Tables~\ref{verttracktab} and~\ref{mounttracktab} also apply to tracks which 
are displayed on another side compared 
to that of the figures~\ref{voievertd} and~\ref{voievertm}.

\vtop{
\begin{tab}\label{verttracktab}
\leurre
Rules forgoing down vertical tracks.
\end{tab}
\vspace{-12pt}
\ligne{\hfill a single locomotive:\hfill}
\vskip 2pt
\ligne{\hfill
\vtop{\leftskip 0pt\parindent 0pt\hsize=76pt
\aff { 17} {W} {WBWBWBBBWW} {W}
\aff { 18} {W} {WBWRBBWWBB} {R}
\aff { 19} {W} {WWWWBRBWWW} {W}
\aff { 20} {R} {WBWWBBWWBB} {W}
\aff { 21} {W} {RWWWBWBWWW} {W}
}\hfill
\vtop{\leftskip 0pt\parindent 0pt\hsize=76pt
\aff { 22} {W} {RBWWBBWWWW} {W}
\aff { 23} {W} {WBWWWWRWWW} {W}
\aff { 24} {B} {WBWWBWWWWW} {B}
\aff { 25} {W} {BWWWWWBWWW} {W}
\aff { 26} {W} {BWWWWBWWWW} {W}
}
\hfill
\vtop{\leftskip 0pt\parindent 0pt\hsize=76pt
\aff { 27} {B} {BWBWWWWRWW} {B}
\aff { 28} {B} {BRBWWWWWWW} {B}
\aff { 29} {B} {BWWWWWRWWW} {B}
\aff { 30} {W} {WBWBWBBWWW} {W}
\aff { 31} {W} {WWWWWBBWWW} {W}
}
\hfill
\vtop{\leftskip 0pt\parindent 0pt\hsize=76pt
\aff { 32} {W} {RBWWBBWWBB} {W}
\aff { 33} {W} {WBWWBBWWWW} {W}
\aff { 34} {B} {BWBWWWRWWW} {B}
\aff { 35} {W} {RBWBWBBWWW} {R}
\aff { 36} {B} {WBWWWWRWWW} {B}
}
\hfill}
\vskip 7pt
\ligne{\hfill a double locomotive:\hfill}
\vskip 2pt
\ligne{\hfill
\vtop{\leftskip 0pt\parindent 0pt\hsize=76pt
\aff { 37} {R} {WBWRBBWWBB} {R}
\aff { 38} {W} {RWWWBRBWWW} {W}
\aff { 39} {R} {RBWWBBWWBB} {W}
\aff { 40} {B} {RWWWWWRWWW} {B}
\aff { 41} {B} {BRBWWWWRWW} {B}
}\hfill
\vtop{\leftskip 0pt\parindent 0pt\hsize=76pt
\aff { 42} {B} {BRBWWWRWWW} {B}
\aff { 43} {R} {RBWBWBBWWW} {R}
\aff { 44} {B} {RBWWWWRWWW} {B}
\aff { 45} {W} {RWWWWBBWWW} {W}
\aff { 46} {W} {BWWWWWRWWW} {W}
}
\hfill}
}
\vskip 10pt
\def\rrr{\hbox{\footnotesize\tt R}}
\def\www{\hbox{\footnotesize\tt W}}
Table~\ref{exvertd} shows which rules where applied to the following cells: 1(1), 0(0)
and~1(3) for the tracks and 1(2) and~1(4) for the milestones when a locomotive goes
down a vertical as illustrated in Figure~\ref{voievertd}.
The basic rules are rules~17, 18, 20 and~32 for a single locomotive
as shown in Table~\ref{exvertd}, rule~17 being the conservative rule.
For a double locomotive, 
rules~37 and~39 are applied instead of rule~20 as when the first locomotive is in the cell
it can see the second one at the entrance of the locomotive and when the second one is in 
the cell, it can see the first locomotive at the exit of the cell. If we compare the
neighbourhood of rule~17 with that of rule~18{} in which we replace the state~\rrr{} by
the state~\www, it may seem that the neighbourhoods are very different. Now, moving the 
neighbourhood of rule~17 by two steps to the right we obtain that of rule~18. Accordingly,
these neighbourhoods are rotated forms of the same one.

   In Table~\ref{mounttracktab}, we have the rules for going down. Table~\ref{exvertm}
shows the rules which where applied in a motion illustrated by Figure~\ref{voievertm},
where the locomotives are going up. This time, rules~47, 50,
51 and~53 allow a single locomotive to run along the tracks, see Table~\ref{exvertm},
rule~47 being the conservative rule. 
When the locomotive is double,
rules~57 and~58 replace rule~51. The reasons are the same as previously.

   Before looking at other tables, remark that the other rules for Tables~\ref{verttracktab}
and~\ref{mounttracktab} deal either with cells which are not changed by the motion of the
locomotive but simply witness that the locomotive is going by them. For instance, rule~24
usually governs the cell 13(2) which is a milestone of the track. When the locomotive
is by the cell, rules~28 or~34 are applied to the cell.

\vtop{
\begin{tab}\label{exvertd}
\leurre
Locomotive going down a vertical.
\end{tab}
\vskip-6pt
\ligne{\hfill\hbox to 180pt{\hfill simple locomotive:\hfill}
\hskip 5pt\hbox to 155pt{\hfill double locomotive:\hfill}\hfill}
\vskip 2pt
\ligne{\hfill
\vtop{\leftskip 0pt\parindent 0pt\hsize=180pt
\ligne{\hfill  \execsept     {}       0       1       2       3       4       5 \hfill}
\ligne{\hfill  \execsept  {1(1)}  {  17}  {  17}  {  17}  {  18}  {  20}  {  32}   \hfill}
\ligne{\hfill  \execsept  {0(0)}  {  17}  {  17}  {  18}  {  20}  {  32}  {  17}   \hfill}
\ligne{\hfill  \execsept  {1(3)}  {  17}  {  18}  {  20}  {  32}  {  17}  {  17}   \hfill}
\ligne{\hfill  \execsept  {1(2)}  {  24}  {  24}  {  27}  {  28}  {  34}  {  24}   \hfill}
\ligne{\hfill  \execsept  {1(4)}  {   2}  {   2}  {   2}  {   7}  {  13}  {   2}   \hfill}
}
\hskip 5pt
\vtop{\leftskip 0pt\parindent 0pt\hsize=155pt
\ligne{\hfill   \execsix       0       1       2       3       4       5 \hfill}
\ligne{\hfill   \execsix   {  17}  {  17}  {  18}  {  37}  {  39}  {  32}   \hfill}
\ligne{\hfill   \execsix   {  17}  {  18}  {  37}  {  39}  {  32}  {  17}   \hfill}
\ligne{\hfill   \execsix   {  18}  {  37}  {  39}  {  32}  {  17}  {  17}   \hfill}
\ligne{\hfill   \execsix   {  24}  {  27}  {  41}  {  42}  {  34}  {  24}   \hfill}
\ligne{\hfill   \execsix   {   2}  {   2}  {   7}  {  40}  {  13}  {   2}   \hfill}
}
\hfill}
}

\vtop{
\begin{tab}\label{mounttracktab}
\leurre
Rules for going up vertical tracks.
\end{tab}
\vspace{-12pt}
\ligne{\hfill a single locomotive:\hfill\hfill\hskip 30pt a double one:\hfill}
\vskip 2pt
\ligne{\hfill
\vtop{\leftskip 0pt\parindent 0pt\hsize=76pt
\aff { 47} {W} {WBWWBWBBBW} {W}
\aff { 48} {W} {WBWWWBWWWW} {W}
\aff { 49} {W} {WWBWBBBWWW} {W}
\aff { 50} {W} {WBRWBWBBBW} {R}
\aff { 51} {R} {WBWWBWBBBW} {W}
}\hfill
\vtop{\leftskip 0pt\parindent 0pt\hsize=76pt
\aff { 52} {W} {WBWWWBRWWW} {W}
\aff { 53} {W} {RBWWBWBBBW} {W}
\aff { 54} {W} {RBWWWBWWWW} {W}
\aff { 55} {W} {WBWWBWBWWW} {W}
\aff { 56} {W} {RWBWBBBWWW} {R}
}\hfill\hskip 30pt
\vtop{\leftskip 0pt\parindent 0pt\hsize=76pt
\aff { 57} {R} {WBRWBWBBBW} {R}
\aff { 58} {R} {RBWWBWBBBW} {W}
\aff { 59} {W} {RBWWWBRWWW} {W}
\aff { 60} {W} {RBWWBWBWWW} {W}
\aff { 61} {B} {WRWWWWRWWW} {B}
}
\hfill}
}
\vskip 10pt
\vtop{
\begin{tab}\label{exvertm}
\leurre
Locomotive going up a vertical.
\end{tab}
\vskip-6pt
\ligne{\hfill\hbox to 180pt{\hfill simple locomotive:\hfill}
\hskip 5pt\hbox to 155pt{\hfill double locomotive:\hfill}\hfill}
\vskip 2pt
\ligne{\hfill
\vtop{\leftskip 0pt\parindent 0pt\hsize=180pt
\ligne{\hfill  \execsept     {}       0       1       2       3       4       5 \hfill}
\ligne{\hfill  \execsept  {1(3)}  {  47}  {  50}  {  51}  {  53}  {  47}  {  47}   \hfill}
\ligne{\hfill  \execsept  {0(0)}  {  47}  {  47}  {  50}  {  51}  {  53}  {  47}   \hfill}
\ligne{\hfill  \execsept  {1(1)}  {  47}  {  47}  {  47}  {  50}  {  51}  {  53}   \hfill}
\ligne{\hfill  \execsept  {1(2)}  {  24}  {  24}  {  34}  {  28}  {  27}  {  24}   \hfill}
\ligne{\hfill  \execsept  {1(5)}  {   2}  {   2}  {   2}  {   7}  {  13}  {   2}   \hfill}
}
\hskip 5pt
\vtop{\leftskip 0pt\parindent 0pt\hsize=155pt
\ligne{\hfill   \execsix       0       1       2       3       4       5 \hfill}
\ligne{\hfill   \execsix   {  50}  {  57}  {  58}  {  53}  {  47}  {  47}   \hfill}
\ligne{\hfill   \execsix   {  47}  {  50}  {  57}  {  58}  {  53}  {  47}   \hfill}
\ligne{\hfill   \execsix   {  47}  {  47}  {  50}  {  57}  {  58}  {  53}   \hfill}
\ligne{\hfill   \execsix   {  24}  {  34}  {  42}  {  41}  {  27}  {  24}   \hfill}
\ligne{\hfill   \execsix   {   2}  {   2}  {   7}  {  61}  {  13}  {   2}   \hfill}
}
\hfill}
}
\vskip 10pt

   However, rules for verticals on both sides of a line of the pentagrid are not enough.
We need also rules for turning around a cell or around several of them. 
Tables~\ref{round1stab}, \ref{round1dtab} and~\ref{round3htab} 
gives us rules for the locomotive turning around 1, 2 or 3 contiguous cells. 
Figures~\ref{rond1h}, \ref{rond1ah}, \ref{rond2h}, \ref{rond2ah}, \ref{rond3h}
and~\ref{rond3ah} illustrate these motions, clockwise and counter-clockwise, with a single
locomotive and with a double one. The trick consists in placing basic elements in such a way
that for some of them, the exit occurs through a cell which, in a vertical tracks would work
as an entry. This is typically the case of the cell 2(1) in Figures~\ref{rond1h} 
and~\ref{rond1ah}, with a simple or a double locomotive clockwise or counter-clockwise
turning around a single cell. The figures illustrate how a simple and a double locomotive
runs around several cells, whatever the direction. Tables~\ref{round1stab} 
and~\ref{round1dtab} give the rules
for the locomotive turning around a single cell in both orientations. Table~\ref{round3htab}
completes these rules when the locomotive turns around two or three contiguous cells.

   Tables~\ref{exrond1h} and~\ref{exrond1ah} show the execution of the rules on a small
part of such a circuit when the locomotive turns around a single cell. Note that the
cells involved are different in each table: this is due to the structure of the computer
program which checked the rules and computed both the figures and the execution tables.
The difference is induced by the opposite orientations which also is taken into account
for defining the time interval which is represented in the tables.
In each table, the rules are given first for a simple locomotive and then for a double one.

\vtop{
\begin{tab}\label{round1stab}
\leurre
Rules for turning around a single cell, with a single locomotive:
\end{tab}
\vspace{-12pt}
\ligne{\hfill clockwise:\hfill\hfill\hskip 30pt counter-clockwise:\hfill}
\vskip 2pt
\ligne{\hfill
\vtop{\leftskip 0pt\parindent 0pt\hsize=76pt
\aff { 62} {R} {BWBWWWWBWB} {W}
\aff { 63} {B} {RBWWWWWWWW} {B}
\aff { 64} {W} {BWBWWWRBWB} {W}
\aff { 65} {W} {WRBWBBBWWW} {W}
\aff { 66} {W} {BWBWWWWBWB} {W}
}\hfill
\vtop{\leftskip 0pt\parindent 0pt\hsize=76pt
\aff { 67} {W} {BWBWWRWBWB} {W}
\aff { 68} {W} {BRBWWWWBWB} {W}
\aff { 69} {R} {WWBWBBBWWW} {W}
\aff { 70} {W} {BWBWRWWBWB} {R}
}
\hfill\hskip 30pt
\vtop{\leftskip 0pt\parindent 0pt\hsize=76pt
\aff { 71} {R} {BWWBWWWBWB} {W}
\aff { 72} {W} {RWBWBWBBWW} {W}
\aff { 73} {W} {BWWBWWRBWB} {W}
\aff { 74} {W} {BWWBWWWBWB} {W}
}\hfill
\vtop{\leftskip 0pt\parindent 0pt\hsize=76pt
\aff { 75} {W} {BWWBWRWBWB} {W}
\aff { 76} {W} {BWWBRWWBWB} {W}
\aff { 77} {W} {BRWBWWWBWB} {R}
\aff { 78} {R} {WWBWBWBBWW} {W}
}
\hfill}
}

\vtop{
\begin{tab}\label{exrond1h}
\leurre
Execution of the rules of Tables~\textrm{\ref{round1stab}}
and~\textrm{\ref{round1dtab}} for a locomotive clockwise running around a single cell: 
\end{tab}
\vskip-6pt
\ligne{\hfill\hbox to 180pt{\hfill simple locomotive:\hfill}
\hskip 5pt\hbox to 155pt{\hfill double locomotive:\hfill}\hfill}
\vskip 2pt
\ligne{\hfill
\vtop{\leftskip 0pt\parindent 0pt\hsize=180pt
\ligne{\hfill  \execsept     {}       0       1       2       3       4       5 \hfill}
\ligne{\hfill  \execsept  {1(5) } {  67}  {  70}  {  62}  {  68}  {  64}  {  66}   \hfill}
\ligne{\hfill  \execsept  {2(5) } {  49}  {  49}  {  56}  {  69}  {  65}  {  49}   \hfill}
\ligne{\hfill  \execsept  {1(4) } {  66}  {  66}  {  67}  {  70}  {  62}  {  68}   \hfill}
\ligne{\hfill  \execsept  {2(4) } {  49}  {  49}  {  49}  {  49}  {  56}  {  69}   \hfill}
\ligne{\hfill  \execsept  {0(0) } {   7}  {  13}  {   7}  {  13}  {   7}  {  13}   \hfill}
\ligne{\hfill  \execsept  {3(5) } {   6}  {   6}  {  63}  {  36}  {   6}  {   6}   \hfill}
\ligne{\hfill  \execsept  {7(5) } {   6}  {   6}  {  29}  { 125}  {   6}  {   6}   \hfill}
\ligne{\hfill  \execsept  {5(5) } {   2}  {   2}  {   2}  {   7}  {  13}  {   2}   \hfill}
}
\hskip 5pt
\vtop{\leftskip 0pt\parindent 0pt\hsize=155pt
\ligne{\hfill   \execsix       0       1       2       3       4       5 \hfill}
\ligne{\hfill   \execsix   {  81}  {  84}  {  79}  {  82}  {  64}  {  66}   \hfill}
\ligne{\hfill   \execsix   {  49}  {  56}  {  80}  {  83}  {  65}  {  49}   \hfill}
\ligne{\hfill   \execsix   {  66}  {  67}  {  81}  {  84}  {  79}  {  82}   \hfill}
\ligne{\hfill   \execsix   {  49}  {  49}  {  49}  {  56}  {  80}  {  83}   \hfill}
\ligne{\hfill   \execsix   {  61}  {  40}  {  61}  {  40}  {  61}  {  40}   \hfill}
\ligne{\hfill   \execsix   {   6}  {  63}  {  44}  {  36}  {   6}  {   6}   \hfill}
\ligne{\hfill   \execsix   {   6}  {  29}  { 142}  { 125}  {   6}  {   6}   \hfill}
\ligne{\hfill   \execsix   {   2}  {   2}  {   7}  {  40}  {  13}  {   2}   \hfill}
}
\hfill}
}
\vskip 10pt
   Note that the rules applied around the cell~0(0) are periodic: the neighbourhood is the
same in each sector around~0(0) and the front of the locomotive alternates in appearing
sometimes through a side-neighbour, sometimes in a vertex-neighbour. Also note that
the sequences of rules are the same, but with a delay, for 1(5) and~1(4) and also for
2(5) and~2(4) in Table~\ref{exrond1h}. Similar remarks can be made for Table~\ref{rond1ah} too.
Due to the opposite orientations, the order of the lines for the tracks and then for the 
milestones in Table~\ref{exrond1ah} are the reverse with respect to Table~\ref{exrond1h}.

\vtop{
\begin{tab}\label{round1dtab}
\leurre
Rules for turning around a single cell, with a double locomotive:
\end{tab}
\vspace{-12pt}
\ligne{\hfill clockwise:\hfill\hfill\hskip 30pt counter-clockwise:\hfill}
\vskip 2pt
\ligne{\hfill
\vtop{\leftskip 0pt\parindent 0pt\hsize=76pt
\aff { 79} {R} {BRBWWWWBWB} {W}
\aff { 80} {R} {RWBWBBBWWW} {R}
\aff { 81} {W} {BWBWRRWBWB} {R}
\aff { 82} {W} {BRBWWWRBWB} {W}
\aff { 83} {R} {WRBWBBBWWW} {W}
\aff { 84} {R} {BWBWRWWBWB} {R}
}\hfill\hskip 30pt
\vtop{\leftskip 0pt\parindent 0pt\hsize=76pt
\aff { 85} {R} {BRWBWWWBWB} {R}
\aff { 86} {R} {RWBWBWBBWW} {W}
\aff { 87} {W} {BWWBRRWBWB} {W}
\aff { 88} {R} {BWWBRWWBWB} {W}
\aff { 89} {W} {BRWBWWRBWB} {R}
}
}
}
\vskip 10pt

\vtop{
\begin{tab}\label{exrond1ah}
\leurre
Execution of the rules of Tables~\textrm{\ref{round1stab}}
and~\textrm{\ref{round1dtab}} for a locomotive counter-clockwise running around a single cell: 
\end{tab}
\vskip-6pt
\ligne{\hfill\hbox to 180pt{\hfill simple locomotive:\hfill}
\hskip 5pt\hbox to 155pt{\hfill double locomotive:\hfill}\hfill}
\vskip 2pt
\ligne{\hfill
\vtop{\leftskip 0pt\parindent 0pt\hsize=180pt
\ligne{\hfill  \execsept     {}       4       5       6       7       8       9 \hfill}
\ligne{\hfill  \execsept  {2(4) } {  35}  {  78}  {  72}  {  30}  {  30}  {  30}   \hfill}
\ligne{\hfill  \execsept  {1(4) } {  73}  {  77}  {  71}  {  76}  {  75}  {  74}   \hfill}
\ligne{\hfill  \execsept  {2(5) } {  30}  {  30}  {  35}  {  78}  {  72}  {  30}   \hfill}
\ligne{\hfill  \execsept  {1(5) } {  74}  {  74}  {  73}  {  77}  {  71}  {  76}   \hfill}
\ligne{\hfill  \execsept  {0(0) } {   7}  {  13}  {   7}  {  13}  {   7}  {  13}   \hfill}
\ligne{\hfill  \execsept  {5(5) } {   6}  {   6}  {  36}  {  63}  {   6}  {   6}   \hfill}
\ligne{\hfill  \execsept  {7(5) } {   2}  {   2}  {   2}  {   7}  {  13}  {   2}   \hfill}
\ligne{\hfill  \execsept  {4(5) } {   6}  {   6}  {   6}  {   6}  { 125}  {  29}   \hfill}
}
\hskip 5pt
\vtop{\leftskip 0pt\parindent 0pt\hsize=155pt
\ligne{\hfill   \execsix       4       5       6       7       8       9 \hfill}
\ligne{\hfill   \execsix   {  35}  {  43}  {  86}  {  72}  {  30}  {  30}   \hfill}
\ligne{\hfill   \execsix   {  73}  {  89}  {  85}  {  88}  {  87}  {  75}   \hfill}
\ligne{\hfill   \execsix   {  30}  {  30}  {  35}  {  43}  {  86}  {  72}   \hfill}
\ligne{\hfill   \execsix   {  74}  {  74}  {  73}  {  89}  {  85}  {  88}   \hfill}
\ligne{\hfill   \execsix   {  61}  {  40}  {  61}  {  40}  {  61}  {  40}   \hfill}
\ligne{\hfill   \execsix   {   6}  {   6}  {  36}  {  44}  {  63}  {   6}   \hfill}
\ligne{\hfill   \execsix   {   2}  {   2}  {   2}  {   7}  {  61}  {  13}   \hfill}
\ligne{\hfill   \execsix   {   6}  {   6}  {   6}  {   6}  { 125}  { 142}   \hfill}
}
\hfill}
}
\vskip 10pt

The execution tables show that the rules for turning around a single cell are not enough
in order to show that the simple or double locomotive can run along a horizontal line.
Table~\ref{round3htab} give additional rules which are needed for that purpose
and also for allowing the locomotive the connection connection between vertical lines 
with a horizontal: whichever end of the vertical and of
the horizontal, whichever direction of the locomotive, whether simple or double the 
locomotive is. Note that Tables~\ref{exrond2hs}, \ref{exrond2hd}, \ref{exrond2ahs}
and~\ref{exrond2ahd} show the application of the rules of~\ref{round3htab} also to
the case of a locomotive running around two consecutive cells. For three consecutive cells,
Tables~\ref{exrond3ahs} and~\ref{exrond3ahd} show the application of the rules
of Table~\ref{round3htab}. It was checked by the computer program that the execution
traces for a clockwise motion corresponding to Figures~\ref{rond3h} involve exactly the same 
rules as those indicated by Tables~\ref{exrond2hs} and~\ref{exrond2hd}.
The rules, the executions and the corresponding figures allow us to establish two properties: the 
rules allow the locomotive, either a simple or a double one, to run along a horizontal. 
It can perform such a motion, whichever the direction. The second property is the 
connection of a vertical with a horizontal as already mentioned.
\vskip 10pt
\vtop{
%\ligne{\hrulefill}
%\vskip 3pt
\ligne{\hfill
%\phantom{
\includegraphics[scale=0.4]{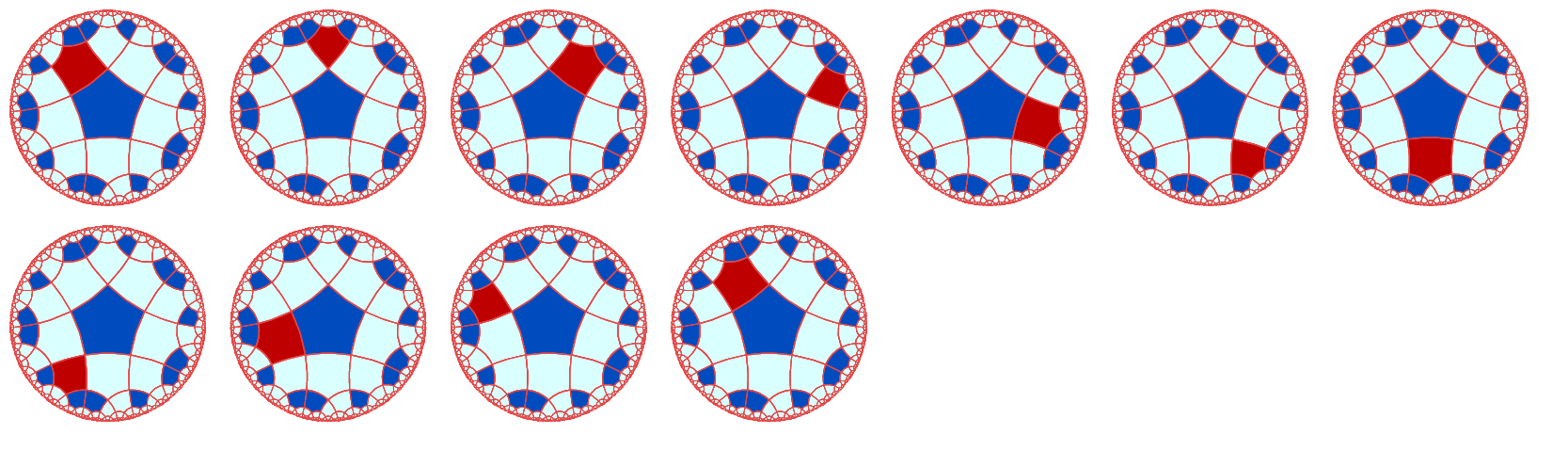} %}
\hfill}
\ligne{\hfill
%\phantom{
\includegraphics[scale=0.4]{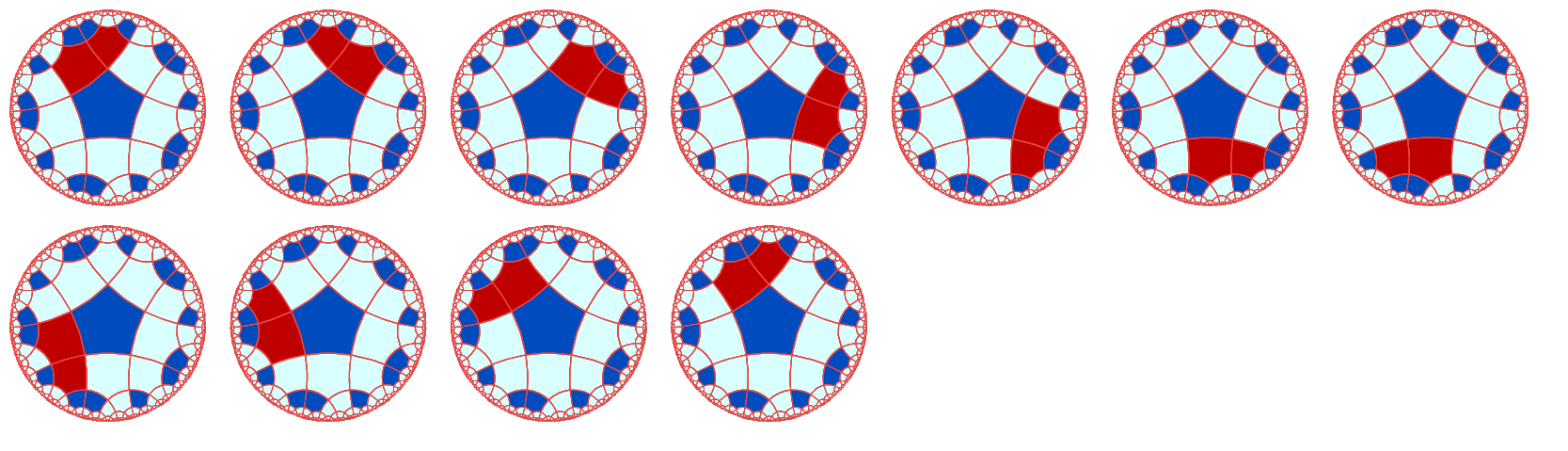} %}
\hfill}
\begin{fig}\label{rond1h}
\leurre
Above : a single locomotive endlessly turning around a cell in a clockwise orientation.
Below:  a double locomotive endlessly turning around a cell in a clockwise orientation.
\end{fig}
}
\vskip 10pt

\vtop{
%\ligne{\hrulefill}
%\vskip 3pt
\ligne{\hfill
%\phantom{
\includegraphics[scale=0.4]{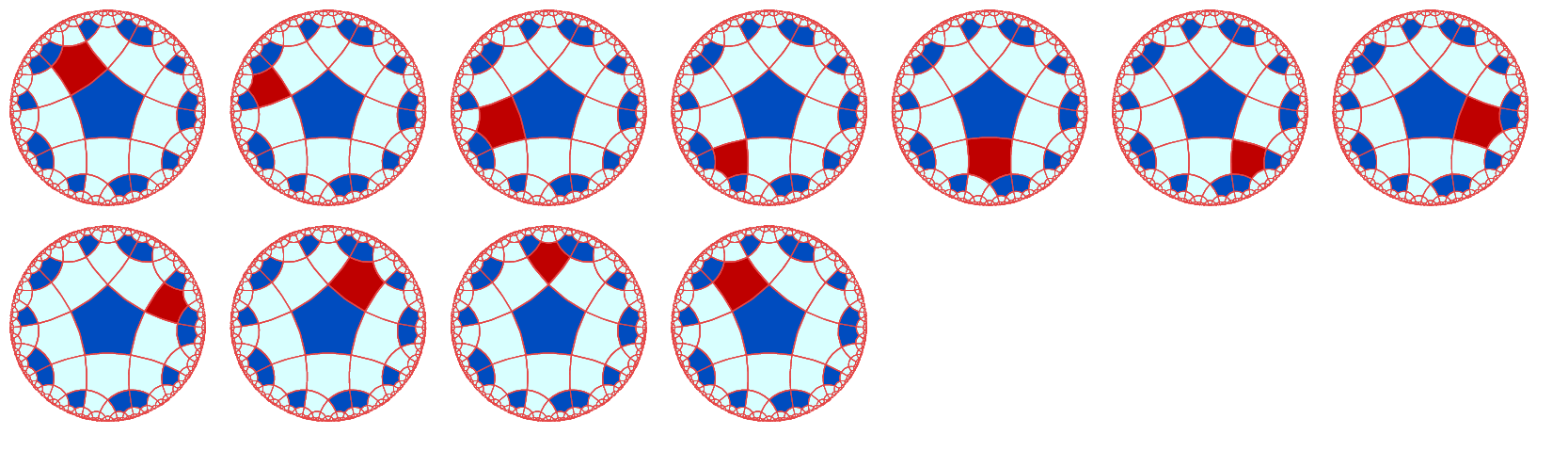} %}
\hfill}
\ligne{\hfill
%\phantom{
\includegraphics[scale=0.4]{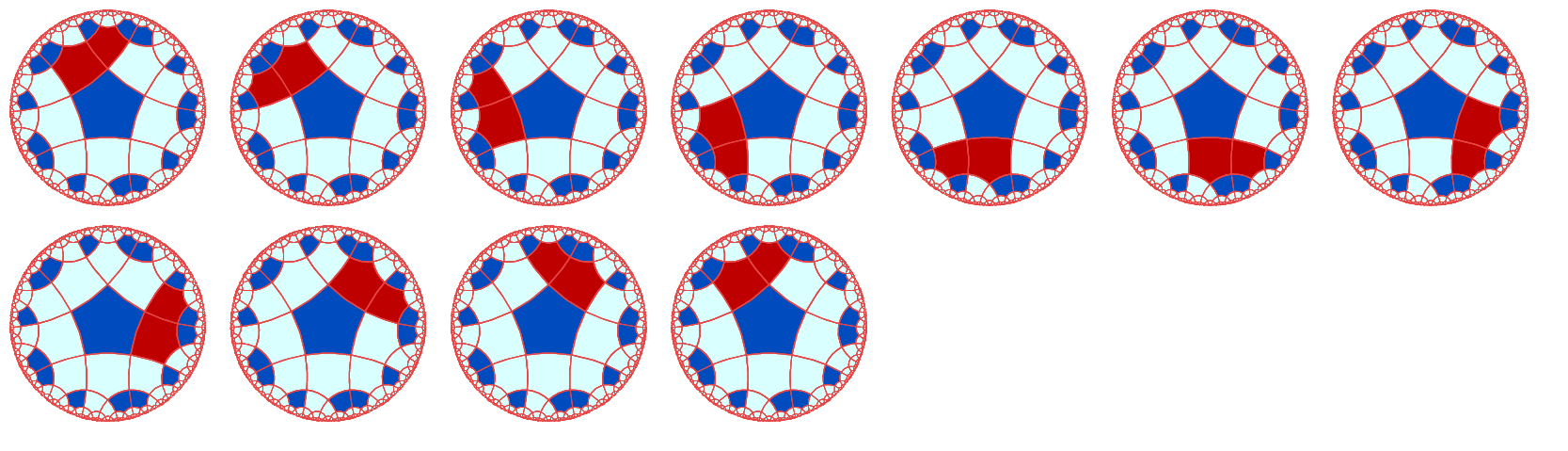} %}
\hfill}
\begin{fig}\label{rond1ah}
\leurre
Above : a single locomotive endlessly turning around a cell in a counter-clockwise orientation.
Below:  a double locomotive endlessly turning around a cell in a counter-clockwise orientation.
\end{fig}
}

\vtop{
%\ligne{\hrulefill}
%\vskip 3pt
\ligne{\hfill
%\phantom{
\includegraphics[scale=0.4]{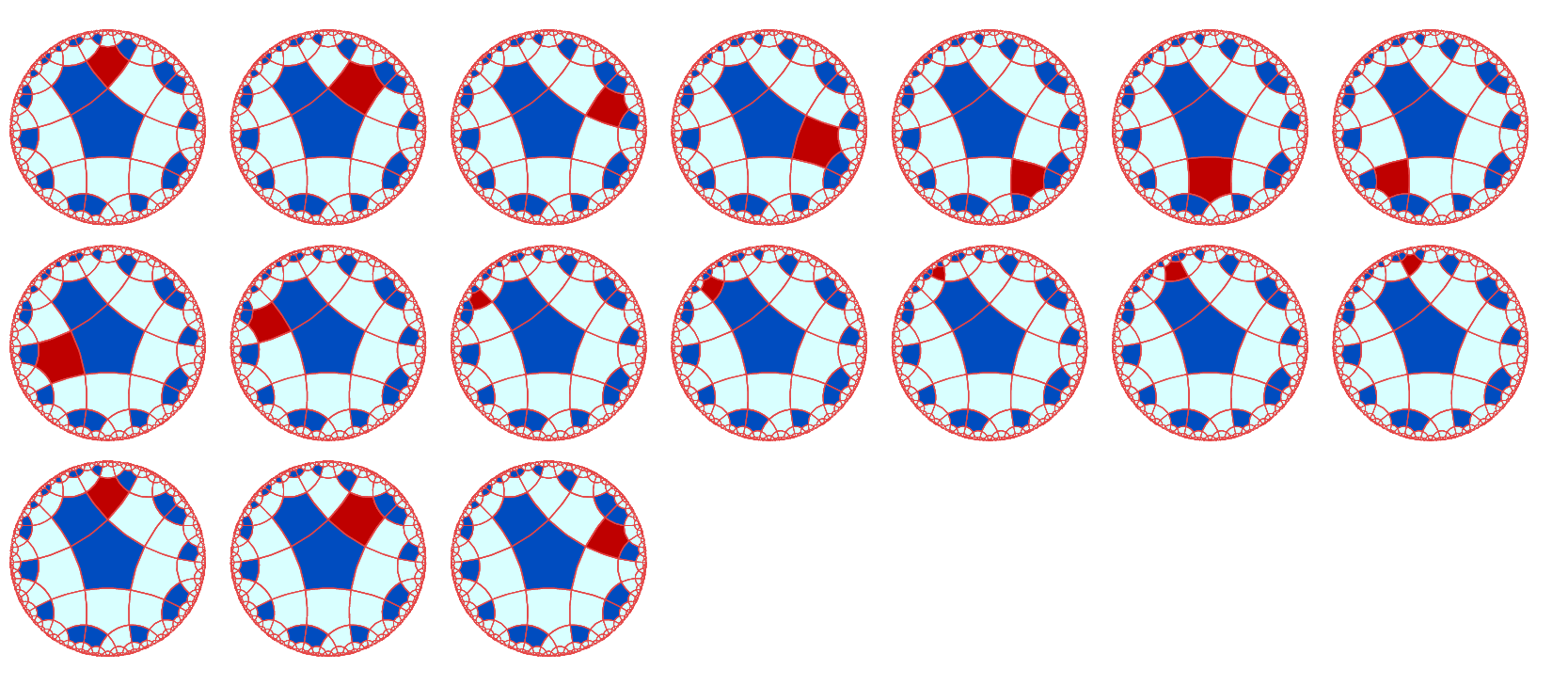} %}
\hfill}
\ligne{\hfill
%\phantom{
\includegraphics[scale=0.4]{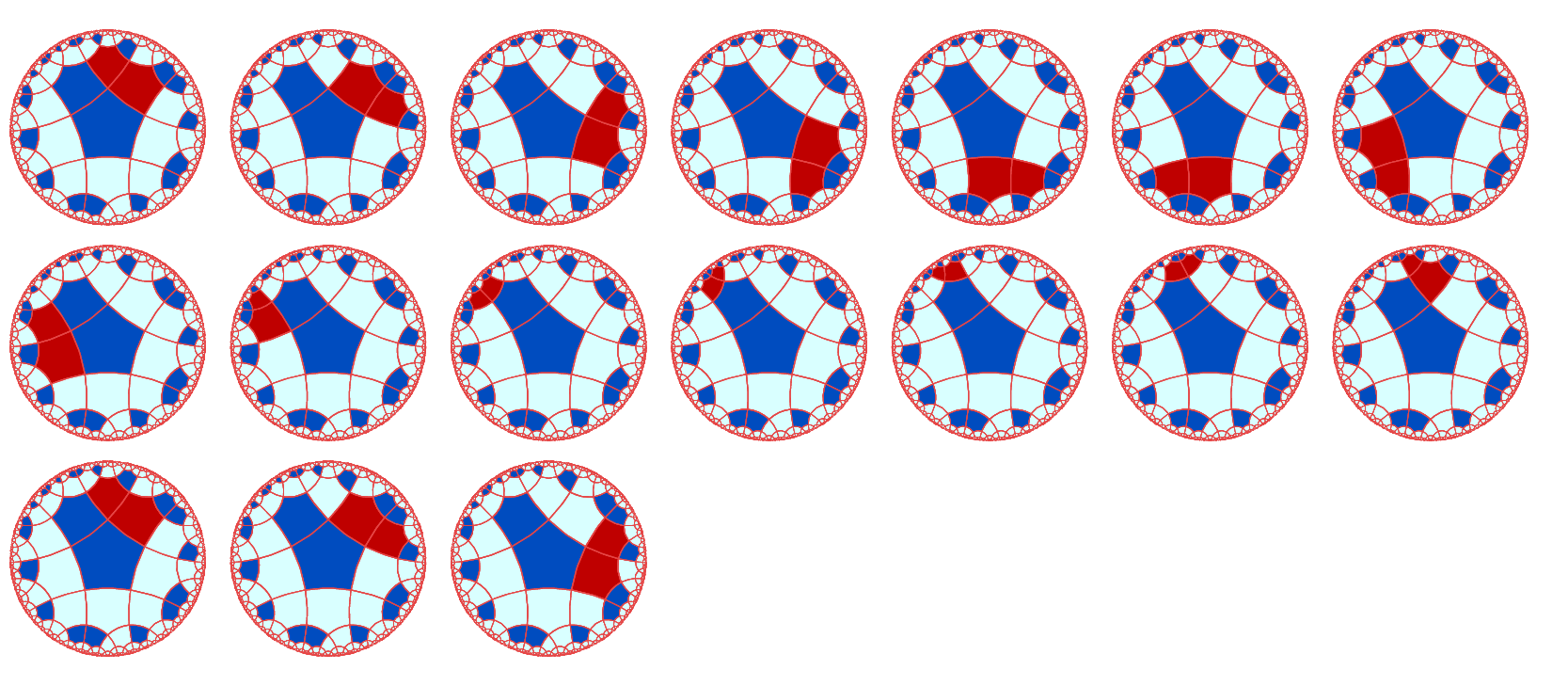} %}
\hfill}
\begin{fig}\label{rond2h}
\leurre
Above : a single locomotive endlessly turning around two cells in a clockwise orientation.
Below:  a double locomotive endlessly turning around two cells in a clockwise orientation.
\end{fig}
}
\vskip 5pt

\vtop{
\begin{tab}\label{exrond2hs}
\leurre
Execution of the circuit of a simple locomotive clockwise running around two contiguous
cells.
\end{tab}
\vskip-6pt
\ligne{\hfill 
\vtop{\leftskip 0pt\parindent 0pt\hsize=210pt
\ligne{\hfill\execneuf    {}      7      8      9     10     11     12     13     14 \hfill}
\ligne{\hfill\execneuf {2(2) } { 91}  { 92}  { 94}  { 95}  { 90}  { 90}  { 90}  { 90} \hfill}
\ligne{\hfill\execneuf {5(2) } { 49}  { 56}  { 69}  { 65}  { 49}  { 49}  { 49}  { 49} \hfill}
\ligne{\hfill\execneuf {4(1) } { 66}  { 67}  { 70}  { 62}  { 68}  { 64}  { 66}  { 66} \hfill}
\ligne{\hfill\execneuf {10(1)} { 49}  { 49}  { 49}  { 56}  { 69}  { 65}  { 49}  { 49} \hfill}
\ligne{\hfill\execneuf {3(1) } { 66}  { 66}  { 66}  { 67}  { 70}  { 62}  { 68}  { 64} \hfill}
\ligne{\hfill\execneuf {7(1) } { 49}  { 49}  { 49}  { 49}  { 49}  { 56}  { 69}  { 65} \hfill}
\ligne{\hfill\execneuf {2(1) } { 74}  { 74}  { 74}  { 74}  { 74}  {277}  { 98}  { 71} \hfill}
\ligne{\hfill\execneuf {8(1) } { 36}  { 63}  { 93}  {245}  { 96}  { 97}  {127}  {125} \hfill}
\ligne{\hfill\execneuf {20(1)} {125}  { 29}  {  6}  {  6}  {  6}  {  6}  {  6}  { 36} \hfill}
\ligne{\hfill\execneuf {18(1)} {  6}  {  6}  {  6}  {  6}  {  6}  { 63}  { 36}  {  6} \hfill}
\ligne{\hfill\execneuf {1(1) } {  6}  {  6}  {  6}  {  6}  {  6}  { 29}  {125}  {  6} \hfill}
\ligne{\hfill\execneuf {0(0) } {  2}  {  2}  {  2}  {  2}  {  2}  {  2}  {  7}  { 13} \hfill}
}
\hfill}
}

\ifnum 1=0 {
% rond3hs
\ligne{\hfill\execneuf    {}      1      2       3       4       5       6       7       8 \hfill}
\ligne{\hfill\execneuf {2(5) } { 91}  { 92}  {  94}  {  95}  {  90}  {  90}  {  90}  {  90} \hfill}
\ligne{\hfill\execneuf {5(5) } { 49}  { 56}  {  69}  {  65}  {  49}  {  49}  {  49}  {  49} \hfill}
\ligne{\hfill\execneuf {4(4) } { 66}  { 67}  {  70}  {  62}  {  68}  {  64}  {  66}  {  66} \hfill}
\ligne{\hfill\execneuf {10(4)} { 49}  { 49}  {  49}  {  56}  {  69}  {  65}  {  49}  {  49} \hfill}
\ligne{\hfill\execneuf {3(4) } { 66}  { 66}  {  66}  {  67}  {  70}  {  62}  {  68}  {  64} \hfill}
\ligne{\hfill\execneuf {7(4) } { 49}  { 49}  {  49}  {  49}  {  49}  {  56}  {  69}  {  65} \hfill}
\ligne{\hfill\execneuf {2(4) } { 74}  { 74}  {  74}  {  74}  {  74}  { 277}  {  98}  {  71} \hfill}
\ligne{\hfill\execneuf {0(0) } { 28}  { 34}  {  24}  {  24}  {  24}  {  24}  {  24}  {  99} \hfill}
\ligne{\hfill\execneuf {1(4) } { 36}  { 63}  {  93}  { 245}  {  96}  {  97}  { 127}  { 125} \hfill}
\ligne{\hfill\execneuf {8(4) } {  6}  {  6}  {   6}  {   6}  {   6}  {  63}  {  36}  {   6} \hfill}
\ligne{\hfill\execneuf {20(4)} {  6}  {  6}  {   6}  {   6}  {   6}  {  29}  { 125}  {   6} \hfill}
\ligne{\hfill\execneuf {18(4)} {  2}  {  2}  {   2}  {   2}  {   2}  {   2}  {   7}  {  13} \hfill}
} \fi

\vskip 10pt
\vtop{
\begin{tab}\label{exrond2hd}
\leurre
Execution of the circuit of a double locomotive clockwise running around two contiguous
cells.
\end{tab}
\vskip-6pt
\ligne{\hfill 
\vtop{\leftskip 0pt\parindent 0pt\hsize=210pt
\ligne{\hfill\execneuf    {}      7      8      9     10     11     12     13     14 \hfill}
\ligne{\hfill\execneuf {2(2) } {104}  {105}  {107}  { 95}  { 90}  { 90}  { 90}  { 90} \hfill}
\ligne{\hfill\execneuf {5(2) } { 56}  { 80}  { 83}  { 65}  { 49}  { 49}  { 49}  { 49} \hfill}
\ligne{\hfill\execneuf {4(1) } { 67}  { 81}  { 84}  { 79}  { 82}  { 64}  { 66}  { 66} \hfill}
\ligne{\hfill\execneuf {10(1)} { 49}  { 49}  { 56}  { 80}  { 83}  { 65}  { 49}  { 49} \hfill}
\ligne{\hfill\execneuf {3(1) } { 66}  { 66}  { 67}  { 81}  { 84}  { 79}  { 82}  { 64} \hfill}
\ligne{\hfill\execneuf {7(1) } { 49}  { 49}  { 49}  { 49}  { 56}  { 80}  { 83}  { 65} \hfill}
\ligne{\hfill\execneuf {2(1) } { 74}  { 74}  { 74}  { 74}  {277}  {109}  {171}  { 88} \hfill}
\ligne{\hfill\execneuf {8(1) } { 44}  {253}  {106}  {111}  {300}  {108}  {137}  {142} \hfill}
\ligne{\hfill\execneuf {20(1)} {142}  { 29}  {  6}  {  6}  {  6}  {  6}  { 36}  { 44} \hfill}
\ligne{\hfill\execneuf {18(1)} {  6}  {  6}  {  6}  {  6}  { 63}  { 44}  { 36}  {  6} \hfill}
\ligne{\hfill\execneuf {1(1) } {  6}  {  6}  {  6}  {  6}  { 29}  {142}  {125}  {  6} \hfill}
\ligne{\hfill\execneuf {0(0) } {  2}  {  2}  {  2}  {  2}  {  2}  {  7}  { 40}  { 13} \hfill}
}
\hfill}
}

\vskip 10pt
\ifnum 1=0 {
% rond3hd
\ligne{\hfill\execneuf    {}      1      2      3      4      5      6      7      8 \hfill}
\ligne{\hfill\execneuf {2(5) } {104}  {105}  {107}  { 95}  { 90}  { 90}  { 90}  { 90} \hfill}
\ligne{\hfill\execneuf {5(5) } { 56}  { 80}  { 83}  { 65}  { 49}  { 49}  { 49}  { 49} \hfill}
\ligne{\hfill\execneuf {4(4) } { 67}  { 81}  { 84}  { 79}  { 82}  { 64}  { 66}  { 66} \hfill}
\ligne{\hfill\execneuf {10(4)} { 49}  { 49}  { 56}  { 80}  { 83}  { 65}  { 49}  { 49} \hfill}
\ligne{\hfill\execneuf {3(4) } { 66}  { 66}  { 67}  { 81}  { 84}  { 79}  { 82}  { 64} \hfill}
\ligne{\hfill\execneuf {7(4) } { 49}  { 49}  { 49}  { 49}  { 56}  { 80}  { 83}  { 65} \hfill}
\ligne{\hfill\execneuf {2(4) } { 74}  { 74}  { 74}  { 74}  {277}  {109}  {171}  { 88} \hfill}
\ligne{\hfill\execneuf {0(0) } { 42}  { 34}  { 24}  { 24}  { 24}  { 24}  { 99}  {110} \hfill}
\ligne{\hfill\execneuf {1(4) } { 44}  {253}  {106}  {111}  {300}  {108}  {137}  {142} \hfill}
\ligne{\hfill\execneuf {8(4) } {  6}  {  6}  {  6}  {  6}  { 63}  { 44}  { 36}  {  6} \hfill}
\ligne{\hfill\execneuf {20(4)} {  6}  {  6}  {  6}  {  6}  { 29}  {142}  {125}  {  6} \hfill}
\ligne{\hfill\execneuf {18(4)} {  2}  {  2}  {  2}  {  2}  {  2}  {  7}  { 40}  { 13} \hfill}
} \fi

\vtop{
\begin{tab}\label{exrond2ahs}
\leurre
Execution of the circuit of a simple locomotive counter-clockwise running around two contiguous
cells.
\end{tab}
\vskip-6pt
\ligne{\hfill 
\vtop{\leftskip 0pt\parindent 0pt\hsize=210pt
\ligne{\hfill\execneuf    {}      0      1      2      3      4      5      6      7 \hfill}
\ligne{\hfill\execneuf {2(1) } {155}  {159}  {164}  {174}  {174}  {174}  {174}  {174} \hfill}
\ligne{\hfill\execneuf {7(1) } { 35}  { 78}  { 72}  { 30}  { 30}  { 30}  { 30}  { 30} \hfill}
\ligne{\hfill\execneuf {3(1) } { 73}  { 77}  { 71}  { 76}  { 75}  { 74}  { 74}  { 74} \hfill}
\ligne{\hfill\execneuf {10(1)} { 30}  { 30}  { 35}  { 78}  { 72}  { 30}  { 30}  { 30} \hfill}
\ligne{\hfill\execneuf {4(1) } { 74}  { 74}  { 73}  { 77}  { 71}  { 76}  { 75}  { 74} \hfill}
\ligne{\hfill\execneuf {5(2) } { 30}  { 30}  { 30}  { 30}  { 35}  { 78}  { 72}  { 30} \hfill}
\ligne{\hfill\execneuf {2(2) } { 66}  { 66}  { 66}  { 66}  {246}  {250}  { 62}  { 68} \hfill}
\ligne{\hfill\execneuf {0(0) } { 36}  {  6}  {  6}  {  6}  {  6}  {  6}  { 29}  {125} \hfill}
\ligne{\hfill\execneuf {1(1) } {125}  {127}  { 97}  { 96}  {245}  { 93}  { 63}  { 36} \hfill}
\ligne{\hfill\execneuf {18(1)} { 36}  { 63}  {  6}  {  6}  {  6}  {  6}  {  6}  {  6} \hfill}
\ligne{\hfill\execneuf {20(1)} {  2}  {  7}  { 13}  {  2}  {  2}  {  2}  {  2}  {  2} \hfill}
\ligne{\hfill\execneuf {8(1) } {  6}  {  6}  {125}  { 29}  {  6}  {  6}  {  6}  {  6} \hfill}
}
\hfill}
}
\vskip 10pt

\vtop{
\begin{tab}\label{exrond2ahd}
\leurre
Execution of the circuit of a double locomotive counter-clockwise running around two contiguous
cells.
\end{tab}
\vskip-6pt
\ligne{\hfill 
\vtop{\leftskip 0pt\parindent 0pt\hsize=210pt
\ligne{\hfill\execneuf    {}      0      1      2      3      4      5      6      7 \hfill}
\ligne{\hfill\execneuf {2(1) } {114}  {115}  {116}  {164}  {174}  {174}  {174}  {174} \hfill}
\ligne{\hfill\execneuf {7(1) } { 35}  { 43}  { 86}  { 72}  { 30}  { 30}  { 30}  { 30} \hfill}
\ligne{\hfill\execneuf {3(1) } { 73}  { 89}  { 85}  { 88}  { 87}  { 75}  { 74}  { 74} \hfill}
\ligne{\hfill\execneuf {10(1)} { 30}  { 30}  { 35}  { 43}  { 86}  { 72}  { 30}  { 30} \hfill}
\ligne{\hfill\execneuf {4(1) } { 74}  { 74}  { 73}  { 89}  { 85}  { 88}  { 87}  { 75} \hfill}
\ligne{\hfill\execneuf {5(2) } { 30}  { 30}  { 30}  { 30}  { 35}  { 43}  { 86}  { 72} \hfill}
\ligne{\hfill\execneuf {2(2) } { 66}  { 66}  { 66}  { 66}  {246}  {117}  {118}  { 79} \hfill}
\ligne{\hfill\execneuf {0(0) } { 44}  { 36}  {  6}  {  6}  {  6}  {  6}  { 29}  {142} \hfill}
\ligne{\hfill\execneuf {1(1) } {142}  {137}  {108}  {300}  {111}  {106}  {253}  { 44} \hfill}
\ligne{\hfill\execneuf {18(1)} { 36}  { 44}  { 63}  {  6}  {  6}  {  6}  {  6}  {  6} \hfill}
\ligne{\hfill\execneuf {20(1)} {  2}  {  7}  { 61}  { 13}  {  2}  {  2}  {  2}  {  2} \hfill}
\ligne{\hfill\execneuf {8(1) } {  6}  {  6}  {125}  {142}  { 29}  {  6}  {  6}  {  6}   \hfill}
}
\hfill}
}
\vskip 10pt

\vtop{
%\ligne{\hrulefill}
%\vskip 3pt
\ligne{\hfill
%\phantom{
\includegraphics[scale=0.4]{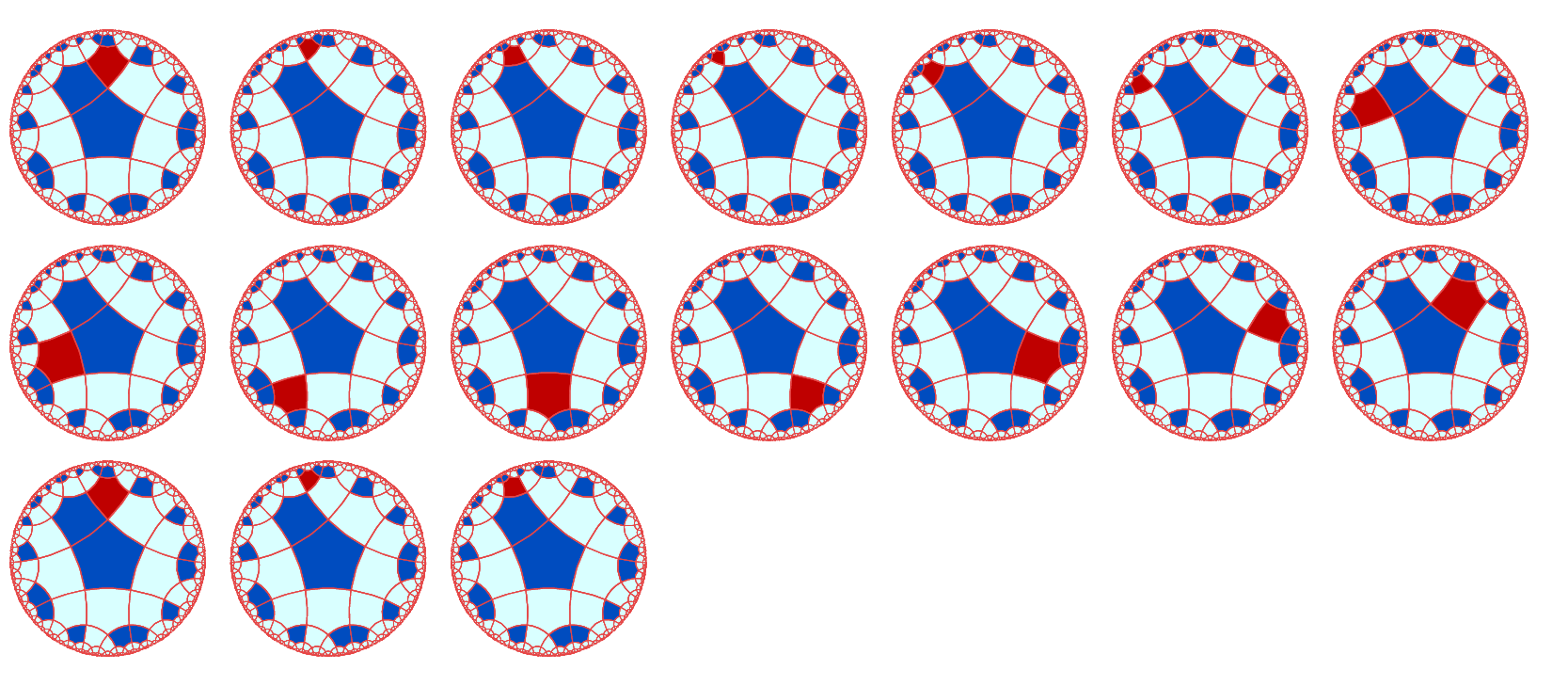} %}
\hfill}
\ligne{\hfill
%\phantom{
\includegraphics[scale=0.4]{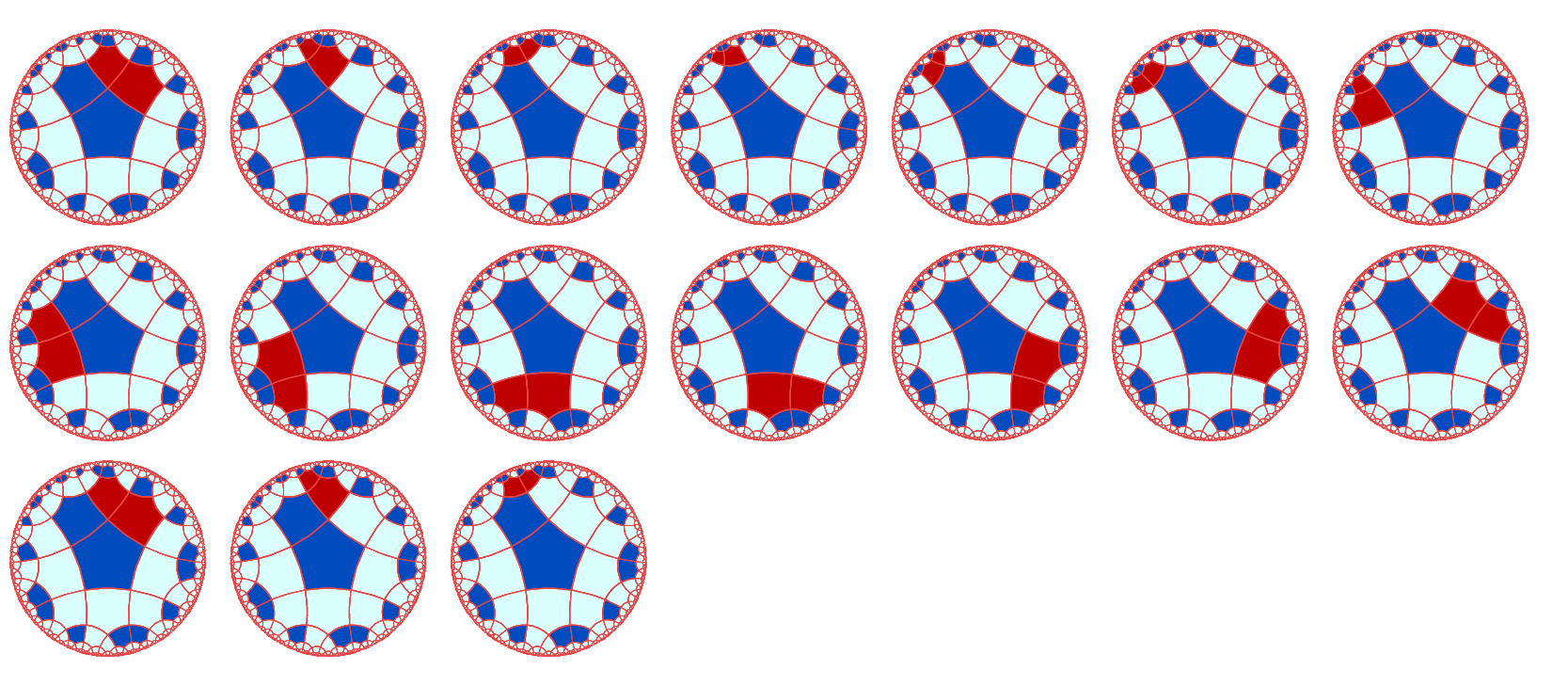} %}
\hfill}
\begin{fig}\label{rond2ah}
\leurre
Above : a single locomotive endlessly turning around two cells in a counter-clockwise orientation.
Below:  a double locomotive endlessly turning around two cells in a counter-clockwise orientation.
\end{fig}
}

\vtop{
\begin{tab}\label{round3htab}
\leurre
Rules for turning around three contiguous cells: 
\end{tab}
\vspace{-12pt}
\ligne{\hfill clockwise:\hfill\hfill counter-clockwise:\hskip 20pt}
\vskip 2pt
\ligne{\hfill single:\hskip 50pt double: \hfill\hfill\hskip 70pt double:\hfill}
\vskip 2pt
\ligne{\hfill
\vtop{\leftskip 0pt\parindent 0pt\hsize=76pt
\aff { 90} {W} {WBWBWBBWBW} {W}
\aff { 91} {W} {RBWBWBBWBW} {R}
\aff { 92} {R} {WBWBWBBWBW} {W}
\aff { 93} {B} {BWWWWWWWWR} {B}
\aff { 94} {W} {WBRBWBBWBW} {W}
\aff { 95} {W} {WBWBWBBRBW} {W}
\aff { 96} {B} {BWWWWWWWRW} {B}
\aff { 97} {B} {BWRWWWWWWW} {B}
\aff { 98} {W} {BWBWRWBWWB} {R}
\aff { 99} {B} {BWWBWWWWRW} {B}
}\hfill
\vtop{\leftskip 0pt\parindent 0pt\hsize=76pt
\aff {100} {B} {BWRBWWWWWW} {B}
\aff {101} {B} {BWWBWWWRWW} {B}
\aff {102} {B} {BRWBWWWWWW} {B}
\aff {103} {B} {BWWBWWRWWW} {B}
}
\hfill\hskip 15pt
\vtop{\leftskip 0pt\parindent 0pt\hsize=76pt
\aff {104} {R} {RBWBWBBWBW} {R}
\aff {105} {R} {WBRBWBBWBW} {W}
\aff {106} {B} {BWWRWWWWWR} {B}
\aff {107} {W} {WBRBWBBRBW} {W}
\aff {108} {B} {BWRWWWWRWW} {B}
\aff {109} {W} {BWBWRRBWWB} {R}
\aff {110} {B} {BWRBWWWWRW} {B}
\aff {111} {B} {BWWRWWWWRW} {B}
\aff {112} {B} {BWRBWWWRWW} {B}
\aff {113} {B} {BRWBWWWRWW} {B}
}
\hfill\hskip 30pt
\vtop{\leftskip 0pt\parindent 0pt\hsize=76pt
\aff {114} {R} {BRWBWWBBWB} {R}
\aff {115} {R} {BWWBRWBBWB} {W}
\aff {116} {W} {BWWBRRBBWB} {W}
\aff {117} {W} {WBRWBWBRBW} {R}
\aff {118} {R} {WBRWBWBWBW} {R}
}
\hfill}
}
\vskip 15pt
\vtop{
%\ligne{\hrulefill}
%\vskip 3pt
\ligne{\hfill
%\phantom{
\includegraphics[scale=0.4]{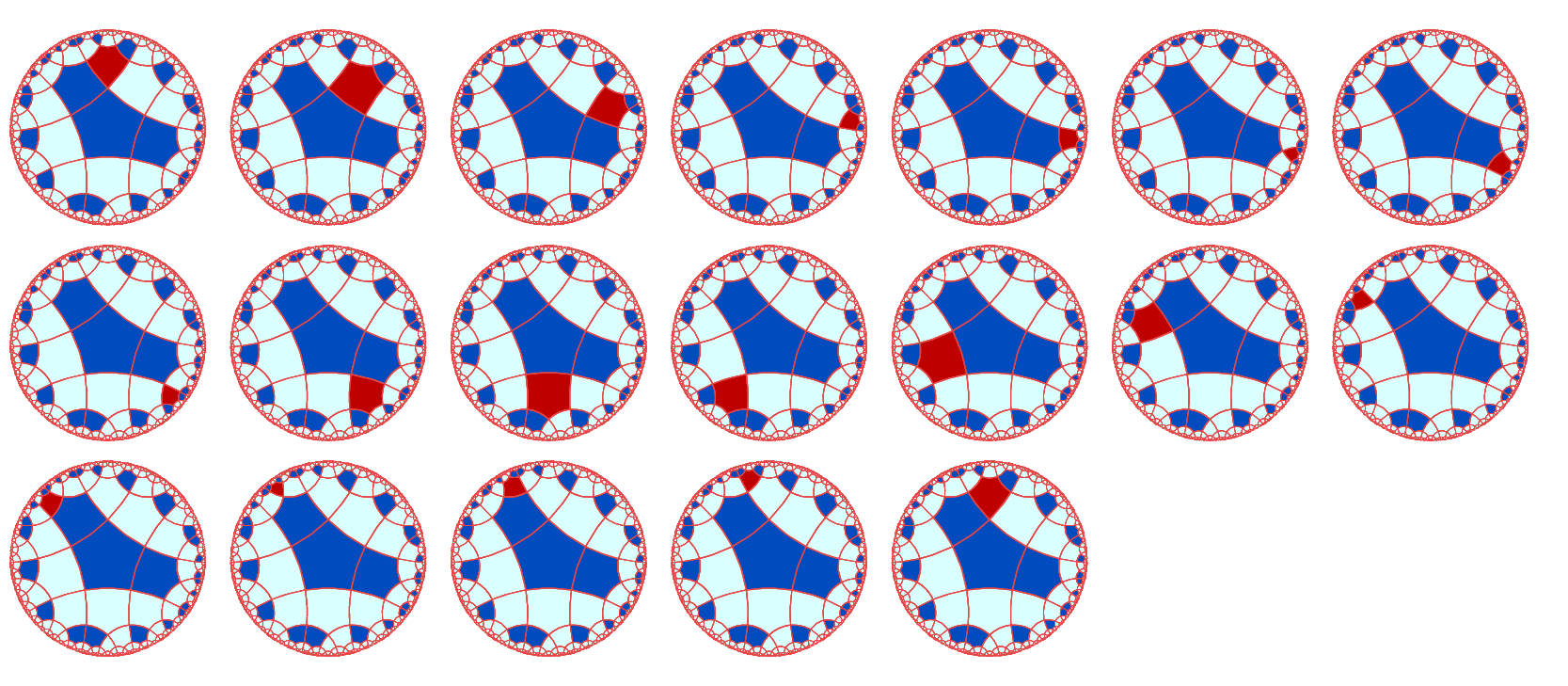} %}
\hfill}
\ligne{\hfill
%\phantom{
\includegraphics[scale=0.4]{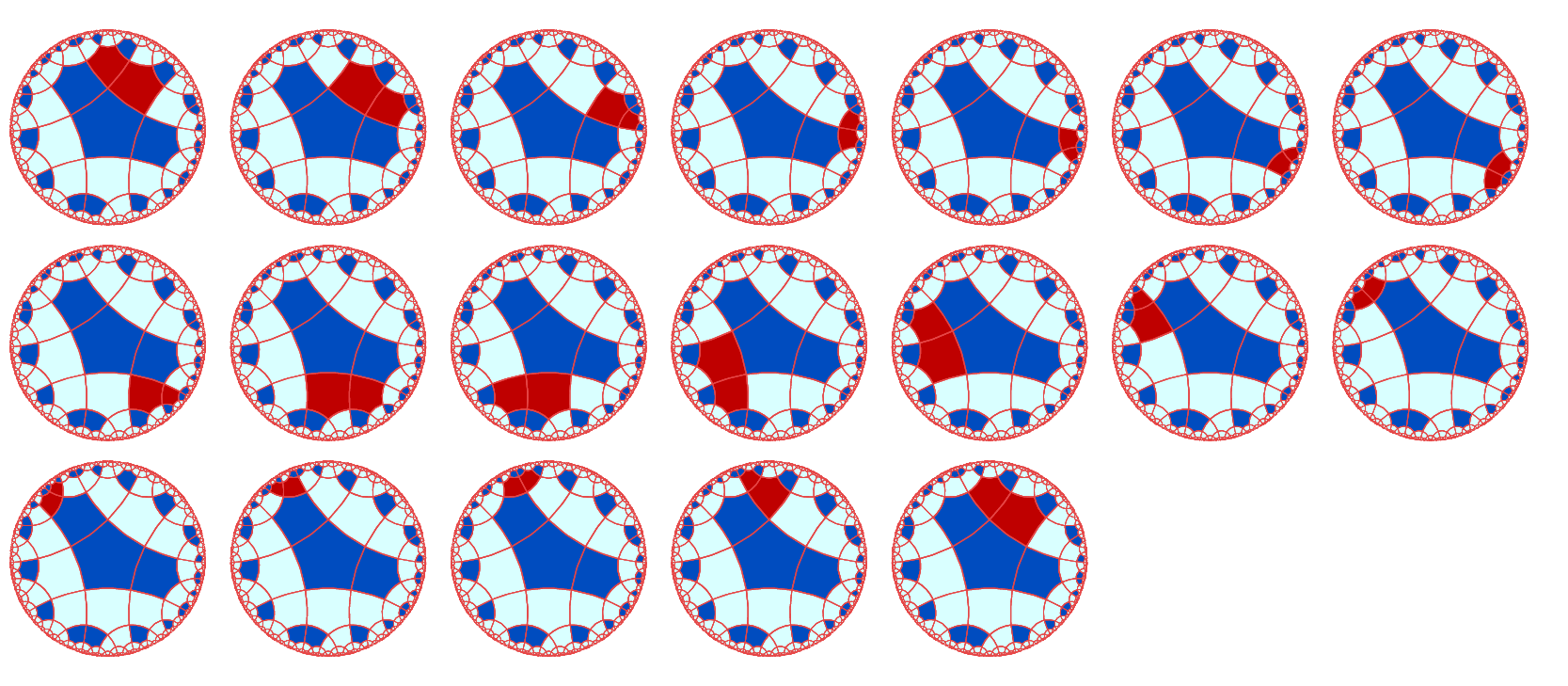} %}
\hfill}
\begin{fig}\label{rond3h}
\leurre
Above : a single locomotive endlessly turning around three cells in a clockwise orientation.
Below:  a double locomotive endlessly turning around three cells in a clockwise orientation.
\end{fig}
}

\vtop{
\begin{tab}\label{exrond3ahs}
\leurre
Execution of the circuit of a simple locomotive counter-clockwise running around 
\textbf{three} contiguous cells.
\end{tab}
\vskip -6pt
\ligne{\hfill
\vtop{\leftskip 0pt\parindent 0pt\hsize=210pt
\ligne{\hfill\execneuf    {}      9     10      11      12      13      14      15      16 \hfill}
\ligne{\hfill\execneuf {2(4) } {145}  {155}  { 159}  { 164}  { 174}  { 174}  { 174}  { 174} \hfill}
\ligne{\hfill\execneuf {7(4) } { 30}  { 35}  {  78}  {  72}  {  30}  {  30}  {  30}  {  30} \hfill}
\ligne{\hfill\execneuf {3(4) } { 74}  { 73}  {  77}  {  71}  {  76}  {  75}  {  74}  {  74} \hfill}
\ligne{\hfill\execneuf {10(4)} { 30}  { 30}  {  30}  {  35}  {  78}  {  72}  {  30}  {  30} \hfill}
\ligne{\hfill\execneuf {4(4) } { 74}  { 74}  {  74}  {  73}  {  77}  {  71}  {  76}  {  75} \hfill}
\ligne{\hfill\execneuf {5(5) } { 30}  { 30}  {  30}  {  30}  {  30}  {  35}  {  78}  {  72} \hfill}
\ligne{\hfill\execneuf {2(5) } { 66}  { 66}  {  66}  {  66}  {  66}  { 246}  { 250}  {  62} \hfill}
\ligne{\hfill\execneuf {18(4)} {  6}  { 36}  {  63}  {   6}  {   6}  {   6}  {   6}  {   6} \hfill}
\ligne{\hfill\execneuf {20(4)} {  2}  {  2}  {   7}  {  13}  {   2}  {   2}  {   2}  {   2} \hfill}
\ligne{\hfill\execneuf {8(4) } { 48}  { 48}  {  52}  {  54}  {  48}  {  48}  {  48}  {  48} \hfill}
\ligne{\hfill\execneuf {1(4) } { 29}  {125}  { 127}  {  97}  {  96}  { 245}  {  93}  {  63} \hfill}
\ligne{\hfill\execneuf {0(0) } {100}  { 99}  {  24}  {  24}  {  24}  {  24}  {  24}  {  34} \hfill}
}
\hfill}
}
\vskip 10pt

\vtop{
%\ligne{\hrulefill}
%\vskip 3pt
\ligne{\hfill
%\phantom{
\includegraphics[scale=0.4]{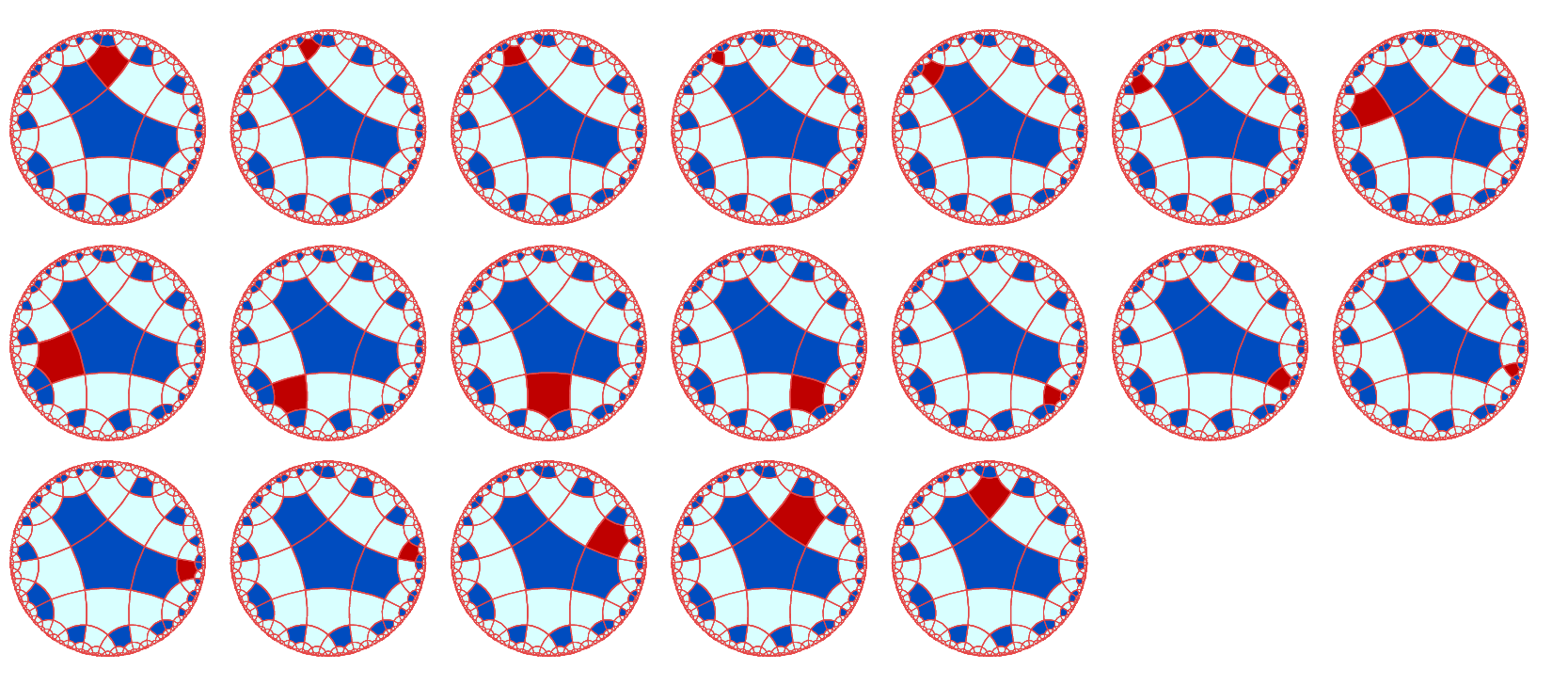} %}
\hfill}
\ligne{\hfill
%\phantom{
\includegraphics[scale=0.4]{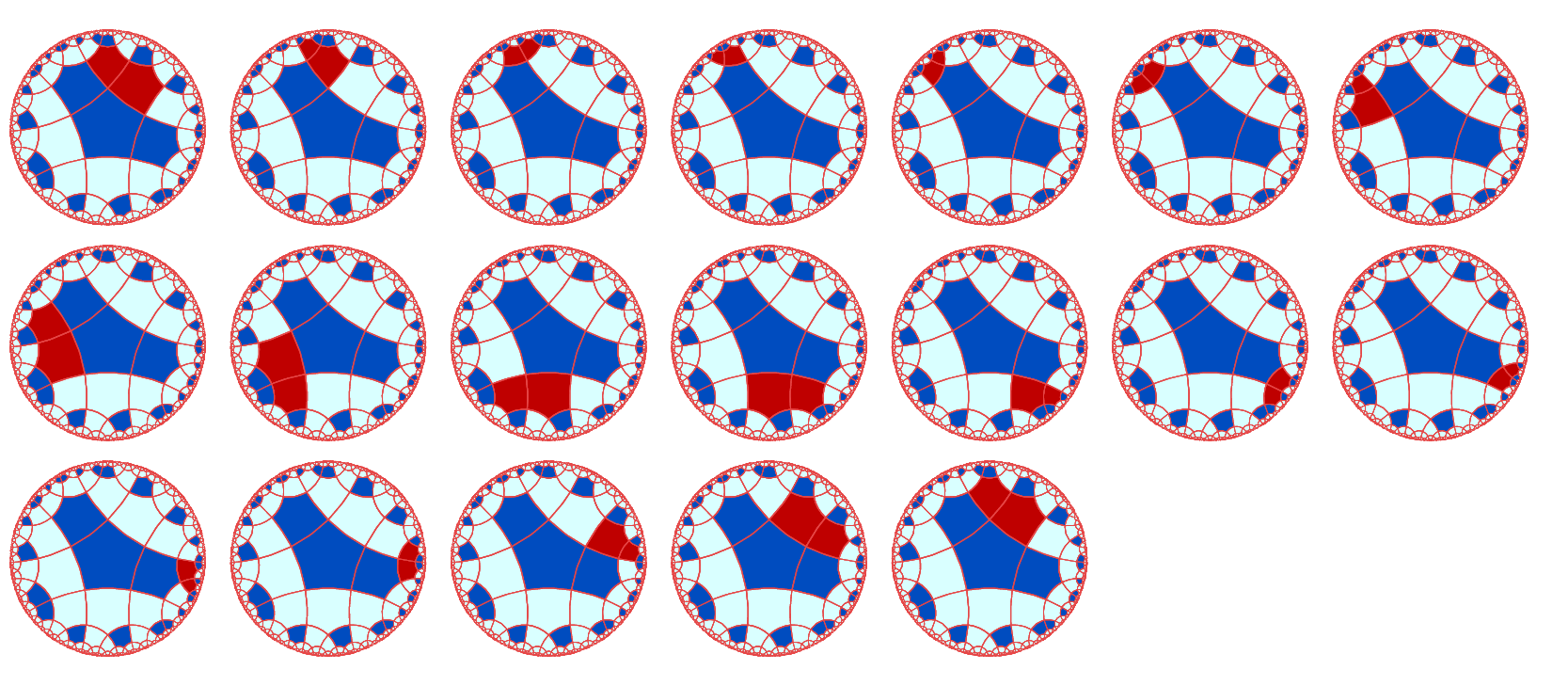} %}
\hfill}
\begin{fig}\label{rond3ah}
\leurre
Above : a single locomotive endlessly turning around three cells in a counter-clockwise orientation.
Below:  a double locomotive endlessly turning around three cells in a counter-clockwise orientation.
\end{fig}
}

\vtop{
\begin{tab}\label{exrond3ahd}
\leurre
Execution of the circuit of a double locomotive counter-clockwise running around 
\textbf{three} contiguous cells.
\end{tab}
\vskip -6pt
\ligne{\hfill
\vtop{\leftskip 0pt\parindent 0pt\hsize=210pt
\ligne{\hfill\execneuf    {}      9     10     11     12     13     14     15     16 \hfill}
\ligne{\hfill\execneuf {2(4) } {145}  {114}  {115}  {116}  {164}  {174}  {174}  {174} \hfill}
\ligne{\hfill\execneuf {7(4) } { 30}  { 35}  { 43}  { 86}  { 72}  { 30}  { 30}  { 30} \hfill}
\ligne{\hfill\execneuf {3(4) } { 74}  { 73}  { 89}  { 85}  { 88}  { 87}  { 75}  { 74} \hfill}
\ligne{\hfill\execneuf {10(4)} { 30}  { 30}  { 30}  { 35}  { 43}  { 86}  { 72}  { 30} \hfill}
\ligne{\hfill\execneuf {4(4) } { 74}  { 74}  { 74}  { 73}  { 89}  { 85}  { 88}  { 87} \hfill}
\ligne{\hfill\execneuf {5(5) } { 30}  { 30}  { 30}  { 30}  { 30}  { 35}  { 43}  { 86} \hfill}
\ligne{\hfill\execneuf {2(5) } { 66}  { 66}  { 66}  { 66}  { 66}  {246}  {117}  {118} \hfill}
\ligne{\hfill\execneuf {18(4)} {  6}  { 36}  { 44}  { 63}  {  6}  {  6}  {  6}  {  6} \hfill}
\ligne{\hfill\execneuf {20(4)} {  2}  {  2}  {  7}  { 61}  { 13}  {  2}  {  2}  {  2} \hfill}
\ligne{\hfill\execneuf {8(4) } { 48}  { 48}  { 52}  { 59}  { 54}  { 48}  { 48}  { 48} \hfill}
\ligne{\hfill\execneuf {1(4) } { 29}  {142}  {137}  {108}  {300}  {111}  {106}  {253} \hfill}
\ligne{\hfill\execneuf {0(0) } {112}  {110}  { 99}  { 24}  { 24}  { 24}  { 24}  { 34} \hfill}
}
\hfill}
}

\subsection{The fixed switch}
\label{fixedswitch}

   Figures~\ref{fixeg} and~\ref{fixed} illustrate how the locomotive passively crosses
a fixed switch. Both figures illustrate the case when the locomotive comes from the
left-hand side and when it comes from the right-hand side. 

\vtop{
\vspace{-10pt}
\begin{tab}\label{fixtab}
\leurre
Rules for the passive fixed switch:
\end{tab}
\vspace{-12pt}
\ligne{\hfill single locomotive:\hfill\hfill double locomotive:\hskip 20pt}
\vskip 2pt
\ligne{\hfill
\vtop{\leftskip 0pt\parindent 0pt\hsize=76pt % simple, de la gauche
\ligne{\hfill from left:\hfill}
\vskip 2pt
\aff {119} {W} {WBWBWBBBWB} {W}
\aff {120} {W} {WBWWWBBWWW} {W}
\aff {121} {W} {WBWWBWBBWW} {W}
\aff {122} {W} {RBWBWBBBWB} {R}
\aff {123} {W} {WBWWBRBBBW} {W}
\aff {124} {R} {WBWBWBBBWB} {W}
\aff {125} {B} {RWWWBWWWWW} {B}
\aff {126} {W} {WBRBWBBBWB} {W}
\aff {127} {B} {WWWWBWRWWW} {B}
}
\hfill
\vtop{\leftskip 0pt\parindent 0pt\hsize=76pt % simple, de la droite
\ligne{\hfill from right:\hfill}
\vskip 2pt
\aff {128} {R} {WBWWBWBBWW} {W}
\aff {129} {W} {RBWWWWBWWW} {W}
\aff {130} {W} {RWWWWBWWWW} {W}
\aff {131} {W} {RBWWBWBBWW} {W}
\aff {132} {W} {WBWBRBBBWB} {R}
\aff {133} {W} {WBWWBBRWBB} {W}
}
\hfill
\vtop{\leftskip 0pt\parindent 0pt\hsize=76pt % double, de la gauche
\ligne{\hfill from left:\hfill}
\vskip 2pt
\aff {134} {R} {RBWBWBBBWB} {R}
\aff {135} {W} {RBWWBRBBBW} {W}
\aff {136} {R} {WBRBWBBBWB} {W}
\aff {137} {B} {RWWWBWRWWW} {B}
\aff {138} {W} {RBWWWBBWWW} {W}
}
\hfill
\vtop{\leftskip 0pt\parindent 0pt\hsize=76pt % double, de la droite
\ligne{\hfill from right:\hfill}
\vskip 2pt
\aff {139} {R} {RBWWBWBBWW} {W}
\aff {140} {R} {WBWBRBBBWB} {R}
\aff {141} {W} {RBWWBBRWBB} {W}
\aff {142} {B} {RWWWBRWWWW} {B}
}
\hfill}
}
\vskip 10pt
Note that the switch is passive
and that it is the same, whether the locomotive comes from the left or from the right. 
The fixed switch makes use the property of the basic element to accept an entering locomotive
from any one of its two consecutive white neighbours through a side. The other parts
of the switch consists of basic elements only. They are assembled in an appropriate way.
Again, these figures are enough to establish the rules, both for a simple and a double
locomotive. Table~\ref{fixtab} gives the rules for the motion of the locomotive through
the switch together with those for the milestones involved in this structure. 
Tables~\ref{exfixs} and~\ref{exfixd} shows traces of execution of the rules displayed by
Table~\ref{fixtab}.

\vtop{
%\ligne{\hrulefill}
%\vskip 3pt
\ligne{\hfill
%\phantom{
\includegraphics[scale=0.4]{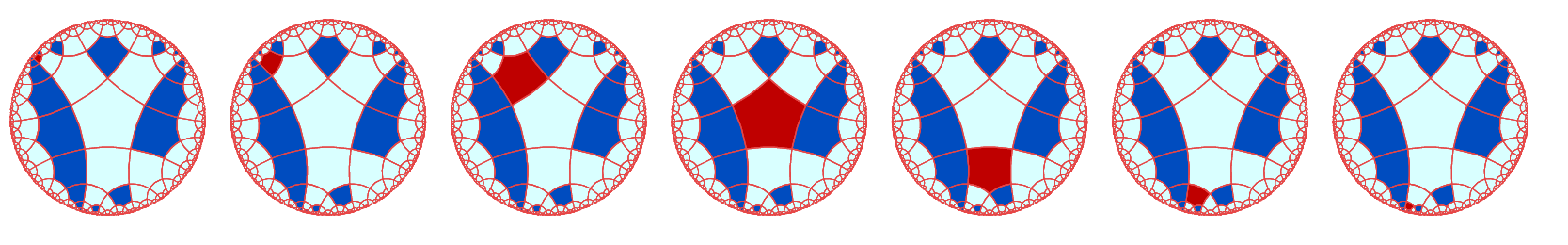} %}
\hfill}
\ligne{\hfill
%\phantom{
\includegraphics[scale=0.4]{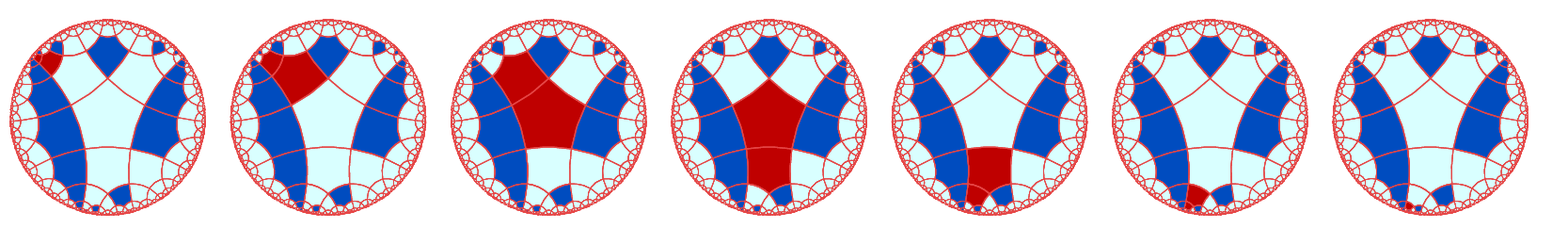} %}
\hfill}
\begin{fig}\label{fixeg}
\leurre
Locomotives passively crossing a fixed switch from the left-hand side. Above : a single
locomotive; below: a double one.
\end{fig}
}
\vskip 10pt
\vtop{
%\ligne{\hrulefill}
%\vskip 3pt
\ligne{\hfill
%\phantom{
\includegraphics[scale=0.4]{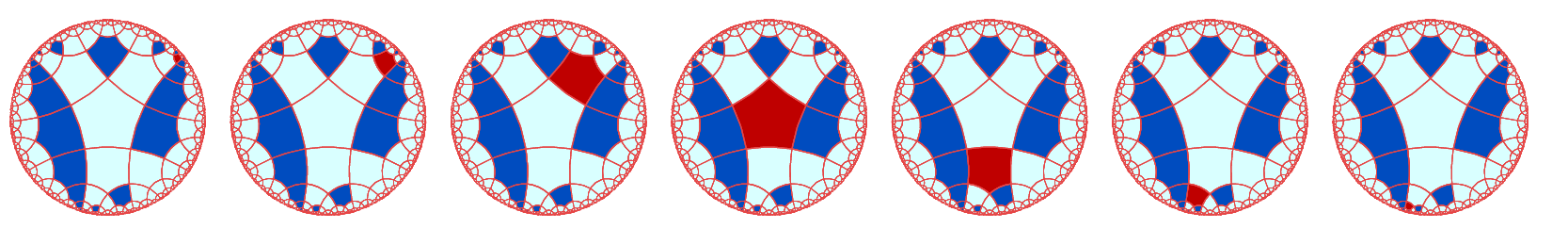} %}
\hfill}
\ligne{\hfill
%\phantom{ 
\includegraphics[scale=0.4]{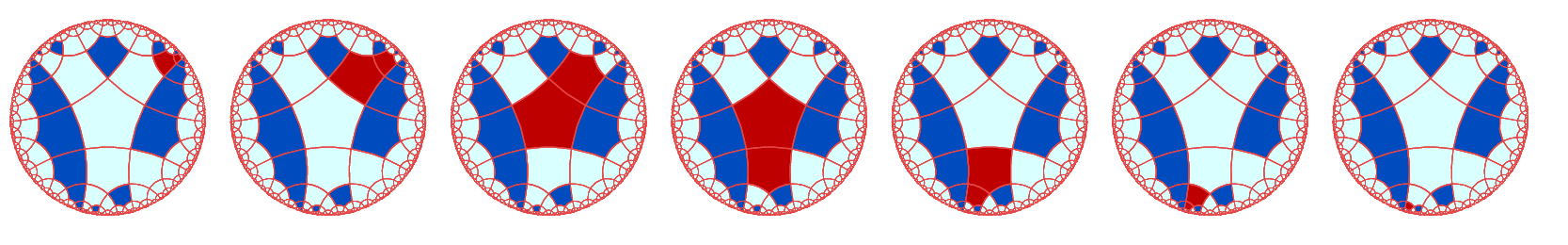} %}
\hfill}
\begin{fig}\label{fixed}
\leurre
Locomotives passively crossing a fixed switch from the right-hand side. Above : a single
locomotive; below: a double one.
\end{fig}
}

\vtop{
\begin{tab}\label{exfixs}
\leurre
Execution of the rules on the cells~$1(1)$, $1(5)$, $0(0)$, $1(3)$, $1(2)$ and~$1(4)$
when a simple locomotive passively crosses the fixed switch.
\end{tab}
\vskip-6pt
\ligne{\hfill\hbox to 180pt{\hfill from the left:\hfill}
\hskip 5pt\hbox to 155pt{\hfill from the right:\hfill}\hfill}
\vskip 2pt
\ligne{\hfill
\vtop{\leftskip 0pt\parindent 0pt\hsize=180pt
\ligne{\hfill  \execsept     {}       0       1       2       3       4       5 \hfill}
\ligne{\hfill  \execsept  {1(1)}  {  17}  {  18}  {  20}  {  32}  {  17}  {  17}   \hfill}
\ligne{\hfill  \execsept  {1(5)}  {  47}  {  47}  { 123}  {  53}  {  47}  {  47}   \hfill}
\ligne{\hfill  \execsept  {0(0)}  { 119}  { 119}  { 122}  { 124}  { 126}  { 119}   \hfill}
\ligne{\hfill  \execsept  {1(3)}  {  17}  {  17}  {  17}  {  18}  {  20}  {  32}   \hfill}
\ligne{\hfill  \execsept  {1(2)}  {  24}  {  24}  {  27}  {  28}  {  34}  {  24}   \hfill}
\ligne{\hfill  \execsept  {1(4)}  {   6}  {   6}  {   6}  { 125}  { 127}  {   6}   \hfill}
}
\hskip 5pt
\vtop{\leftskip 0pt\parindent 0pt\hsize=155pt
\ligne{\hfill   \execsix       0       1       2       3       4       5 \hfill}
\ligne{\hfill   \execsix   {  17}  {  17}  { 133}  {  32}  {  17}  {  17}   \hfill}
\ligne{\hfill   \execsix   {  47}  {  50}  {  51}  {  53}  {  47}  {  47}   \hfill}
\ligne{\hfill   \execsix   { 119}  { 119}  { 132}  { 124}  { 126}  { 119}   \hfill}
\ligne{\hfill   \execsix   {  17}  {  17}  {  17}  {  18}  {  20}  {  32}   \hfill}
\ligne{\hfill   \execsix   {  24}  {  24}  {  24}  {  28}  {  34}  {  24}   \hfill}
\ligne{\hfill   \execsix   {   6}  {   6}  {  29}  { 125}  { 127}  {   6}   \hfill}
}
\hfill}
}
\vskip 10pt
These traces deal with the cells 1(1), 1(5), 0(0) and~1(3) for the tracks and
1(2) together with 1(4) for the particular milestones which are not in the situation of
milestones in a vertical track or around a set of cells. 

In Table~\ref{exfixs}, we can see that rule~17 applies to~1(1). Note that the neighbourhood
which is given in Table~\ref{verttracktab} is that to cell~0(0) for which it is also the 
conservative rule. In Table~\ref{exfixs}, the cells 1(1) and~1(3) are applied the same rules
with a delay of two ticks of the clock.

\vtop{
\begin{tab}\label{exfixd}
\leurre
Execution of the rules on the cells~$1(1)$, $1(5)$, $0(0)$, $1(3)$, $1(2)$ and~$1(4)$
when a double locomotive passively crosses the fixed switch.
\end{tab}
\vskip-6pt
\ligne{\hfill\hbox to 180pt{\hfill from the left:\hfill}
\hskip 5pt\hbox to 155pt{\hfill from the right:\hfill}\hfill}
\vskip 2pt
\ligne{\hfill
\vtop{\leftskip 0pt\parindent 0pt\hsize=180pt
\ligne{\hfill  \execsept     {}       0       1       2       3       4       5 \hfill}
\ligne{\hfill  \execsept  {1(1)}  {  18}  {  37}  {  39}  {  32}  {  17}  {  17}   \hfill}
\ligne{\hfill  \execsept  {1(5)}  {  47}  { 123}  { 135}  {  53}  {  47}  {  47}   \hfill}
\ligne{\hfill  \execsept  {0(0)}  { 119}  { 122}  { 134}  { 136}  { 126}  { 119}   \hfill}
\ligne{\hfill  \execsept  {1(3)}  {  17}  {  17}  {  18}  {  37}  {  39}  {  32}   \hfill}
\ligne{\hfill  \execsept  {1(2)}  {  24}  {  27}  {  41}  {  42}  {  34}  {  24}   \hfill}
\ligne{\hfill  \execsept  {1(4)}  {   6}  {   6}  { 125}  { 137}  { 127}  {   6}   \hfill}
}
\hskip 5pt
\vtop{\leftskip 0pt\parindent 0pt\hsize=155pt
\ligne{\hfill   \execsix       0       1       2       3       4       5 \hfill}
\ligne{\hfill   \execsix   {  17}  { 133}  { 141}  {  32}  {  17}  {  17}   \hfill}
\ligne{\hfill   \execsix   {  50}  {  57}  {  58}  {  53}  {  47}  {  47}   \hfill}
\ligne{\hfill   \execsix   { 119}  { 132}  { 140}  { 136}  { 126}  { 119}   \hfill}
\ligne{\hfill   \execsix   {  17}  {  17}  {  18}  {  37}  {  39}  {  32}   \hfill}
\ligne{\hfill   \execsix   {  24}  {  24}  {  28}  {  42}  {  34}  {  24}   \hfill}
\ligne{\hfill   \execsix   {   6}  {  29}  { 142}  { 137}  { 127}  {   6}   \hfill}
}
\hfill}
}
\vskip 10pt
Note that in Table~\ref{exfixd}, 1(1) and~1(3) are also applied the same rules with a
delay when the locomotive comes from the left. This is no more the case when it comes
from the right. Also note that the rules applied to~1(3) are the same whenever the locomotive
comes from the left or from the right.

   Note that in both tables, we can see rules of Table~\ref{verttracktab} as well as rules
of Table~\ref{fixtab}.

\subsection{The round-about}

   In Section~\ref{scenar}, we have seen the idle configurations of the other pieces of the
round-about: the doubler, see Sub-subsection~\ref{subsdoubler} and the selector,
see Sub-subsection~\ref{subssel}.

\subsubsection{The doubler}
\label{subsdoubler}

   Figure~\ref{doubleur} illustrates the motion of the locomotive through the doubler.
The locomotive arrives in the form of a simple one. It enters the central cell of the 
doubler, the cell~$0(0)$ in the figure, after passing through the cell 1(1). When the 
locomotive is in~$0(0)$, the milestone at the cell 1(2) becomes white. This makes the cell~0(0)
to remain red. Now, the cell~1(3) also becomes red by the simple application of the
rules on the white cell of a basic element. Now we have two consecutive red cells which
have a common side. Now, when the cell~$0(0)$ is red and when it can see a red neighbour
through its side~3 and a white one through its side~2, it becomes white. From this moment
a double locomotive goes on its way along the tracks.

Table~\ref{doubltab} gives the rule which are involved by the doubler. We 
remind the reader that the rules of previous tables may also apply to this 
situation, particularly the rules of Table~\ref{consrules} 
and those for the motion of the locomotive on the tracks: 
Tables~\ref{verttracktab}, \ref{mounttracktab}, \ref{round1stab},
\ref{round1dtab} and~\ref{round3htab}, as these tables contain rules for the
milestones which witness the motion of the locomotive. Rules~143, 153, 157, 162,
169 and 173 apply to the central cell~$d_0$ of the doubler, rule~143 being the conservative
rule.

\vskip 10pt
\vtop{
%\ligne{\hrulefill}
%\vskip 3pt
\ligne{\hfill
%\phantom{
\includegraphics[scale=0.4]{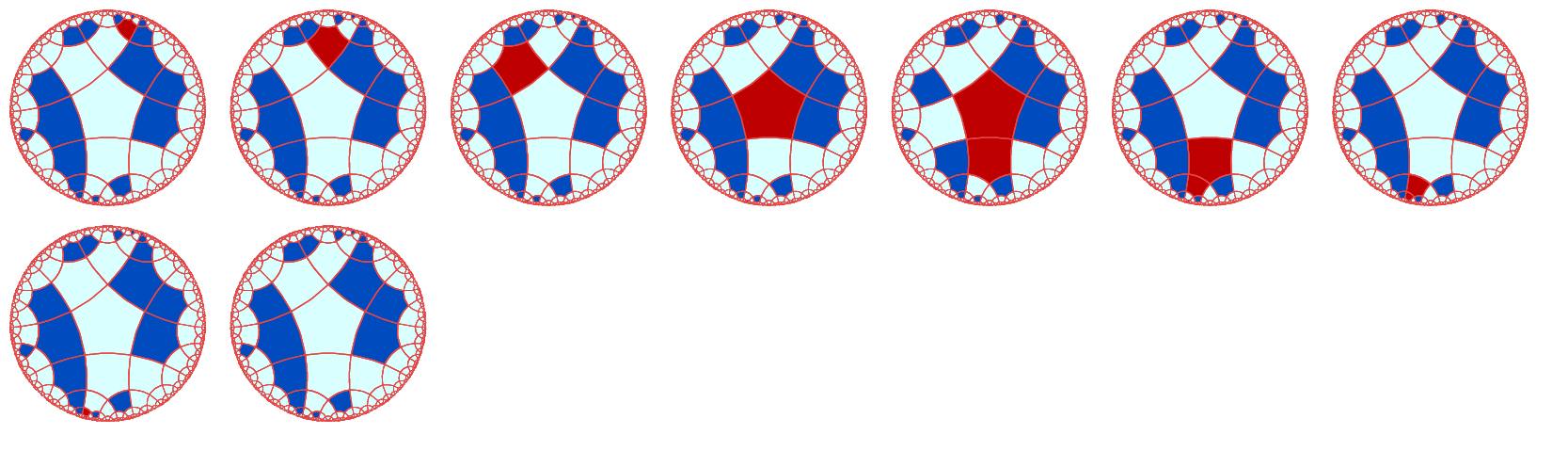} %}
\hfill}
\begin{fig}\label{doubleur}
\leurre
The structure which doubles a locomotive: a single locomotive enters the structure;
a double one leaves it.
\end{fig}
}

\vtop{
\begin{tab}\label{doubltab}
\leurre
Rules for the doubler.
\end{tab}
\vspace{-12pt}
\ligne{\hfill
\vtop{\leftskip 0pt\parindent 0pt\hsize=76pt 
\aff {143} {W} {WBWBBWBBWB} {W} 
\aff {144} {W} {WWBWBBBBWW} {W} 
\aff {145} {W} {WBRWBBWBBW} {R} 
\aff {146} {B} {WBWWBWWWBW} {B} 
\aff {147} {W} {BWWWBWBWWW} {W} 
\aff {148} {W} {BBWWWBWWWW} {W} 
\aff {149} {B} {WWWWBBWWWW} {B} 
\aff {150} {B} {WBWBWWBWWR} {B} 
\aff {151} {B} {BBWWWWWWWW} {B} 
}
\hfill
\vtop{\leftskip 0pt\parindent 0pt\hsize=76pt 
\aff {152} {W} {BWWWWBBWWW} {W} 
\aff {153} {W} {WBWBBRBBWB} {W} 
\aff {154} {W} {WRBWBBBBWW} {R} 
\aff {155} {R} {WBWWBBWBBW} {W} 
\aff {156} {B} {WBWBRWBWWW} {B} 
\aff {157} {W} {RBWBBWBBWB} {R} 
\aff {158} {R} {WWBWBBBBWW} {W} 
\aff {159} {W} {RBWWBBWBBW} {W} 
\aff {160} {B} {WBWWBWRWBW} {B} 
}
\hfill
\vtop{\leftskip 0pt\parindent 0pt\hsize=76pt 
\aff {161} {B} {WBWBWRBWWW} {B} 
\aff {162} {R} {WBWBBWBBWB} {R} 
\aff {163} {W} {RWBWBBBBWW} {W} 
\aff {164} {W} {WBWWBBRBBW} {W} 
\aff {165} {B} {RBWWBWWWBW} {W} 
\aff {166} {B} {RWWWBBWWWW} {B} 
\aff {167} {B} {RBWBWWBWWW} {B} 
\aff {168} {B} {BBWWWWRWWW} {B} 
\aff {169} {R} {WWRBBWBBWB} {W} 
}
\hfill
\vtop{\leftskip 0pt\parindent 0pt\hsize=76pt 
\aff {170} {W} {RBWWBRWWBW} {B} 
\aff {171} {R} {RBWBWBWBWW} {R} 
\aff {172} {B} {RWWWBBRWWW} {B} 
\aff {173} {W} {WBRBBWBBWB} {W} 
\aff {174} {W} {WBWWBBWBBW} {W} 
\aff {175} {B} {WBWWBRWWBW} {B} 
\aff {176} {B} {WWWWBBRWWW} {B} 
\aff {177} {B} {WBWBWWBWWW} {B} 
}
\hfill
}
}
\vskip 10pt
\vtop{
\begin{tab}\label{exdoubl}
\leurre
Execution of a run through the doubler.
\end{tab}
\vskip-6pt
\ligne{\hfill
\vtop{\leftskip 0pt\parindent 0pt\hsize=180pt
\ligne{\hfill  \execsept     {}       0       1       2       3       4       5 \hfill}
\ligne{\hfill  \execsept  {2(1)}  { 145}  { 155}  { 159}  { 164}  { 164}  { 174}   \hfill}
\ligne{\hfill  \execsept  {1(1)}  { 144}  { 154}  { 158}  { 163}  {  72}  { 144}   \hfill}
\ligne{\hfill  \execsept  {0(0)}  { 143}  { 153}  { 157}  { 162}  { 169}  { 173}   \hfill}
\ligne{\hfill  \execsept  {3(1)}  {  17}  {  17}  {  17}  {  18}  { 171}  {  39}   \hfill}
\ligne{\hfill  \execsept  {2(2)}  {   6}  {   6}  { 125}  {  29}  {  13}  {   6}   \hfill}
\ligne{\hfill  \execsept  {1(2)}  { 146}  { 146}  { 160}  { 165}  { 170}  { 175}   \hfill}
\ligne{\hfill  \execsept  {2(3)}  {  24}  {  24}  {  24}  {  27}  { 137}  {  42}   \hfill}
\ligne{\hfill  \execsept  {1(5)}  { 150}  { 156}  { 161}  { 167}  { 167}  { 177}   \hfill}
\ligne{\hfill  \execsept  {2(5)}  { 151}  { 151}  { 151}  { 168}  { 168}  { 151}   \hfill}
}
\hfill}
}
\vskip 10pt

The arriving locomotive
is detected by rule~153 through a vertex-neighbour. When side~1 can see the 
locomotive, rule~157 makes it becomes red. Rule~162 keeps the locomotive which arrived
to~$d_0$ which, accordingly, remains red. 
Now, rule~169 can see
the locomotive in the neighbour~3 of~$d_0$ and it also noticed that
the neighbour~2 became white. Rule~173 witnesses that the second cell of
the double locomotive is in neighbour~3 and that neighbour~2 returned to blue. 
As neighbour~2, \textit{i.e.} the
cell 1(2) plays an important role. Table~\ref{exdoubl} also lists the rules concerning that
cell: rules~146, 160, 165, 170 and~175, rule~146 begin conservative. Note the 
occurrence of a blue neighbour~9. The cell can see the arriving locomotive 
through a vertex as shown by the neighbourhood of rule~160 and rule~165 makes 
the cell 1(2) become white as the locomotive
is seen from its neighbour~1. Rule~170 restores the blue colour of the 
milestone and rule~175 can see through a vertex that the second red cell of 
the double locomotive is leaving. We leave the checking of the other
rules to the reader: Table~\ref{doubltab}, Figure~\ref{doubleur} and Table~\ref{exdoubl} are 
enough for that purpose.

   Table~\ref{exdoubl} gives the traces of execution of the rules given in Table~\ref{doubltab}.

\subsubsection{The selector}
\label{subssel}

    Figure~\ref{select} illustrates the working of the selector of the 
round-about. The case of a single locomotive is illustrated by the first row 
of the figure while the case of a double locomotive is illustrate by the
second row. Tables~\ref{selectstab} and~\ref{selectdtab} give the rules
for the locomotive and for the cells of the selector.

\vskip 10pt
\vtop{
%\ligne{\hrulefill}
%\vskip 3pt
\ligne{\hfill
%\phantom{
\includegraphics[scale=0.4]{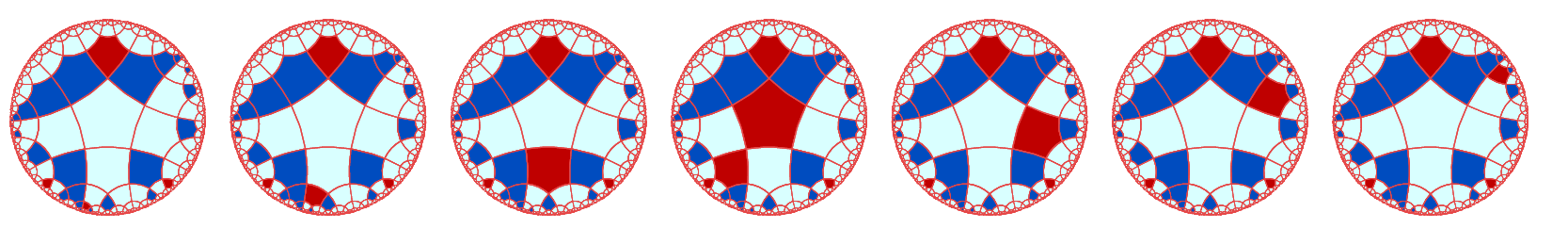} %}
\hfill}
\ligne{\hfill
%\phantom{
\includegraphics[scale=0.4]{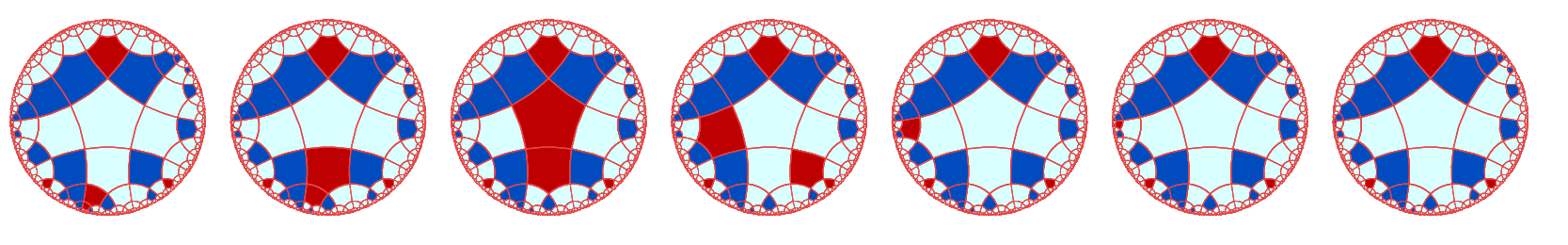} %}
\hfill}
\begin{fig}\label{select}
\leurre
The selector of the round about. Above: a single locomotive enters the selector. It leaves
it to go on its way on the right track. Below: a double locomotive enters the selector. 
It leaves it on the way going to the next selector.
\end{fig}
}

\vtop{
\begin{tab}\label{exsels}
\leurre
Execution of the rules %on the cells $0(0)$, $2(3)$, $2(4)$, $1(2)$, $1(3)$ and~$1(4)$ 
of the selector when it is crossed by a simple locomotive. %crosses the structure.
\end{tab}
\vspace{-9pt}
\ligne{\hfill
\vtop{\leftskip 0pt\parindent 0pt
\ligne{\hfill\execsept     {}     0      1      2      3      4      5 \hfill}
\ligne{\hfill\execsept {1(3)}  { 49}  { 56}  { 69}  {204}  {265}  { 49} \hfill}
\ligne{\hfill\execsept {0(0)}  {178}  {178}  {194}  {200}  {210}  {215} \hfill}
\ligne{\hfill\execsept {1(2)}  {182}  {182}  {195}  {203}  {182}  {182} \hfill}
\ligne{\hfill\execsept {1(4)}  {241}  {241}  {243}  {244}  {248}  {252} \hfill}
\ligne{\hfill\execsept {2(5)}  { 66}  { 66}  { 66}  {246}  {250}  { 62} \hfill}
\ligne{\hfill\execsept {2(3)}  {183}  {192}  {196}  {205}  {183}  {183} \hfill}
\ligne{\hfill\execsept {2(4)}  {187}  {187}  {198}  {208}  {212}  {187} \hfill}
}
\hfill}
}
\vskip 10pt
   Tables~\ref{exsels} and~\ref{exseld} list the rules used at a few cells of
the selector: cells 0(0), 1(2), 1(3) and 1(4) for the tracks and cells 2(3) and~2(4) 
for the working of the selector. For each cell we distinguish at most four consecutive 
times outside the long period in which the single conservative rule of the cell is 
applied to it. In the tables, the conservative rule lies in the column headed by~0.

    Note that cells 1(3), 0(0), 1(2), and~1(4) are cells of the track, but each one has a 
neighbourhood which is not obtained from that of an element of the tracks by rotation
of the neighbours. Note that the first two times for cell~1(3) are applied rules of the
vertical tracks. Remember that the cells 2(3) and~2(4) are milestones of
the cell 1(3). These cells are not permanently blue and they determine whether the 
locomotive is simple or double as they can see both the cell~0(0) and the cell~1(3). 
Note that the neighbourhood of the cells 2(3) and~2(4) is not that of a milestone of 
an element of the tracks. 

\vtop{
\begin{tab}\label{exseld}
\leurre
Execution of the rules %on the cells $0(0)$, $2(3)$, $2(4)$, $1(2)$, $1(3)$ and~$1(4)$ 
of the selector when it is crossed by a double locomotive. % crosses the structure.
\end{tab}
\vspace{-9pt}
\ligne{\hfill
\vtop{\leftskip 0pt\parindent 0pt
\ligne{\hfill\execsept    {}      0      1      2      3      4      5 \hfill}
\ligne{\hfill\execsept {1(3)}  { 56}  { 80}  {224}  {232}  { 49}  { 49} \hfill}
\ligne{\hfill\execsept {0(0)}  {178}  {194}  {222}  {228}  {178}  {178} \hfill}
\ligne{\hfill\execsept {1(2)}  {182}  {195}  {223}  {230}  {240}  {182} \hfill}
\ligne{\hfill\execsept {1(4)}  {241}  {243}  {226}  {236}  {241}  {241} \hfill}
\ligne{\hfill\execsept {2(5)}  { 66}  { 66}  {246}  { 66}  { 66}  { 66} \hfill}
\ligne{\hfill\execsept {2(3)}  {192}  {220}  {225}  {233}  {183}  {183} \hfill}
\ligne{\hfill\execsept {2(4)}  {187}  {198}  {227}  {237}  {187}  {187} \hfill}
}
\hfill}
}
\vskip 10pt
Usually, for a milestone of the track, there are five times when a simple locomotive 
passes: time~1, the locomotive is seen from a vertex-neighbour; time~2, it is seen from
the side shared with the cell of the track where the locomotive is; time~3, it is
again seen from another vertex-neighbour. When a double locomotive passes, there is one 
more time when the locomotive leaves the window on the track seen by the cell.  

For the cells~2(3) and~2(4), it can be seen that, due to the configuration illustrated
by Figure~\ref{select}, the cell~2(3) can see that the locomotive is simple or double
one time before the cell 2(4) can see the same thing. However, the cells 2(3) and 2(4) 
both turn to red at time~2, when time~1 is the first one when they can see the 
approaching locomotive. Note that the conservative rule applies at time~5 for the 
cell~2(3) when the locomotive is double, while the conservative rule applies at time~4 
for the cell 2(4) when the locomotive is simple.

%\vskip 10pt
\ifnum 1=0 {
Denote by~$s_0$ the central cell in Figure~\ref{select}. Rule~190 is its
conservative rule when no locomotive is around. Note the red neighbour~6: the 
first case of a red milestone. We shall look the rules for~$s_0$ and also
for its neighbours~$2$ and~(4), the cells 1(2) and~1(4) respectively and
also for its neighbours~8 and~9, the cells 2(3) and 2(4) respectively.
As vertex neighbours of~$s_0$, the cells 2(3) and~2(4) can also see which
state is in the neighbour~3 of~$s_0$, the cell 1(3). Indeed, 2(3) and 2(4)
are the neighbours~2 and~5 respectively of the cell~1(3). Accordingly, the
cells 2(3) and 2(4) can both check whether the locomotive which just arrived 
at~$s_0$ is simple or double. 

First, consider the
case of a single locomotive. Rules~206, 212, 222 and 227 apply to~$s_0$. 
Rule~206 can see the locomotive in the neighbour~3 of~$s_0$, so that the cell
becomes red. Rule~212 Makes $s_0$ becomes white again witnesses that its
neighbour~9 became red, preventing the locomotive to enter cell~1(2). 
Rule~222 witnesses that the locomotive is now in the cell~1(4), the 
neighbour~4 of~$s_0$. Rule~227 can see the locomotive going out through a
vertex-neighbour. 

Cell 2(3) is applied rules~195, 204, 208 and 217. Rule~195 is the conservative rule. 
The vertex-neighbourhood \hbox{\footnotesize\tt WWBRB} is a characteristic of this cell. 
Rule 204 witness that a locomotive is in its vertex-neighbour~6. Rule 208 makes the
cell turning to red in order to block the entrance of the locomotive into~1(2) 
as the cell detected that the locomotive is simple: it is in 1(3) and cell 3(3) is white.
Both cells 1(3) and~3(3) can be seen from~2(3). Rule 217 turns the cell back to blue.

Cell 2(4) is applied rules~199, 210, 220 and~224, Rule~199 is conservative. Its
vertex-neighbourhood \hbox{\footnotesize\tt WWWBR} is a characteristic of this cell.
Rule 210 can see the locomotive when it is in cell~1(3). It cannot see it before and
at this moment, the cell cannot know whether the locomotive is simple or not. Accordingly,
the cell remains blue. Rule 220 can see the locomotive when it is in~$s_0$ as its 
vertex-neighbour~7. As the cell can see that cell 1(3) is white, which means that the 
locomotive is simple, it remains blue in order the locomotive enters the cell 1(4).

Cell 1(2) is applied rules 194, 207 and 215. Rule 194 is conservative and it can be 
noticed that its neighbourhood is not that of an element of the tracks. Rule 207 can see 
the approaching locomotive through the neighbour~6 of the cell. Rule~215 can see the 
locomotive in~$s_0$ and it can also see that the cell 2(3) is now red. From that, the
cell knows that it must remains white, which is performed by rule~215. After that, 
the cell 2(3) returns to blue and the locomotive, a simple one, is no more visible.   

Cell~1(4) is applied rules~143, 145, 146, 150 and~154. Rule~143 is conservative. Rule~145
can see an approaching locomotive through its neighbour~7. Rule~146 can see the locomotive
in~$s_0$. It can also see that the locomotive is simple, so that the cell turns to red.
Rule~150 turns the cell back to white and rule~154 can see the locomotive as left the
cell as it is in its neighbour~5.  
} \fi

\vskip -10pt
\vtop{
\begin{tab}\label{selectstab}
\leurre
Rules for the selector: the case of a simple locomotive.  % loco simple
\end{tab}
\vspace{-12pt}
\ligne{\hfill
\vtop{\leftskip 0pt\parindent 0pt\hsize=76pt 
\aff {178} {W} {BWWWBRBBBW} {W} 
\aff {179} {B} {WRWWBWBWWW} {B} 
\aff {180} {R} {BBWWWWWWWW} {R} 
\aff {181} {W} {RWWWWWBWWW} {W} 
\aff {182} {W} {WBWBBWBBWW} {W} 
\aff {183} {B} {WWWWBWWBRB} {B} 
\aff {184} {W} {BBWWRWWWWW} {W} 
\aff {185} {W} {BRWWBBWWWW} {W} 
\aff {186} {B} {WBBWBRWWWW} {B} 
\aff {187} {B} {WWWWWWWWBR} {B} 
\aff {188} {W} {BWWWBWWWWW} {W} 
}
\hfill
\vtop{\leftskip 0pt\parindent 0pt\hsize=76pt 
\aff {189} {W} {WBRWWWWWWW} {W} 
\aff {190} {B} {WWBWRBWWWW} {B} 
\aff {191} {W} {BWWWWRBWWW} {W} 
\aff {192} {B} {WWWWBRWBRB} {B} 
\aff {193} {B} {RBBWBWWWWW} {B} 
\aff {194} {W} {BWRWBRBBBW} {R} 
\aff {195} {W} {WBWBBRBBWW} {W} 
\aff {196} {B} {RWWWBWWBRB} {R} 
\aff {197} {B} {WBBWBWRWWW} {B} 
\aff {198} {B} {WRWWWWWWBR} {B} 
\aff {199} {W} {BWWWBWRWWW} {W} 
}
\hfill
\vtop{\leftskip 0pt\parindent 0pt\hsize=76pt 
\aff {200} {R} {BWWWBRBRBW} {W} 
\aff {201} {B} {RRWWBWBWWW} {B} 
\aff {202} {R} {BBWWWWRWWW} {R} 
\aff {203} {W} {RBWBRWBBWW} {W} 
\aff {204} {W} {RRWWBWWBBW} {W} 
\aff {205} {R} {WWWWBWRBRB} {B} 
\aff {206} {W} {RBWWRWWWWW} {W} 
\aff {207} {W} {RRWWBBWWWW} {W} 
\aff {208} {B} {WWWWWWRWBR} {B} 
\aff {209} {B} {RWBWRBWWWW} {B} 
\aff {210} {W} {BWWRBRBBBW} {W} 
}
\hfill
\vtop{\leftskip 0pt\parindent 0pt\hsize=76pt 
\aff {211} {B} {WBBWBWWWWW} {B} 
\aff {212} {B} {RWWWWWWWBR} {B} 
\aff {213} {W} {WBRWWWRWWW} {W} 
\aff {214} {B} {WWBWRBRWWW} {B} 
\aff {215} {W} {BWWWBRBBBR} {W} 
\aff {216} {B} {WRBWRBWWWW} {B} 
\aff {217} {B} {WWBWRBWRWW} {B} 
\aff {218} {W} {BRWWWWBBWW} {W} 
}
\hfill
}
}

\vtop{
\begin{tab}\label{selectdtab}
\leurre
Rules for the selector: the case of double locomotive.  % loco double
\end{tab}
\vspace{-12pt}
\ligne{\hfill
\vtop{\leftskip 0pt\parindent 0pt\hsize=76pt 
\aff {219} {B} {RBBWBRWWWW} {B} 
\aff {220} {B} {RWWWBRWBRB} {B} 
\aff {221} {B} {RBBWBWRWWW} {B} 
\aff {222} {R} {BWRWBRBBBW} {W} 
\aff {223} {W} {RBWBBRBBWW} {R} 
\aff {224} {R} {RBWWBWWBBW} {W} 
}
\hfill
\vtop{\leftskip 0pt\parindent 0pt\hsize=76pt 
\aff {225} {B} {RWWWBWRBRB} {B} 
\aff {226} {W} {RBWBWBRWWW} {W} 
\aff {227} {B} {WRWWWWRWBR} {R} 
\aff {228} {W} {BRWWBRBBRW} {W} 
\aff {229} {B} {WRWWBRBWWW} {B} 
\aff {230} {R} {WBWBBWBBWW} {W} 
}
\hfill
\vtop{\leftskip 0pt\parindent 0pt\hsize=76pt 
\aff {231} {B} {RWWWWBWWWW} {B} 
\aff {232} {W} {WBWWRWRBBW} {W} 
\aff {233} {B} {WRWWBWWBRB} {B} 
\aff {234} {W} {WBWWWRWWWW} {W} 
\aff {235} {W} {BBWWRWRWWW} {W} 
\aff {236} {W} {WRWBWBWWWW} {W} 
}
\hfill
\vtop{\leftskip 0pt\parindent 0pt\hsize=76pt 
\aff {237} {R} {WWWWWWWWBR} {B} 
\aff {238} {W} {WWWWWBRWWW} {W} 
\aff {239} {W} {WRRWWWWWWW} {W} 
\aff {240} {W} {WBRBBWBBWW} {W} 
}
\hfill
}
}
\vskip 10pt
\ifnum 1=0 {
    Consider the case of a double locomotive.

Cell 0(0) is applied rules~190 and~206 as in the case of a simple one: $s_0$ does not know
yet that the locomotive is double. Then it is applied rules~234 and~240 before again
rule~190, the conservative one. Rule~234 turns the locomotive to white: it transforms the
locomotive into a simple one. Rule~240 witnesses that the locomotive is in the cell~1(2).

Cell 2(3)

Cell 2(4)

Cell 1(2)

Cell 1(4)
} \fi

Tables~\ref{exsels} and~\ref{exseld} show us which rules apply to the cells of the selector.
We can see that when the locomotive is double, rule 227 makes the cell become red, 
preventing the locomotive to enter the cell~1(4), see rule~226. At this moment, the cell~2(3)
remains blue, see rule~220, which allows the locomotive to enter the cell~1(2). When the
locomotive is simple, the cell~2(4) remains blue, rule~198, so that the locomotive enters
the cell~1(4) and the cell~2(3) turns to red, rule~196, preventing the locomotive to
enter the cell~1(2), rule~203.

\subsubsection{The fork}

   The fork is a structure which, in some sense, is the exact reverse of the passive 
fixed switch. The difference is that the fork produces two simple locomotives, each
one going in a different direction, see Figure~\ref{fourche}, while a single simple or 
double locomotive enters the passive fixed switch. Table~\ref{forktab} gives the rules 
for the fork. We can note that a few of them were already used by the cell~1(4) in the 
selector when it is crossed by a simple locomotive.

   Table~\ref{forktab} shows the rules for the cells of the fork~: 0(0), 1(3), 1(1) 
and~1(4). The cells involved in this situation are those of the tracks, but they are
assembled in a slightly unusual way. They use the trick we already used for the rules
allowing a locomotive to run around a single cell or a few of them, pairwise sharing a 
common side.
  
\vtop{
\begin{tab}\label{forktab}
\leurre
Rules for the fork.
\end{tab}
\vspace{-12pt}
\ligne{\hfill
\vtop{\leftskip 0pt\parindent 0pt\hsize=76pt 
\aff {241} {W} {WBWBWBWWWW} {W} 
\aff {242} {W} {BWWWWWBBWW} {W} 
\aff {243} {W} {WBWBWBRWWW} {W} 
}
\hfill
\vtop{\leftskip 0pt\parindent 0pt\hsize=76pt 
\aff {244} {W} {RBWBWBWWWW} {R} 
\aff {245} {B} {RWBWWWWWWW} {B} 
\aff {246} {W} {BWWBWBRBWW} {W} 
}
\hfill
\vtop{\leftskip 0pt\parindent 0pt\hsize=76pt
\aff {247} {W} {RBWRBWBBBW} {W} 
\aff {248} {R} {WBWBWBWWWW} {W} 
\aff {249} {B} {WWBWWRRWWW} {B} 
}
\hfill
\vtop{\leftskip 0pt\parindent 0pt\hsize=76pt 
\aff {250} {W} {BRWBWBWBWW} {R} 
\aff {251} {W} {WBWWBWBBBR} {W} 
\aff {252} {W} {WBWBRBWWWW} {W} 
}
\hfill
}
}
\vskip 10pt

In the first three times, the rules applied to cell~1(1) are those of the motion on a
vertical track. This is also the case for the conservative rule. But when the cell
turns to white, the locomotive is leaving the neighbourhood of the cell, it makes use
of a rule for the fixed switch, for the cell 1(5), when a simple locomotive arrives
to the centre.  For the cell~1(1), the rules are those of a vertical track.
For the cell~1(4), the neighbourhood is a bit unusual: it is due to the right-angle turn 
in the path of the locomotive. For this reason, the locomotive is seen from an unusual
neighbour, whence rule~252. Of course, a simpler implementation was possible, but the
one illustrated by Figures~\ref{control} and~\ref{controls} show that the rules allow
to apply to any path for the tracks.

\vtop{
\begin{tab}\label{exfork}
\leurre
Execution of the rules %on the cells $0(0)$, $1(3)$, $1(1)$ and~$1(4)$ 
of the fork. %when a double locomotive crosses the structure.
\end{tab}
\vspace{-9pt}
\ligne{\hfill
\vtop{\leftskip 0pt\parindent 0pt
\ligne{\hfill  \execsept     {}       0       1       2       3       4       5 \hfill}
\ligne{\hfill  \execsept  {1(3)}  {  47}  {  50}  {  51}  {  53}  { 123}  {  47}   \hfill}
\ligne{\hfill  \execsept  {0(0)}  {  47}  {  47}  {  50}  {  51}  { 247}  { 251}   \hfill}
\ligne{\hfill  \execsept  {1(1)}  {  47}  {  47}  {  47}  {  50}  {  51}  {  53}   \hfill}
\ligne{\hfill  \execsept  {1(4)}  { 241}  { 241}  { 243}  { 244}  { 248}  { 252}   \hfill}
\ligne{\hfill  \execsept  {2(5)}  {  66}  {  66}  {  66}  { 246}  { 250}  {  62}   \hfill}
\ligne{\hfill  \execsept  {1(5)}  {   6}  {   6}  {   6}  { 245}  { 249}  {  63}   \hfill}
}
\hfill}
}
\vskip 10pt
Table~\ref{exfork} gives the traces of execution for the rules of Table~\ref{forktab}. The rules
apply to the cells 1(3), 0(0), 1(1), 1(4) and 2(5) for the tracks and to the cell 1(5) for the
milestones. Note that rule~47 is the conservative rule for 1(3), 0(0) and~1(1): they are standard
elements of the track assembled as in a vertical track. The cell~1(4) is a bit different: it has
three milestones only, all of them blue ones, and one of them is a vertex-neighbour. This is
why the conservative rule is a new one: rule~241.

\vskip 10pt
\vtop{
%\ligne{\hrulefill}
%\vskip 3pt
\ligne{\hfill
%\phantom{
\includegraphics[scale=0.4]{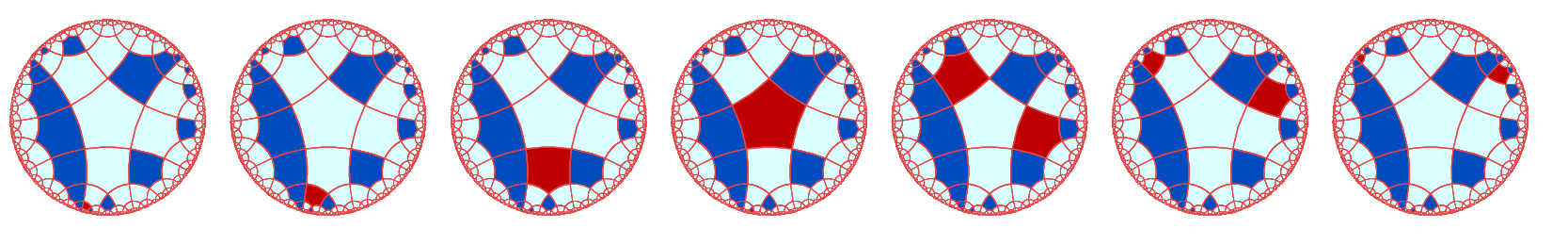} %}
\hfill}
\begin{fig}\label{fourche}
\leurre
The fork: a single locomotive enters the fork. Two single one leave it in different directions.
\end{fig}
}

\subsection{Flip-flop and active memory switch}
\label{flflp}

    In Section~\ref{scenar} we have seen that we can implement both a flip-flop and
the active memory switch by using the same basic devices provided that they are
suitable arranged. Figure~\ref{control} shows how the controller is working, depending
on the state of its cell~1(1) which we call the \textbf{colour} of the controller. 
Figure~\ref{controls} shows how the colour of controller is changed by the arrival 
of the appropriate signal: we remind the reader that the signal has the form of a 
simple locomotive. Table~\ref{controltab} gives the rules corresponding to 
the illustration given by Figure~\ref{control} while Table~\ref{controlstab} does the 
same thing for Figure~\ref{controls}. Also, Table~\ref{excontrol} gives the traces of executions
of the rules applied to significant cells when the locomotive arrives to the controller while
Table~\ref{excontrols} gives similar traces when the signal arrives to the controller in
order to change its colour.

\vskip 10pt
\vtop{
%\ligne{\hrulefill}
%\vskip 3pt
\ligne{\hfill
%\phantom{
\includegraphics[scale=0.4]{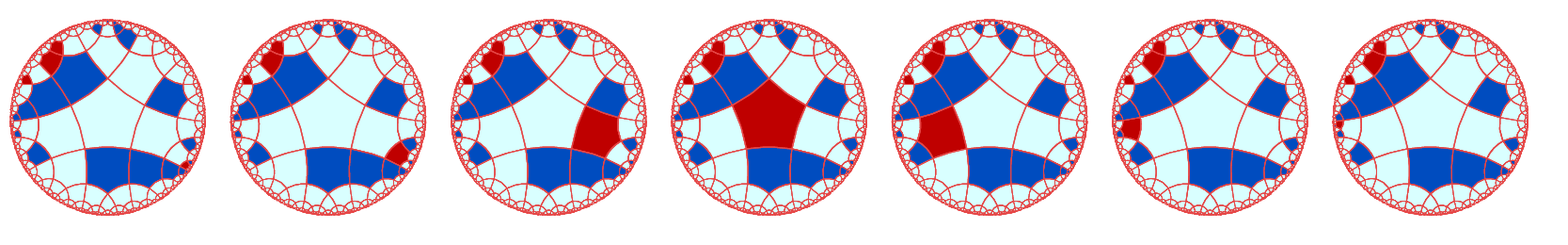} %}
\hfill}
\ligne{\hfill
%\phantom{
\includegraphics[scale=0.4]{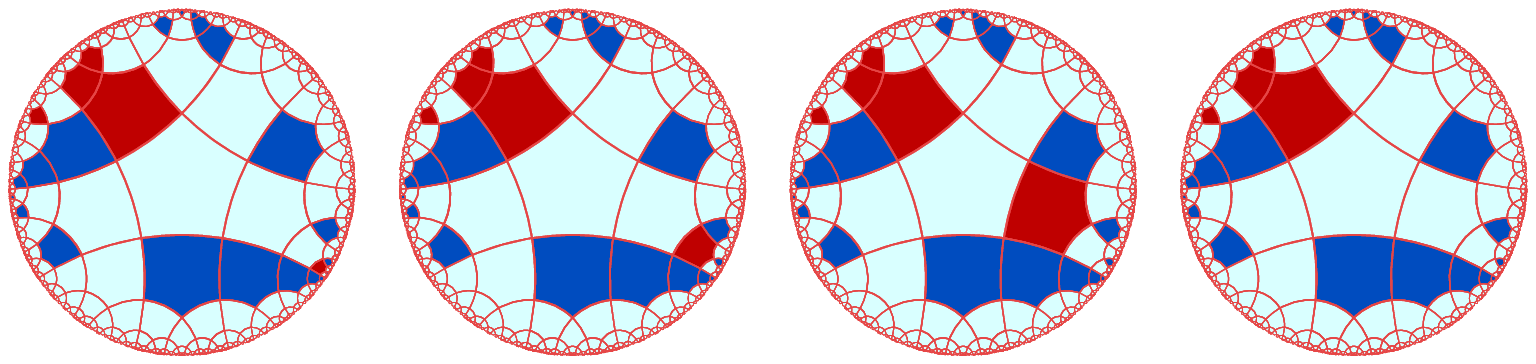} %}
\hfill}
\begin{fig}\label{control}
\leurre
The controller of the flip-flop and of the active memory switch. Above, the
controller is blue: the locomotive passes without problem. Below, the controller is
red: it stops the locomotive which vanishes.
\end{fig}
}

   In Figure~\ref{control}, the locomotive, always a simple one, crosses the cells
8(4), 3(4), 1(4), 0(0), 1(2), 3(2) and 8(2) in this order, where 0(0) is the central tile
of the picture.  From this figure and 
from Table~\ref{excontrol}, we can see that when the cell~1(1) is blue, the locomotive 
can pass: cell~0(0) looks like an element of the 
tracks. When the cell~1(1) is red, the neighbourhood of~0(0) is no more that of an 
element of the tracks: the cell 0(0) remains white and from the rules of the track 
applied to 1(4), the locomotive vanishes.

Table~\ref{excontrol} gives the traces of execution for the rules which apply to
the cells 1(4), 0(0) and 1(2) for the tracks, the cells 2(2) and 1(1) for milestones and 
the cell 2(1) for a witness of transformations induced by the passage of the locomotive.
   
\vtop{
\begin{tab}\label{controltab}
\leurre
Rules for the controllers of the flip-flop and the active memory switch.
\end{tab}
\vspace{-9pt}
\ligne{\hfill crossing through the selected track \hfill}
\vskip 3pt
\ligne{\hfill
\vtop{\leftskip 0pt\parindent 0pt\hsize=76pt 
\aff {253} {B} {WWWRBWWWRW} {B}
\aff {254} {W} {BWWWRRWWWW} {W}
\aff {255} {R} {BRWWWBWWWW} {R}
\aff {256} {R} {RWWWWWBWWW} {R}
\aff {257} {W} {RWWWWWRWWW} {W}
}
\hfill
\vtop{\leftskip 0pt\parindent 0pt\hsize=76pt 
\aff {258} {W} {WRWWWWRWWW} {W}
\aff {259} {B} {WBWWBWWRRW} {B}
\aff {260} {W} {BRWWRWBWWW} {W}
\aff {261} {W} {BRWWWBWWWW} {W}
\aff {262} {W} {WBWWWBWWWB} {W}
}
\hfill
\vtop{\leftskip 0pt\parindent 0pt\hsize=76pt 
\aff {263} {W} {WBWWWBRWWB} {W}
\aff {264} {B} {RWWRBWWWRW} {B}
\aff {265} {W} {BWBWWWRWBB} {W}
\aff {266} {B} {WBWWBWRRRW} {B}
}
\hfill
\vtop{\leftskip 0pt\parindent 0pt\hsize=76pt 
\aff {267} {W} {RBWWWBWWWB} {W}
\aff {268} {B} {WWWRBRWWRW} {B}
\aff {269} {B} {RBWWBWWRRW} {B}
\aff {270} {B} {WBWWBRWRRW} {B}
}
\hfill}
\vskip 7pt
\ligne{\hfill forbidding the non-selected track \hfill}
\vskip 3pt
\ligne{\hfill
\vtop{\leftskip 0pt\parindent 0pt\hsize=76pt 
\aff {271} {W} {RWBWWWBWBB} {W}
\aff {272} {R} {WWWRBWWWRW} {R}
\aff {273} {W} {RWBWWWWWBB} {W}
\aff {274} {W} {RWWWRRWWWW} {W}
}
\hfill
\vtop{\leftskip 0pt\parindent 0pt\hsize=76pt 
\aff {275} {R} {RRWWWBWWWW} {R}
\aff {276} {R} {RWWWWWRWWW} {R}
\aff {277} {W} {WBWBWBRBWW} {W}
\aff {278} {B} {WRWWBWWRRW} {B}
}
\hfill
\vtop{\leftskip 0pt\parindent 0pt\hsize=76pt 
\aff {279} {W} {BRWWRWRWWW} {W}
\aff {280} {W} {WBWWWRWWWB} {W}
\aff {281} {W} {RWBRWWBWBB} {W}
\aff {282} {W} {WBWWWRRWWB} {W}
}
\hfill}
}
\vskip 10pt
\def\exectabff #1 #2 #3 #4 {%
\hbox to 25pt{\hfill#1\hskip 5pt}
\hbox to 25pt{\hfill#2\hskip 5pt}
\hbox to 25pt{\hfill#3\hskip 5pt}
\hbox to 25pt{\hfill#4\hskip 5pt}
}

\vtop{
\begin{tab}\label{excontrol}
\leurre
Execution of the rules %on the cells $1(4)$, $0(0)$, $1(2)$, $1(1)$ and~$2(1)$ 
of the controller when a simple locomotive crosses the structure.
\end{tab}
\vspace{-6pt}
\ligne{\hfill\hbox to 150pt{\hskip 10pt 2(1) is blue:\hfill}
\hskip 20pt\hbox to 75pt{\hskip 10pt 2(1) is red:\hfill}
\hfill}
\vskip 2pt
\ligne{\hfill
\vtop{\leftskip 0pt\parindent 0pt\hsize=150pt
\ligne{\hfill\execsix    {}      0      1      2      3      4 \hfill}
\ligne{\hfill\execsix {1(2)}  { 17}  { 17}  { 17}  { 18}  { 20}   \hfill}
\ligne{\hfill\execsix {0(0)}  {174}  {174}  {145}  {155}  {159}   \hfill}
\ligne{\hfill\execsix {1(4)}  { 47}  { 50}  { 51}  { 53}  { 47}   \hfill}
\ligne{\hfill\execsix {2(2)}  {259}  {259}  {259}  {266}  {269}   \hfill}
\ligne{\hfill\execsix {1(1)}  {253}  {253}  {253}  {264}  {268}   \hfill}
\ligne{\hfill\execsix {2(1)}  { 49}  { 49}  { 49}  {265}  { 49}   \hfill}
}
\hskip 20pt
\vtop{\leftskip 0pt\parindent 0pt\hsize=75pt
\ligne{\hfill\exectroi    0      1      2 \hfill}
\ligne{\hfill\exectroi {277}  {277}  {277} \hfill}
\ligne{\hfill\exectroi {271}  {271}  {281} \hfill}
\ligne{\hfill\exectroi { 47}  { 50}  { 51} \hfill}
\ligne{\hfill\exectroi {278}  {278}  {278} \hfill}
\ligne{\hfill\exectroi {272}  {272}  {272} \hfill}
\ligne{\hfill\exectroi {273}  {273}  {273} \hfill}
}
\hfill}
}
\vskip 10pt
\vtop{
%\ligne{\hrulefill}
%\vskip 3pt
\ligne{\hfill
%\phantom{
\includegraphics[scale=0.4]{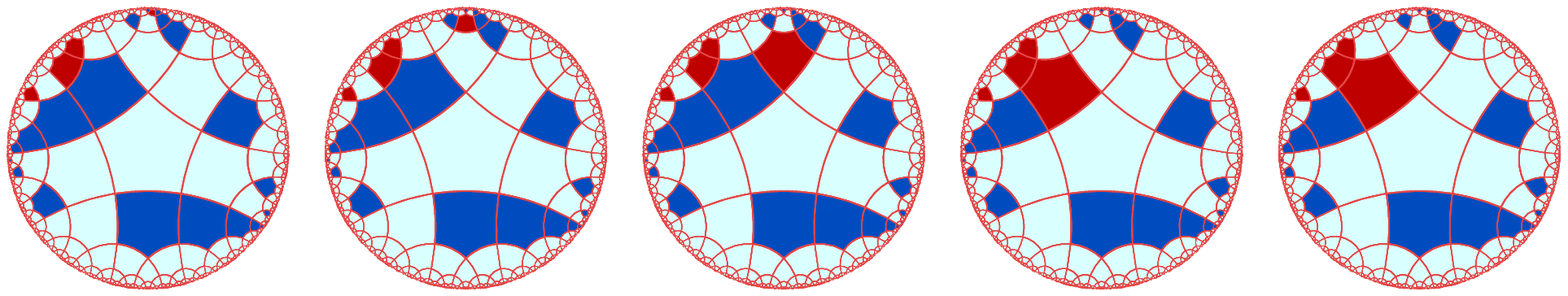} %}
\hfill}
\ligne{\hfill
%\phantom{
\includegraphics[scale=0.4]{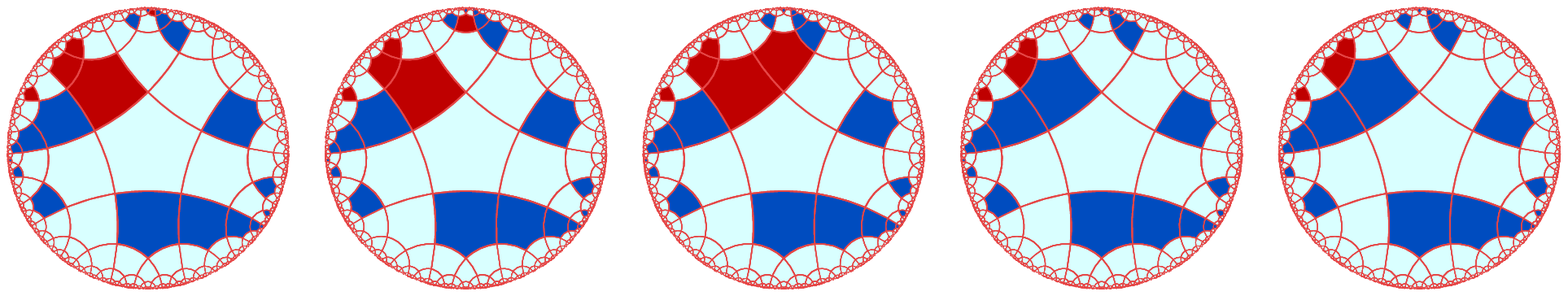} %}
\hfill}
\begin{fig}\label{controls}
\leurre
The controller of the flip-flop and of the active memory switch. The signal arrives to
change the selection. Above, the controller is changed to 'red'. Below, it is changed to 'blue'.
\end{fig}
}

When the colour of the controller is blue, the rules 253, 264 and 268 apply to the cell~1(1).
The rule 253 is conservative, the rule 264 can see the locomotive through its side~1
and the rule~268 can see it through its vertex-neighbour~6 as the locomotive is leaving the
central cell of the controller. When the controller is red, the conservative rule is 272
and no other rule applies to the cell as the locomotive vanishes before being noticed
by the cell.

\vtop{
\begin{tab}\label{controlstab}
\leurre
Rules for changing the selection in the flip-flop and the active memory switch.
\end{tab}
\vspace{-9pt}
\ligne{\hfill\hskip 40pt from blue to red \hfill\hskip 70pt red to blue\hfill}
\vskip 3pt
\ligne{\hfill
\vtop{\leftskip 0pt\parindent 0pt\hsize=76pt 
\aff {283} {W} {BWBWWRBWBB} {W}
\aff {284} {B} {WRWRBWWWRW} {R}
\aff {285} {W} {BWWWRRRWWW} {W}
\aff {286} {W} {WBWWRBWWWB} {W}
}
\hfill
\vtop{\leftskip 0pt\parindent 0pt\hsize=76pt 
\aff {287} {W} {RWBRWWWWBB} {R}
\aff {288} {W} {WWBWWWRRWW} {W}
\aff {289} {W} {RWBWWRBWBB} {W}
\aff {290} {R} {WRWRBWWWRW} {B}
}
\hfill\hskip 30pt
\vtop{\leftskip 0pt\parindent 0pt\hsize=76pt 
\aff {291} {R} {RWBWWWWWBB} {W}
\aff {292} {W} {RWWWRRRWWW} {W}
\aff {293} {W} {WRBWWWRWWW} {W}
\aff {294} {W} {WBWWRRWWWB} {W}
}
\hfill
}
}

\def\exectabffs #1 #2 #3 #4 {%
\hbox to 25pt{\hfill#1\hskip 5pt}
\hbox to 25pt{\hfill#2\hskip 5pt}
\hbox to 25pt{\hfill#3\hskip 5pt}
\hbox to 25pt{\hfill#4\hskip 5pt}
}

\vtop{
\begin{tab}\label{excontrols}
\leurre
Execution of the rules %on the cells $0(0)$, $1(1)$, $2(1)$, $6(1)$ and~$1(5)$ 
of the controller when a signal comes for changing the colour of the controller.
%a simple locomotive crosses the structure.
\end{tab}
\vspace{-6pt}
\ligne{\hfill\hbox to 130pt{\hskip 10pt from blue to red:\hfill}
\hskip 20pt\hbox to 95pt{\hskip 10pt from red to blue:\hfill}\hfill}
\vskip 2pt
\ligne{\hfill
%  from blue to red
\vtop{\leftskip 0pt\parindent 0pt\hsize=130pt
\ligne{\hfill\execcinq    {}      0      1      2      3 \hfill}
\ligne{\hfill\execcinq {6(1)}  { 50}  { 51}  { 53}  { 47}   \hfill}
\ligne{\hfill\execcinq {2(1)}  { 49}  { 56}  { 69}  {273}   \hfill}
\ligne{\hfill\execcinq {0(0)}  {174}  {174}  {283}  {271}   \hfill}
\ligne{\hfill\execcinq {2(2)}  {259}  {259}  {259}  {278}   \hfill}
\ligne{\hfill\execcinq {1(1)}  {253}  {253}  {284}  {272}   \hfill}
\ligne{\hfill\execcinq {1(5)}  {262}  {262}  {286}  {280}   \hfill}
}
\hskip 20pt
%  from red to blue
\vtop{\leftskip 0pt\parindent 0pt\hsize=95pt
\ligne{\hfill\execquat     0      1      2      3 \hfill}
\ligne{\hfill\execquat {  50}  { 51}  { 53}  { 47} \hfill}
\ligne{\hfill\execquat { 273}  {287}  {291}  { 49} \hfill}
\ligne{\hfill\execquat { 271}  {271}  {289}  {174} \hfill}
\ligne{\hfill\execquat { 278}  {278}  {278}  {259} \hfill}
\ligne{\hfill\execquat { 272}  {272}  {290}  {253} \hfill}
\ligne{\hfill\execquat { 280}  {280}  {294}  {262} \hfill}
}
\hfill}
}
\vskip 10pt
   Figure~\ref{controls} and Table~\ref{excontrols} illustrate the action of a signal.
The figure illustrates the fact that the signal operates like a second simple 
locomotive arriving to the controller. This new locomotive comes from another path
and can be seen by cell~1(1) trough its neighbour~2 while the ordinary locomotive is
seen from its neighbours~1 and one time later by its neighbour~6, when the locomotive
leaves the controller which was blue. The cells involved in Table~\ref{excontrols}
are the cells 6(1) and 2(1) for the track of the signal, the cells 1(1) and 2(2) as
milestones and the cell 1(5) as a witness of the changes: the cell 1(5) can see 1(1)
through it vertex-neighbour~6 and the cell 2(2) from its neighbour~5.
Accordingly, rule~286 witnesses that the signal arrived close to~1(1) which is still blue
and rule~280 that 1(1) changed to red and that the signal vanished. Rule 280 is also
the conservative rule of~1(5) when the cell is red. It witnesses the arrival of the signal
with rule~294 and it witnesses that 1(1) turned to blue by the rule 262 which is also
the conservative rule of 1(5) when the controller is blue.

\subsection{The passive memory switch}

    As we said in Section~\ref{scenar}, the controller of the passive memory switch
works in a different way than that of the active switches.  For this reason, we call it 
the \textbf{controller-sensor} as it does not stop the locomotive when it runs through 
the non-selected track. As in the case of the active switches, this controller is 
characterized by the state of a cell, here the cell~1(2), see Figure~\ref{controlsign}.
Here, the cell 1(2) is called the \textbf{sensor} as its role is to detect a passage,
not to stop it. 
The tracks goes along the cells 6(1), 2(1), 1(1), 4(1) and 11(1), followed in this order by 
the simple locomotive when it is present. As in the case of the active switch, we say that 
the controller is blue or red according to the state of the cell 1(2). Also as in the case 
of the active switches, this colour is changed by the arrival of a signal in the form of a 
simple locomotive too. In the figure, the way of this signal passes through the cells 13(3), 
5(3) and 2(5). When the signal arrived in the cell 2(5), the cell 1(2) changes its colour as 
illustrated by Figure~\ref{controlsign}. Note that this change occurs when the controller is 
blue only. When it is red, the change is triggered by the passage of the simple locomotive 
on its ordinary way as illustrated by Figure~\ref{controlsign} too.

\vskip 10pt
\vtop{
%\ligne{\hrulefill}
%\vskip 3pt
\ligne{\hfill
%\phantom{
\includegraphics[scale=0.4]{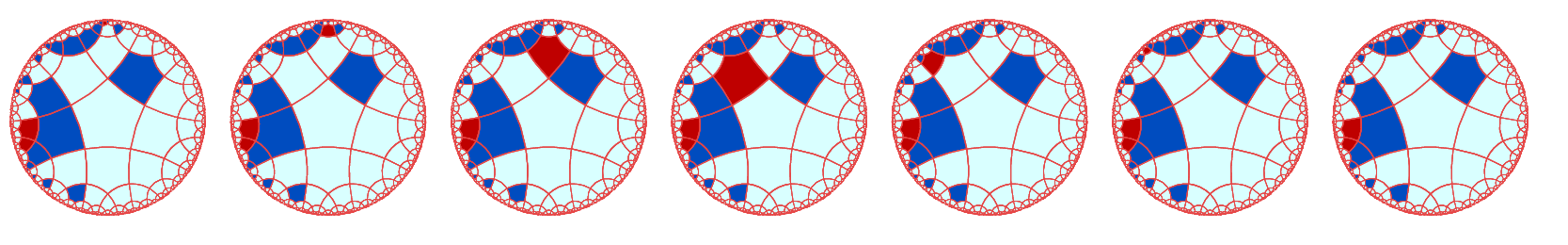} %}
\hfill}
\ligne{\hfill
%\phantom{
\includegraphics[scale=0.4]{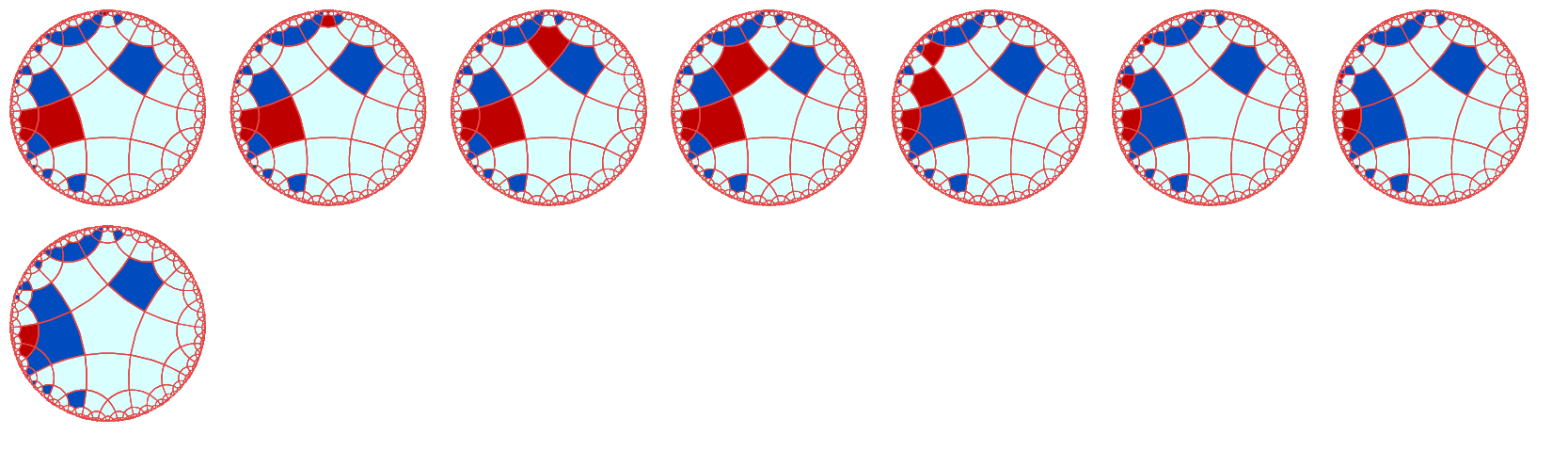} %}
\hfill}
\ligne{\hfill
%\phantom{
\includegraphics[scale=0.4]{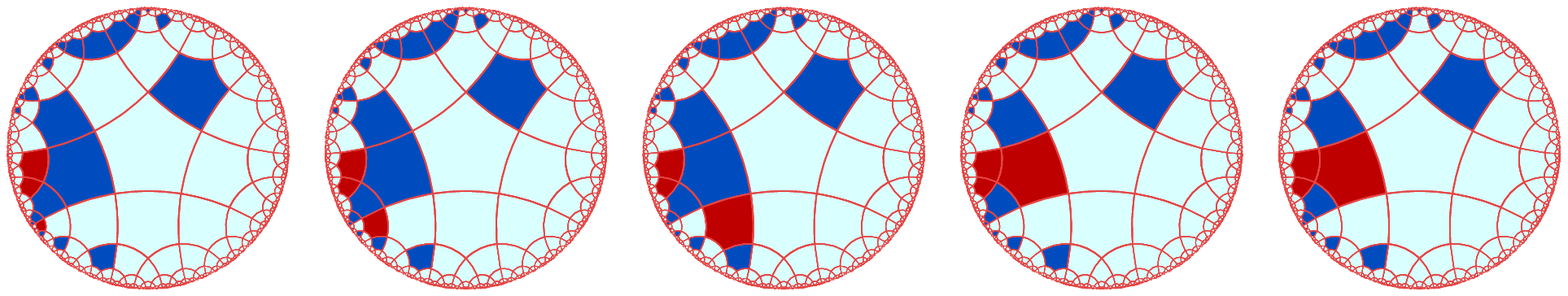} %}
\hfill}
\begin{fig}\label{controlsign}
\leurre
The controller-sensor of the passive memory switch. Above: a single locomotive passively
crossed the switch through its selected track. Middle row: the locomotive crossed the passive
switch through the non-selected track. Below: the change of signal induced by a passive crossing
through the non-selected track.
\end{fig}
}

\vtop{
\begin{tab}\label{passmemotab1}
\leurre
Rules for the passive memory switch.
\end{tab}
\vspace{-9pt}
\ligne{\hfill rules for the control-sensor\hfill}
\vskip 3pt
\ligne{\hfill non- selected track\hfill\hbox to 76pt{\hfill selected track\hfill}\hskip 5pt}
\vskip 3pt
\ligne{\hfill
\vtop{\leftskip 0pt\parindent 0pt\hsize=76pt 
\aff {295} {W} {WRWWBWBWWW} {W}
\aff {296} {W} {WWBWBRBBBW} {W}
\aff {297} {R} {WBRBWWWWWW} {R}
\aff {298} {B} {RWWWWRWWBW} {B}
\aff {299} {R} {RWWWWBBWWW} {R}
\aff {300} {B} {RWWBWWRWWW} {B}
\aff {301} {W} {BWWWBWWBWW} {W}
\aff {302} {W} {BBWBWWWBWW} {W}
\aff {303} {W} {BRWWWWRWWW} {W}
\aff {304} {B} {WWWWBWBWWW} {B}
\aff {305} {W} {WWWWWWRBWW} {W}
\aff {306} {W} {WRWWBWWBBW} {W}
\aff {307} {W} {WBWWBWRBBW} {W}
}
\hfill
\vtop{\leftskip 0pt\parindent 0pt\hsize=76pt 
\aff {308} {W} {WRWWBRBWWW} {W}
\aff {309} {W} {WRBWBRBBBW} {R}
\aff {310} {W} {RRWWBWBWWW} {W}
\aff {311} {R} {WWBWBRBBBW} {W}
\aff {312} {R} {WBRBWWRWRW} {B}
\aff {313} {B} {RRWWWRWWBW} {R}
\aff {314} {W} {BWWWBWRBWW} {W}
\aff {315} {W} {WBWWBWRWWW} {W}
\aff {316} {W} {WWBRRBBBBW} {W}
\aff {317} {R} {WBWBWRBBWW} {W}
\aff {318} {B} {WRRBWWWWRW} {B}
\aff {319} {R} {BWWWWRWRBW} {B}
\aff {320} {R} {BWWWRBRWWW} {R}
}
\hfill
\vtop{\leftskip 0pt\parindent 0pt\hsize=76pt 
\aff {321} {B} {BRWBWWRWWW} {B}
\aff {322} {W} {RRWWBWWBWW} {W}
%}
%\hfill
%\vtop{\leftskip 0pt\parindent 0pt\hsize=76pt 
%\ligne{\hfill selected track\hfill}
\aff {323} {W} {RBWBWWWBWW} {R}
\aff {324} {W} {RRWWWWBWWW} {W}
\aff {325} {W} {BRWWWWBWWW} {W}
\aff {326} {B} {BWWRWRWWBW} {B}
\aff {327} {R} {BWWWWBBWWW} {R}
\aff {328} {W} {BWWWBRWBWW} {W}
\aff {329} {R} {BBWBWWWBWW} {W}
\aff {330} {W} {RBWWWWBRWW} {W}
\aff {331} {B} {RWWWBWBWWW} {B}
}
\hfill
\vtop{\leftskip 0pt\parindent 0pt\hsize=76pt 
\aff {332} {B} {BWWWWRWWBW} {B}
\aff {333} {W} {WBWWBRBWWW} {W}
\aff {334} {B} {WBRBWWRWRW} {B}
\aff {335} {B} {BRWWWRWWBW} {B}
\aff {336} {B} {BWWWWRWRBW} {B}
\aff {337} {W} {BRWWBWWBWW} {W}
}
\hfill
}
}
\vskip 10pt
Table~\ref{passmemotab1} gives the rules involved by the passage of the locomotive through
the controller-sensor while Table~\ref{passmemotab2} gives those which are connected with the
change of the sensor from blue to red. Table~\ref{exctrlblue} gives traces of execution for
the rules applied to the cells concerned when the sensor is blue while Table~\ref{exctrlred}
does the same for the case when the sensor is red.

%  2(1), 1(1), 4(1), 1(2), 2(2), 6(2) 
%   2(1),  choisi : 17, 18, 20, 32, 17, 17
%      non-choisi : 17, 18, 20, 32, 17, 17
%          change : 17, 17, 17
%   1(1),  choisi : 119, 119, 122, 124, 126, 119
%      non-choisi : 296, 

\vtop{
\begin{tab}\label{exctrlblue}
\leurre
Executions of the rules for the control of the passive memory switch when the sensor is blue.
\end{tab}
\vskip-6pt
\ligne{\hfill\hbox to 170pt{\hfill the locomotive passes\hfill}\hskip 15pt
\hbox to 120pt{\hfill change from blue to red\hfill}\hfill}
\ligne{\hfill
\vtop{\leftskip 0pt\parindent 0pt\hsize=176pt
% quand le contrôleur est bleu
\ligne{\hfill  \execsept     {}       0       1       2       3       4       5 \hfill}
\ligne{\hfill  \execsept  {6(1)}  {  18}  {  20}  {  32}  {  17}  {  17}  {  17}   \hfill}
\ligne{\hfill  \execsept  {2(1)}  {  17}  {  18}  {  20}  {  32}  {  17}  {  17}   \hfill}
\ligne{\hfill  \execsept  {1(1)}  { 119}  { 119}  { 122}  { 124}  { 126}  { 119}   \hfill}
\ligne{\hfill  \execsept  {4(1)}  {  17}  {  17}  {  17}  {  18}  {  20}  {  32}   \hfill}
\ligne{\hfill  \execsept  {2(2)}  { 332}  { 332}  { 332}  { 335}  { 336}  { 332}   \hfill}
}
\hskip 15pt
\vtop{\leftskip 0pt\parindent 0pt\hsize=125pt 
% contrôleur : de bleu à rouge
\ligne{\hfill  \execcinq     {}       0       1       2       3 \hfill}
\ligne{\hfill  \execcinq  {5(3)}  {  50}  {  51}  {  53}  { 307}   \hfill}
\ligne{\hfill  \execcinq  {2(3)}  {  49}  {  56}  {  69}  { 306}   \hfill}
\ligne{\hfill  \execcinq  {1(2)}  {  41}  { 342}  { 345}  { 347}   \hfill}
\ligne{\hfill  \execcinq  {1(3)}  {  31}  {  31}  {  45}  { 305}   \hfill}
}
\hfill}
}
\vskip 10pt
In the left-hand side sub-table of Table~\ref{exctrlblue}, the involved cells 
are 6(1), 2(1), 1(1) and 4(1) for the tracks and 2(2) as milestone. In the right-hand side
sub-table, the involved cells are 5(3) and 2(3) for the path which conveys the signal, the cell
1(2) as milestone and the cell 1(3) as witness of the changes. The cells of the tracks involve
cells which we already noticed in other parts of the circuit. The cell 2(2)~is a witness of the
passage of the locomotive: rule~332 is conservative, rule~335 can see two red cells: one is always
present as vertex-neighbour~6. It is a marker for the sensor. The other red cell noticed 
by rule~335 as neighbour~2 is the locomotive. Rule~336 can see the locomotive in 
vertex-neighbour~8, going further on its tracks.

In the right-hand side sub-table of Table~\ref{exctrlblue}, the cells which convey the signal
involve rules that we already seen. The cell~1(3) is a simple witness of the change in the cells
1(2), its vertex-neighbour~6, and in the cell 2(3), its neighbour~2. Rule~305 can see that
the cell 1(2) is now red. In 1(2), rule~41 is conservative: this rule is also used when 
two locomotives go down a vertical track. Rule 342 can see the approaching signal through
its vertex-neighbour~10. Rule~345 can see it through its neighbour~5 which makes it turn to red.
Rule~347 is the conservative rule for a red sensor as shown in Table~\ref{exctrlred}.

\vtop{
\begin{tab}\label{exctrlred}
\leurre
Execution of the rules for the controller of the passive memory switch when the sensor is red.
\end{tab}
\vskip -6pt
\ligne{\hfill
\vtop{\leftskip 0pt\parindent 0pt\hsize=186pt
\ligne{\hfill  \execsept     {}       0       1       2       3       4       5 \hfill}
\ligne{\hfill  \execsept  {6(1)}  {  18}  {  20}  {  32}  {  17}  {  17}  {  17}   \hfill}
\ligne{\hfill  \execsept  {2(1)}  {  17}  {  18}  {  20}  {  32}  {  17}  {  17}   \hfill}
\ligne{\hfill  \execsept  {1(1)}  { 296}  { 296}  { 309}  { 311}  { 316}  { 119}   \hfill}
\ligne{\hfill  \execsept  {4(1)}  {  17}  {  17}  {  17}  {  18}  { 317}  {  32}   \hfill}
\ligne{\hfill  \execsept  {2(2)}  { 298}  { 298}  { 298}  { 313}  { 319}  { 326}   \hfill}
\ligne{\hfill  \execsept  {6(2)}  { 302}  { 302}  { 302}  { 302}  { 323}  { 329}   \hfill}
\ligne{\hfill  \execsept  {1(2)}  { 347}  { 347}  { 347}  { 312}  { 318}  {  41}   \hfill}
}
\hfill}
}
\vskip 10pt
   In Table~\ref{exctrlred}, we can see the rules applied to the cells of the tracks,
the same as in Table~\ref{exctrlred} but we can also see the rules applied to the cells 2(2) 
and 6(2). The cell 2(2) is a blue milestone, but when it becomes red, rule~313, it plays the role
of a locomotive seen by the cell 6(2) to which rule~323 applies, making the locomotive enter
the cell 6(2). Rule~319 makes the cell 2(2) return to blue and it can also see the locomotive 
of the tracks through its vertex-neighbour~8, when the locomotive goes on its way on the tracks. 
Rule~326 can see the signal-locomotive in the cell 6(2) which is the neighbour~4 of the cell 2(2).
The cell 6(2) is applied the rule~302, a conservative rule. Rule~323 makes it turn to red,
as already seen. Rule~329 makes it turn back to white. Rule~351 witnesses that the 
signal-locomotive is in neighbour~3 of the cell 6(2), going further on its path to the
other controller-sensor.

\vtop{
\begin{tab}\label{passmemotab2}
\leurre
Rules for changing the selection in the passive memory switch.
\end{tab}
\vspace{-9pt}
\ligne{\hfill
\vtop{\leftskip 0pt\parindent 0pt\hsize=76pt 
\aff {338} {R} {BWWWRBBWWW} {R}
\aff {339} {B} {BRWBWWRWWR} {B}
\aff {340} {R} {BRWWWWBWWW} {R}
\aff {341} {W} {BWWWWBRWWW} {W}
}
\hfill
\vtop{\leftskip 0pt\parindent 0pt\hsize=76pt 
\aff {342} {B} {WBRBWWWWRR} {B}
\aff {343} {B} {BRWBRWRWWW} {B}
\aff {344} {W} {WBWWBWBRWW} {W}
\aff {345} {B} {WBRBRWWWRW} {R}
}
\hfill
\vtop{\leftskip 0pt\parindent 0pt\hsize=76pt 
\aff {346} {B} {BRWBWRRWWW} {B}
\aff {347} {R} {WBRBWWWWRW} {R}
\aff {348} {R} {RWWWRBBWWW} {R}
\aff {349} {B} {RRWBWWRWWW} {B}
}
\hfill
\vtop{\leftskip 0pt\parindent 0pt\hsize=76pt 
\aff {350} {R} {BRWWWWRWWW} {R}
\aff {351} {W} {BBRBWWWBWW} {W}
\aff {352} {B} {WWWWBRBWWW} {B}
}
\hfill}
}
\vskip 15pt
The tables of this section completes the proof of Theorem~\ref{letheo}.\hfill\boxempty

\section*{Conclusion}

There are several questions raised by this result. Is it possible to reduce the number
of states to two of them in this context? If it is possible, is it impossible to do so
in the pentagrid with \textit{\`a la} von Neumann neighbourhood, \textit{i.e.} with
considering as neighbours the cells which share an edge only? Still hard work ahead.
Nobody will complain at that...

\end{document}